\documentclass[12pt,a4paper]{article}
\usepackage{xcolor,paralist}
\usepackage[pdftex,colorlinks=true,hypertexnames=false]{hyperref}
\definecolor{darkblue}{rgb}{0,0,.6}
\hypersetup{citecolor=darkblue,linkcolor=darkblue,urlcolor=darkblue}
\usepackage{amssymb,enumitem,booktabs,subfig,setspace,bm,longtable,dsfont,orcidlink}
\usepackage[final,nopatch=footnote]{microtype}
\usepackage[top=1in, bottom=1in, left=1in, right=1in]{geometry}
\usepackage{natbib}
\usepackage{comment}
\usepackage{url}
\usepackage{amsmath}
\usepackage{amsfonts}
\usepackage{epsfig}
\usepackage{graphics}
\usepackage[normalem]{ulem}
\usepackage{palatino,eulervm}
\usepackage{graphicx}
\usepackage{lscape}
\usepackage{rotating}
\usepackage[linewidth=1pt]{mdframed}
\allowdisplaybreaks[4]
\DeclareMathOperator*{\argmin}{arg\,min}

\providecommand{\U}[1]{\protect\rule{.1in}{.1in}}

\graphicspath{{plots/}}
\newsavebox\CBox

\usepackage{amsthm,thmtools}
\usepackage{mathrsfs}

\makeatletter
\def\th@newremark{\th@remark\thm@headfont{\bfseries}}
\makeatletter
\theoremstyle{newremark}

\declaretheoremstyle[
  spaceabove=6pt, spacebelow=6pt,
  headfont=\bfseries,
  notefont=\mdseries, notebraces={(}{)},
bodyfont=\normalfont,
  postheadspace=0.5em
]{mystyle}

\newcommand{\X}{\mathcal{X}}

\bibpunct{(}{)}{;}{a}{,}{ and }

\makeatletter
\newcommand*{\addFileDependency}[1]{
\typeout{(#1)}
\@addtofilelist{#1}
\IfFileExists{#1}{}{\typeout{No file #1.}}
}\makeatother

\newcommand{\Rlogo}{\protect\includegraphics[height=1.8ex,keepaspectratio]{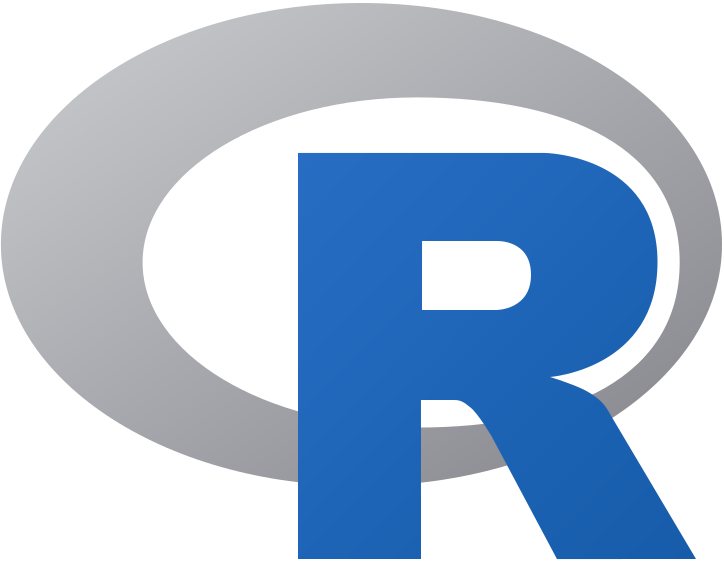}}

\begin{document}

\title{Modeling and forecasting subnational age distribution of death counts}
\author{\normalsize Han Lin Shang \orcidlink{0000-0003-1769-6430} \\
\normalsize Department of Actuarial Studies and Business Analytics \\
\normalsize Macquarie University \\
\\
\normalsize Cristian F. Jim\'{e}nez-Var\'{o}n \orcidlink{0000-0001-7471-3845}\footnote{Corresponding postal address: Department of Mathematics, Heslington, York YO10 5DD, United Kingdom; Email: cristian.jimenezvaron@york.ac.uk} \\
\begin{minipage}{0.42\textwidth}
\centering
\normalsize Department of Mathematics \\
\normalsize University of York\\
\end{minipage}
}

\date{}
%\date{\today}
%\date{\normalsize This version: \today}

\maketitle

\centerline{\bf Abstract}

\medskip

Existing mortality forecasting methods focus on age-specific mortality rates, which lie in an unconstrained space and overlook the distributional nature of life-table death counts. Few studies have developed and compared forecasting methods that model the shape and dynamics of the age distribution of deaths, especially at the subnational level, where data quality varies greatly. This paper presents several forecasting methods to model and forecast the subnational age distribution of death counts. The age distribution of death counts has many similarities to probability density functions, which are non-negative and have a constrained integral, and thus live in a constrained nonlinear space. To address the nonlinear nature of objects, we implement a cumulative distribution function transformation that is scale-free and has additional monotonicity. Using subnational Japanese life-table death counts from the \cite{JMD25}, we evaluate the forecast accuracy of the transformation and forecasting methods. The improved forecast accuracy of life-table death counts implemented here will be of great interest to demographers in estimating regional age-specific survival probabilities and life expectancy, and to actuaries as a foundation for exploring potential applications in determining annuity prices for various ages and maturities.

\medskip


\newpage
\setstretch{1.5}

\section{Introduction}\label{sec:1}

Accurate prediction of subnational mortality rates plays a crucial role in both actuarial studies and demographic planning. From a demographic perspective, subnational mortality forecasts inform regional health planning, pension sustainability assessments, and population projections that guide policy decisions on healthcare infrastructure and resource allocation \citep[see, e.g.,][]{BT08}. In actuarial science, understanding spatial variations in mortality and their evolution over time is highly relevant for pricing life insurance products, designing annuities, and managing longevity risk in pension schemes \citep[e.g.,][]{CBD06, HR11}. Subnational mortality prediction allows actuaries and policymakers to capture heterogeneity across regions, reflecting socio-economic, environmental, and healthcare disparities, improving the reliability of longevity projections \citep{LL05}.

Although many multipopulation techniques exist for modeling and forecasting age-specific mortality rates, such as \citeauthor{LL05}'s \citeyearpar{LL05} method, the age distribution of death counts is a mortality instrument with unique advantages over mortality rates. For example, the age distribution of death counts preserves distributional information, for instance, the age distribution of deaths retains the shape of how deaths are spread across ages. Due to \textit{non-negativity} and \textit{summability}, the changes are relative in age. The age distribution of death counts captures shifts in longevity more directly, for instance, the mode and spread of the distribution can be measured. It is less sensitive to population size estimates, which can be unreliable at the subnational level; this robustness facilitates more consistent comparisons across different populations and time periods. Furthermore, while mortality rates at older ages can occasionally fall outside the plausible $(0, 1)$ range, the age distribution of death counts remains inherently constrained within a valid probability space. The age distribution of death counts is an explicit way of measuring the probability of dying $q_x$, commonly studied in \citeauthor{CBD06}'s \citeyearpar{CBD06} model.

Following the early works of \cite{LHV+20, LHW21} on subnational mortality rates and \cite{BSO+18} on subnational age distribution of death counts, we present several forecasting methods to model and predict the subnational age distribution of deaths, which may offer useful insights into potential applications in actuarial science, such as fixed-term or lifetime annuity pricing \citep[see, e.g.,][]{SH20}. Apart from subnational age distributions of deaths in demography \citep{JSS25}, density-valued objects for multiple groups are common, with examples including income distributions across different populations \citep{KU01}, financial return distributions for multiple stocks \citep{PZK22}, and distributions of bidding times in online auctions for various items \citep{WJS08}, among others.

Since age distributions of deaths naturally take the form of density functions, they are well-suited to modeling using advanced forecasting techniques from compositional and functional data analyses. \cite{Oeppen08} demonstrates that the use of compositional data analysis to forecast the age distribution of death counts performs similarly in accuracy to forecast age-specific mortality rates. Within compositional data analysis, the Lee-Carter model is frequently applied to model and forecast unconstrained data. \cite{SH20} extend this approach by incorporating a functional data-analytic perspective, including multiple functional principal components and nonparametric smoothing. Recently, \cite{SH24} introduce a cumulative distribution function (CDF) transformation, converting each year's age distribution of deaths into a probability bounded within the unit interval to address the presence of zero values. Through cumulative summation, a probability density function can be transformed into a CDF, providing the additional benefit of monotonicity \citep[see also][]{MS13}. With a time series of CDFs, it is common to model and extrapolate the pattern using a logistic transformation.

Multiple density-valued objects observed over time are related to high-dimensional functional time series (HDFTS). In the statistical literature, \cite{ZD23} derive Gaussian and multiplier bootstrap approximations for the sums of HDFTS. With these approximations, they construct joint simultaneous confidence bands for the mean functions and develop a hypothesis to test whether the mean functions in the cross-sectional dimension exhibit parallel behavior. \cite{HNT231} investigate the representation of HDFTS using a functional factor model and determine conditions on the eigenvalues of the covariance operator that are crucial for establishing the existence and uniqueness of the factor model. \cite{GSY19} adopt a two-stage approach, combining truncated principal component analysis and a separate scalar factor model for the resulting panels of scores. \cite{HNT232} introduce a functional factor model with a functional factor loading and a vector of real-valued factors, while \cite{GQW24} propose a functional factor model with a real-valued factor loading and a functional factor. \cite{LLS+24} introduce a unified framework that accommodates both types of factor models. \cite{TSY22} study clustering for age-specific subnational mortality rates, which is an example of HDFTS. \cite{LLS24} introduce hypothesis tests and estimation procedures for testing and locating (common) change points. \cite{CFQ+25} develop a two-stage procedure for modeling and forecasting HDFTS.

Numerous studies have addressed the modelling and forecasting of age distributions of death counts at the national level \citep[see, e.g.,][]{Shang25, SH25, SH24}. However, there remains a notable gap in extending such forecasts to the subnational level. Our contributions are twofold. First, we present several visualization techniques to capture patterns in the age distribution of death counts and to display contrasts between subnational and national data. Second, we revisit several forecasting methods and compare their point and interval forecast accuracy with applications to subnational life-table death counts. For different sexes and prefectures, we provide some recommendations on the optimal multipopulation models, which can vary in terms of their point and interval forecast accuracies. While the primary focus of this paper is on statistical accuracy, an improvement in forecast accuracy of life-table death counts represents a preliminary step that could eventually lead to more informed annuity pricing and life expectancy calculations. In turn, these insights can inform policymakers on the sustainability of pension systems and aged care facilities.

Our paper is structured as follows. In Section~\ref{sec:2}, we describe our motivating data set and introduce a series of image plots for displaying important features in HDFTS. In Section~\ref{sec:3}, we present the CDF transformations. Within the CDF transformation, we consider a number of forecasting methods in Section~\ref{sec:4} to model and forecast unconstrained functional time series (FTS) data. In Section~\ref{sec:5}, we propose two general strategies for constructing pointwise prediction intervals for the age distribution of death counts. In Section~\ref{sec:6}, we evaluate and compare point forecast accuracy using the Kullback-Leibler divergence (KLD) and Jensen-Shannon divergence (JSD), and interval forecast accuracy using the coverage probability difference (CPD) and mean interval score. We show a relationship between life-table death count and survival probability and life expectancy. We present a model diagnostic analysis using functional analysis of variance. Section~\ref{sec:7} concludes with discussion of how the methodology presented can be extended.

\section{Subnational age distribution of death in Japan}\label{sec:2}

In many developed countries, such as Japan, increases in longevity and an aging population have led to concerns about the sustainability of pension, health, and aged care systems \citep[see, e.g.,][]{Coulmas07}. These concerns have led to a surge of interest among government policymakers in accurately modeling and forecasting age-specific mortality. Subnational forecasts of age-specific mortality are useful for informing policy at the local level, and any improvement in forecast accuracy is beneficial for allocating current and future resources at national and subnational levels.

In demography, the age distribution of deaths provides important insights into longevity and lifespan variability that cannot be grasped directly from the central mortality rate or survival function. As pointed out by \cite{Oeppen08}, the age distribution of death counts is more suitable than central mortality rates for computing life expectancy and annuity premia. In addition to providing an informative description of the mortality experience of a population, life-table death counts yield readily available information on the ``central longevity indicators" \citep[e.g., mean, median, and modal age at death, see][]{CRT+05, CR10} and lifespan variability \citep[e.g.,][]{VC13, AV18}. The life-table death count is less sensitive to population size estimate, especially at the subnational level. Because of this, it is insensitive to population size estimates, which facilitate comparisons across populations and over time. The life-table death count is directly linked to the probability of dying~$q_x$, which is the mortality instrument in \citeauthor{CBD06}'s \citeyearpar{CBD06} model.

We study Japanese life-table death counts from 1975 to 2023, obtained from the \cite{JMD25}. We observe complete life-table death counts for ages from 0 to 109 in single years of age, with the last age group being age 110+. A distinct advantage of life-table death counts lies in the availability of data observed across all ages.
\begin{figure}[!htb]
\centering
\subfloat[Japanese female data]
{\includegraphics[width=8.3cm]{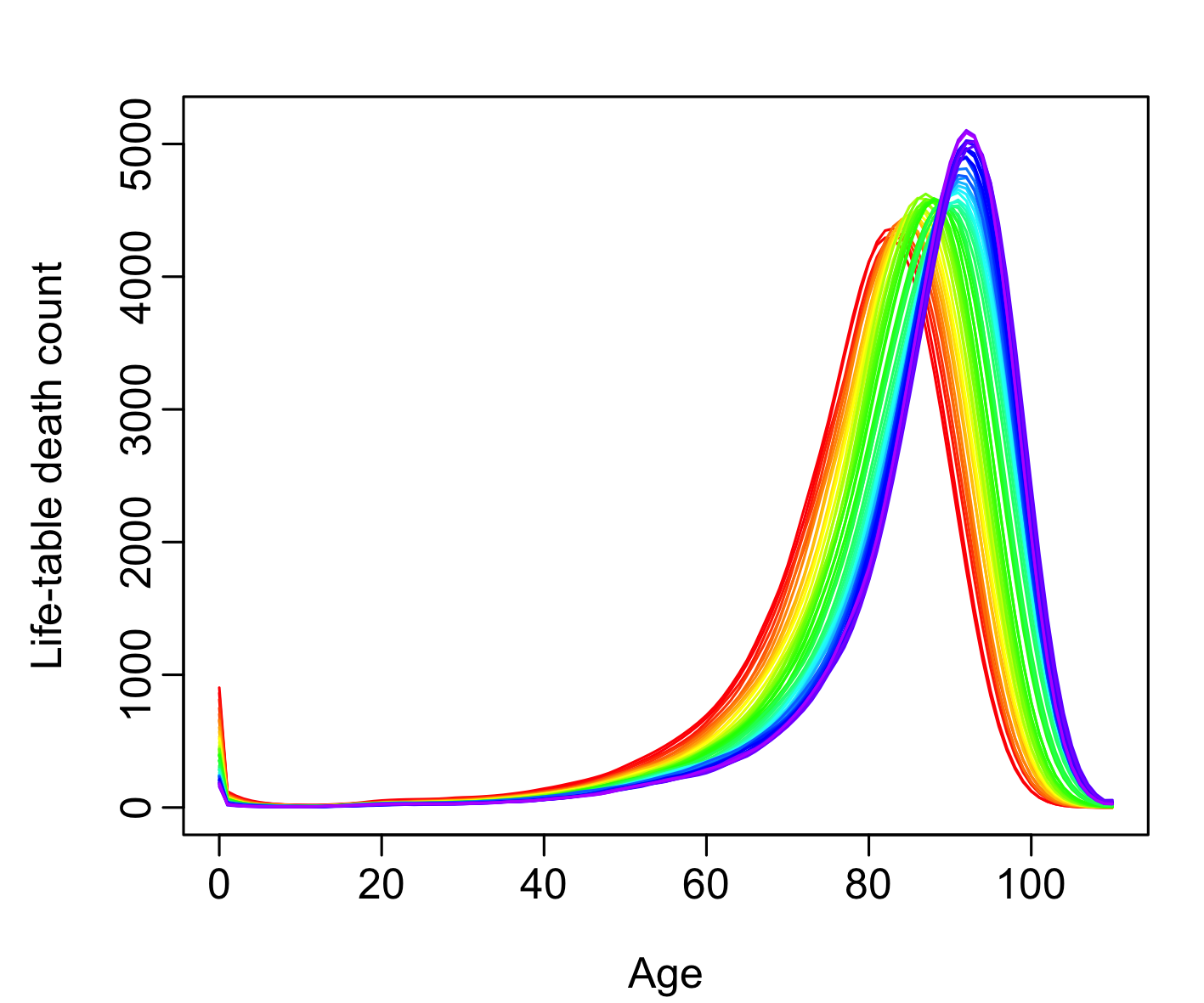}}
\quad
\subfloat[Japanese male data]
{\includegraphics[width=8.3cm]{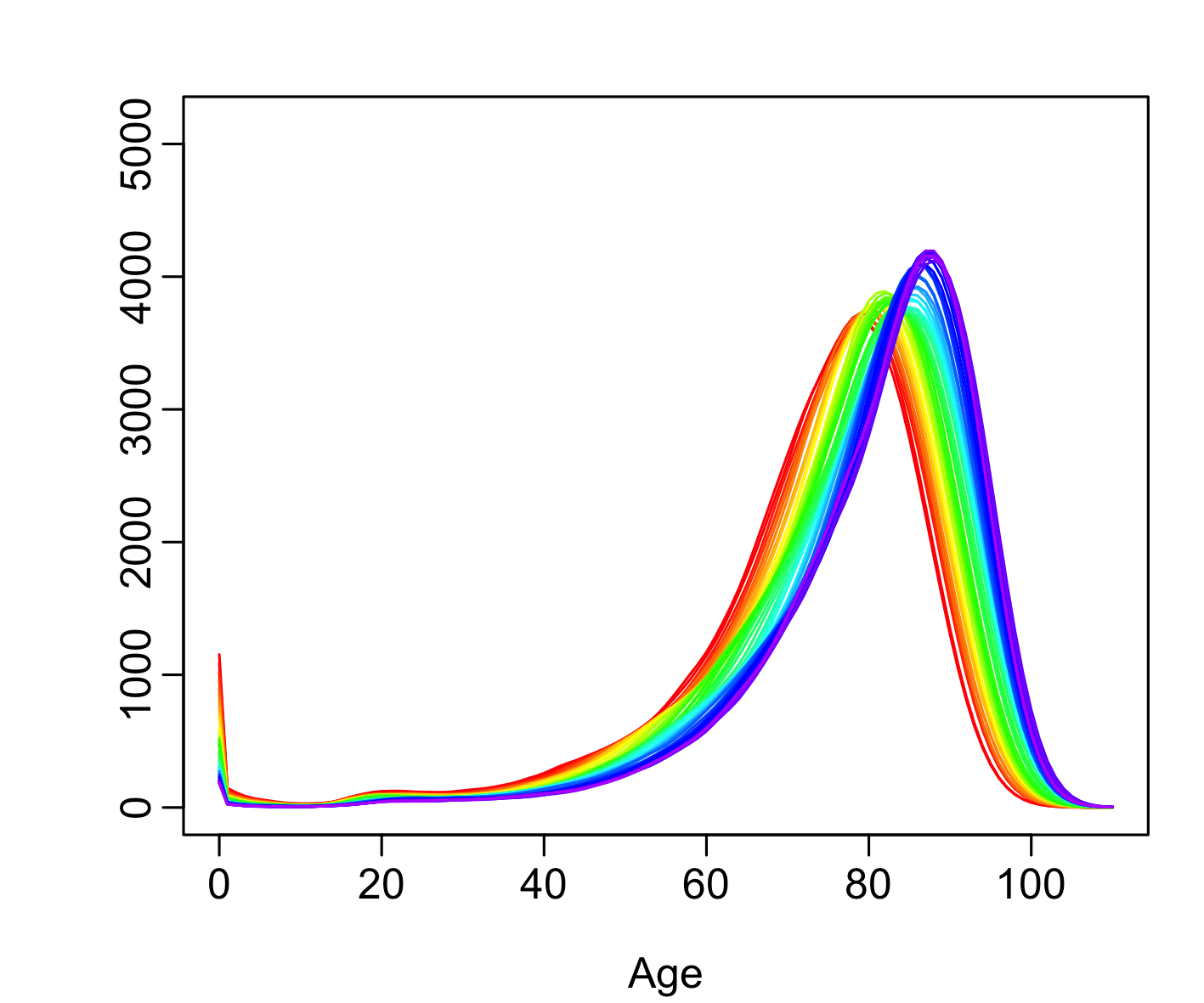}}
\caption{\small{Functional time series graphical displays of age-specific life-table death counts from 1975 to 2023 in a single-year group. The oldest years are shown in red, with the most recent years in violet. Curves are ordered chronologically according to the colors of the rainbow.}}\label{fig:1}
\end{figure}

Figure~\ref{fig:1} shows rainbow plots of the female and male age-specific life-table death counts in Japan from 1975 to 2023. The time ordering of the curves follows the color order of a rainbow, where curves from the distant past are shown in red, and the more recent curves are shown in violet. The figures show typical mortality curves for a developed country, with a declining trend in infant mortality. A typical negatively skewed distribution of life-table death counts is apparent, with the peaks shifting to higher ages for both females and males. This shift highlights the importance of accurate mortality forecasting for managing longevity risk, a core concern for the risk management of annuity products \citep[see][for a discussion]{DDG07}. In addition, the spread of the distribution indicates the variability in lifespan. A decrease in variability over time can be directly observed and quantified using the Gini coefficient \citep[]{WH99, DCH+17}.
\begin{figure}[!htb]
\centering
\includegraphics[width=0.36\linewidth]{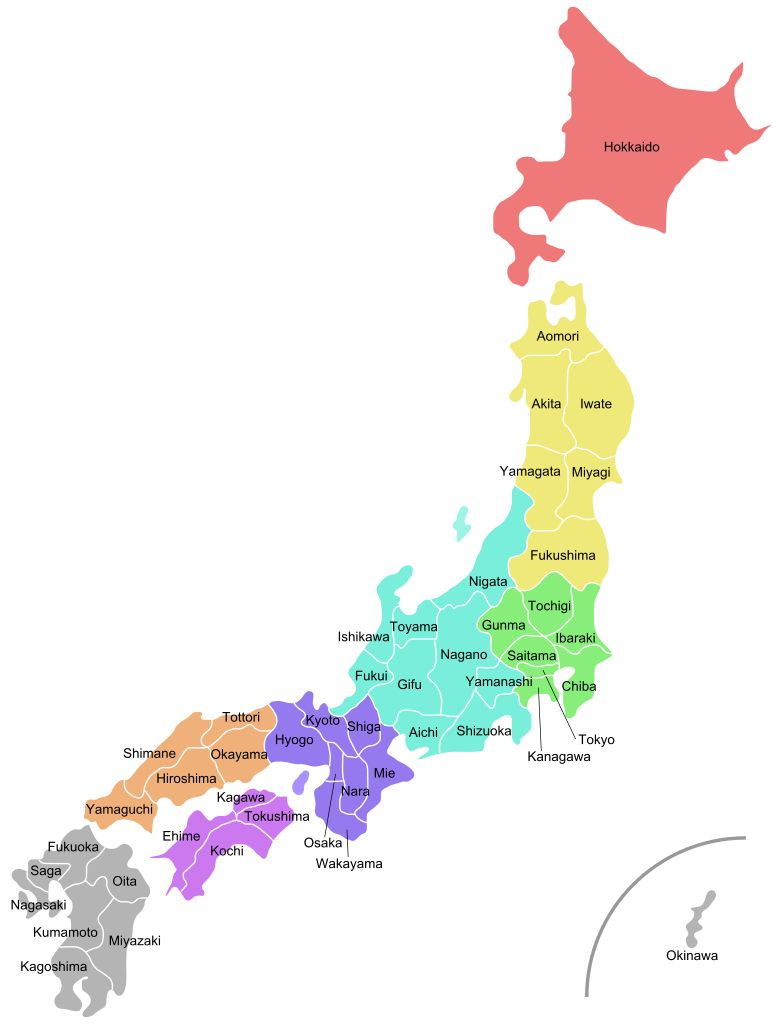}
\caption{\small A geographical map of 47 prefectures in Japan.}\label{fig:3}
\end{figure}

In Figure~\ref{fig:3}, we present a map of 47 Japanese prefectures from the northern region, Hokkaido, to the southern region, Okinawa. Due to geography and socio-economic status, the age distributions of death counts tend to be heterogeneous. Differences between prefectures or years within the same prefecture can be visualized using image plots \citep[see also][]{SH17}.

To measure the distance between two probability densities, such as the subnational and national age distributions of death counts, we consider the symmetric KLD \citep[see][for examples of using the KLD to compare country-specific lifespan distributions]{ET05, DEI14}. For a given prefecture and gender, the symmetric KLD can be expressed as
\begin{align*}
KLD_{s}^{g} = \ & D_{KL}\left[d_{t,s}^{g}(u)|d_{t,\text{national}}^{g}(u)\right] + D_{\text{KL}}\left[d_{t,\text{national}}^{g}(u)|d_{t,s}^{g}(u)\right] \\
=\ &\frac{1}{111\times T}\sum^T_{t=1}\sum_{i=1}^{111}d_{t,s}^{g}(u_i)\times [\ln d_{t,s}^{g}(u_i) - \ln d_{t,\text{national}}^{g}(u_i)] + \\
&\frac{1}{111\times T}\sum^T_{t=1}\sum_{i=1}^{111}d_{t,\text{national}}^{g}(u_i)\times [\ln d_{t,\text{national}}^{g}(u_i) - \ln d_{t,s}^{g}(u_i)],
\end{align*}
where $d_{t,s}^g(u)$ denotes the life-table death count at age $u$, prefecture $s=1,\dots,S$ for gender $g$ in year $t=1,\dots,T$, $S$ denotes the number of prefectures and $T$ denotes the number of years in the data set. 

\begin{figure}[!htb]
\centering
\includegraphics[width=5.9cm]{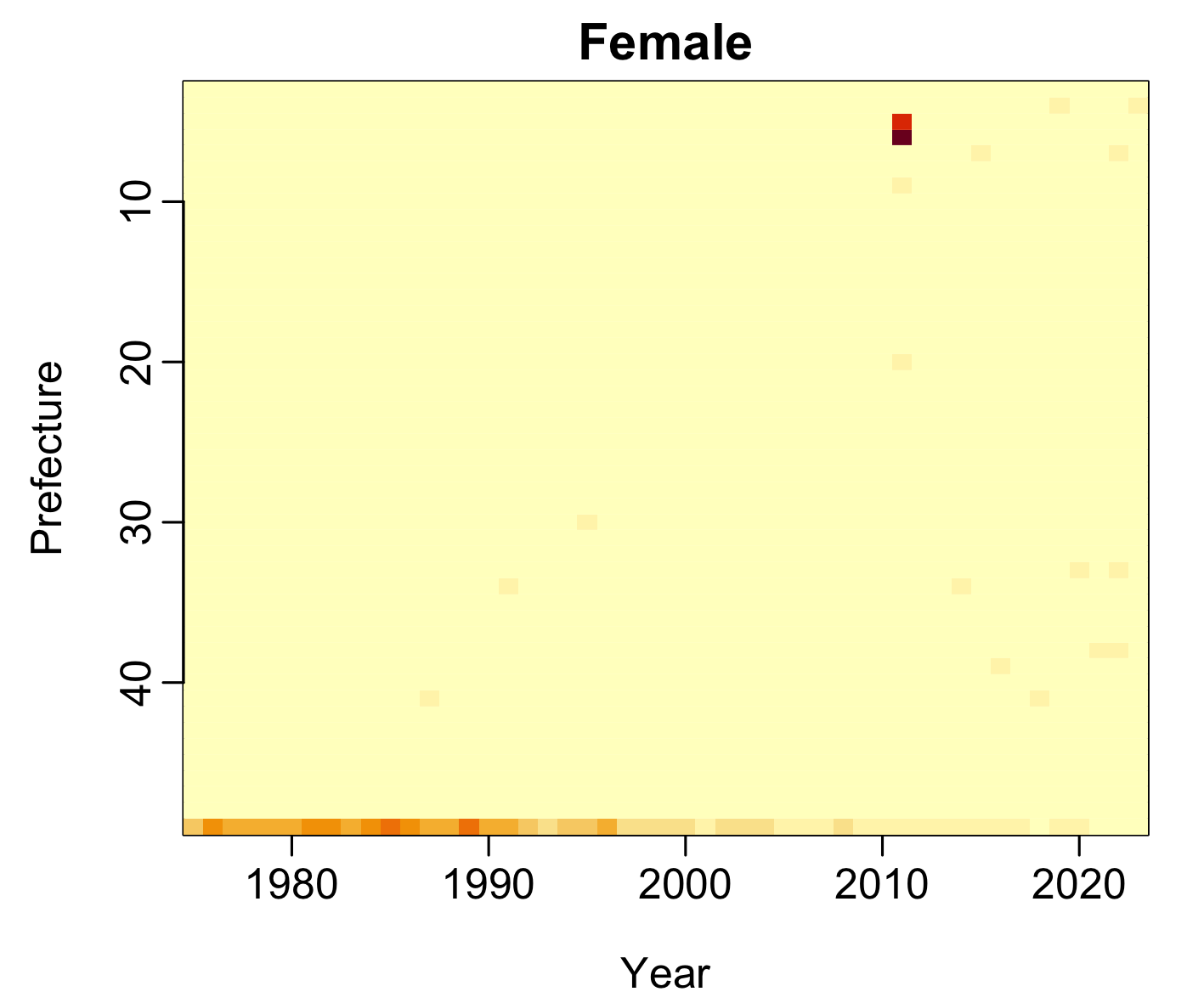}
\includegraphics[width=5.9cm]{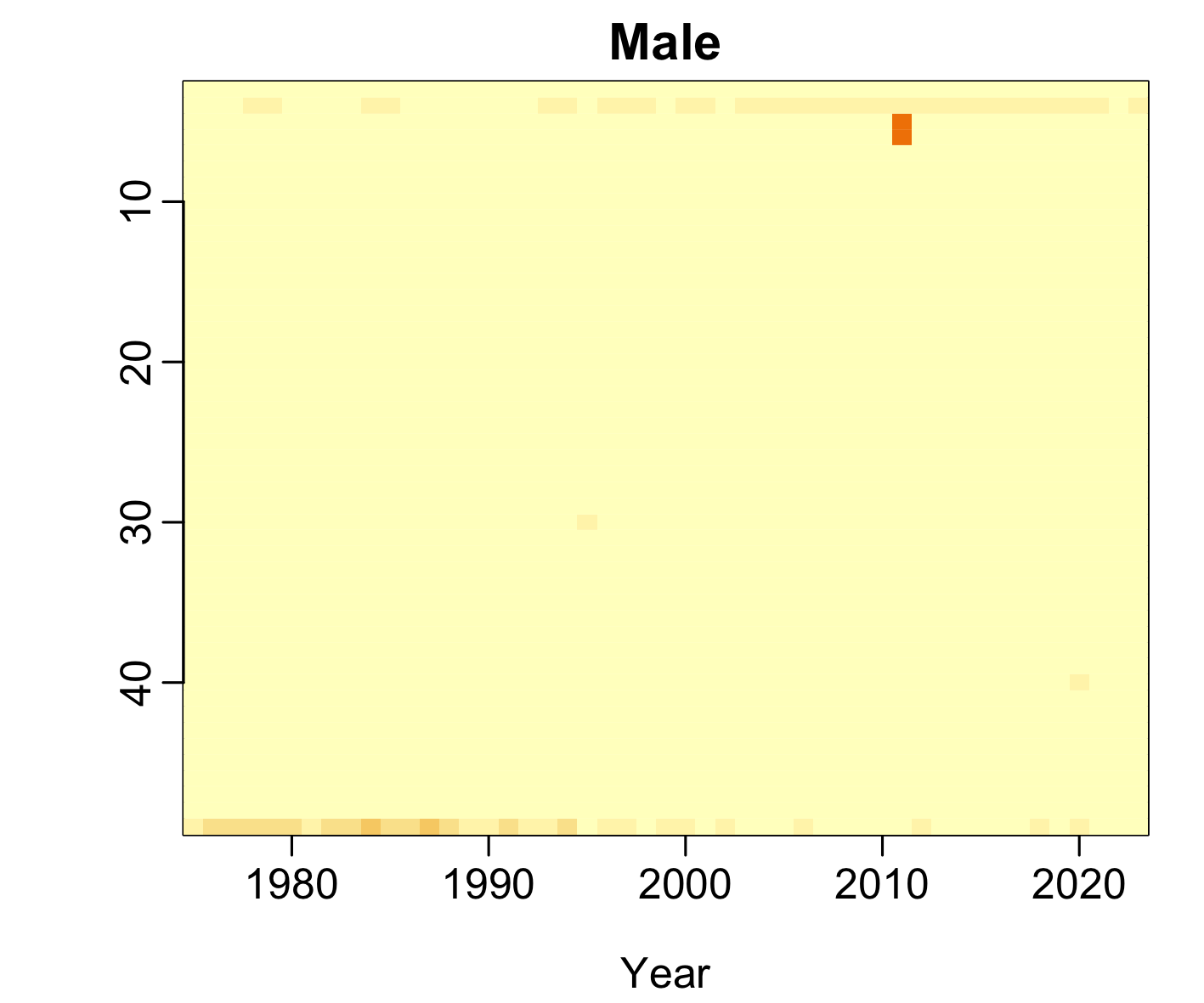}
\includegraphics[width=5.9cm]{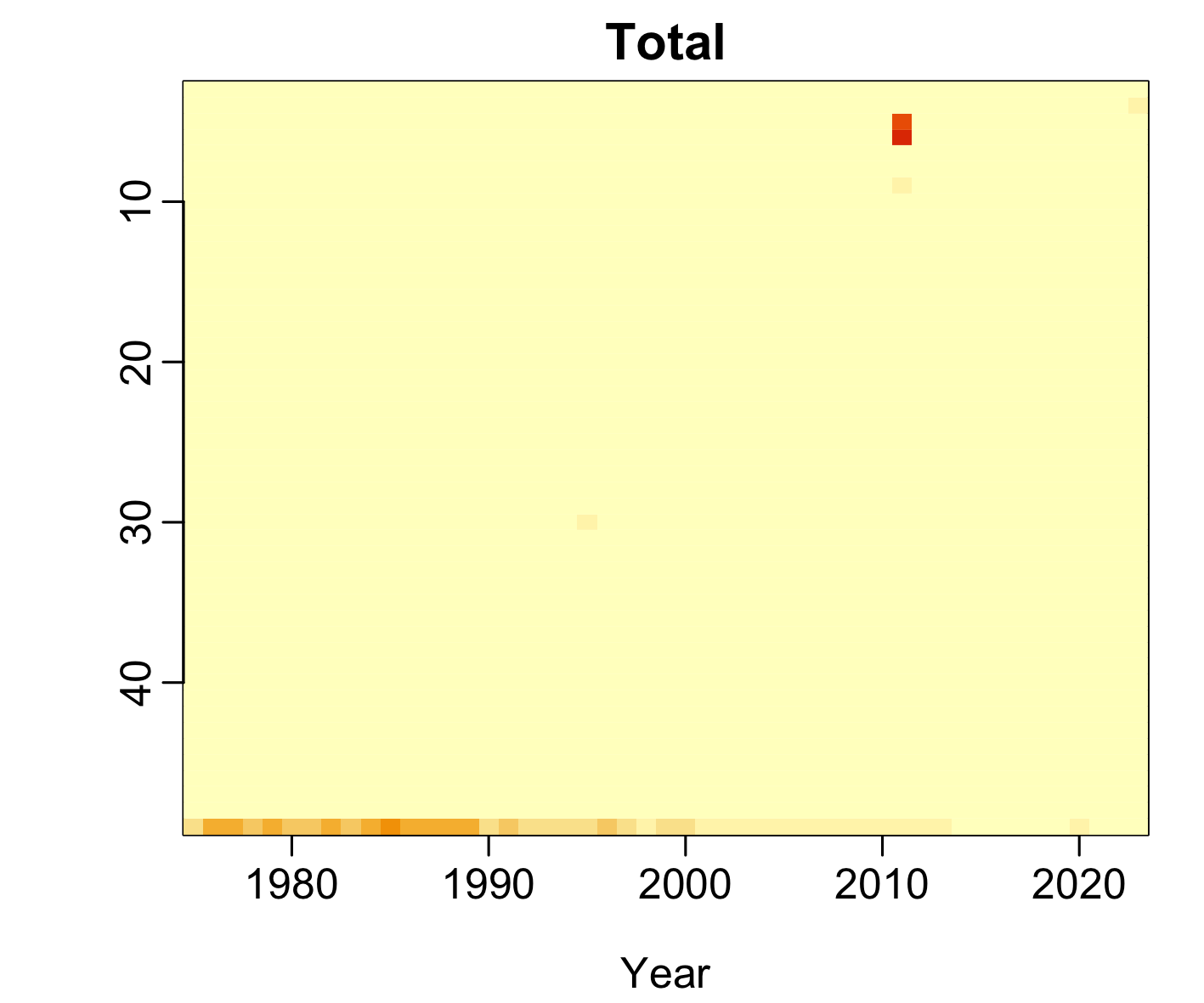}
\\
\includegraphics[width=5.9cm]{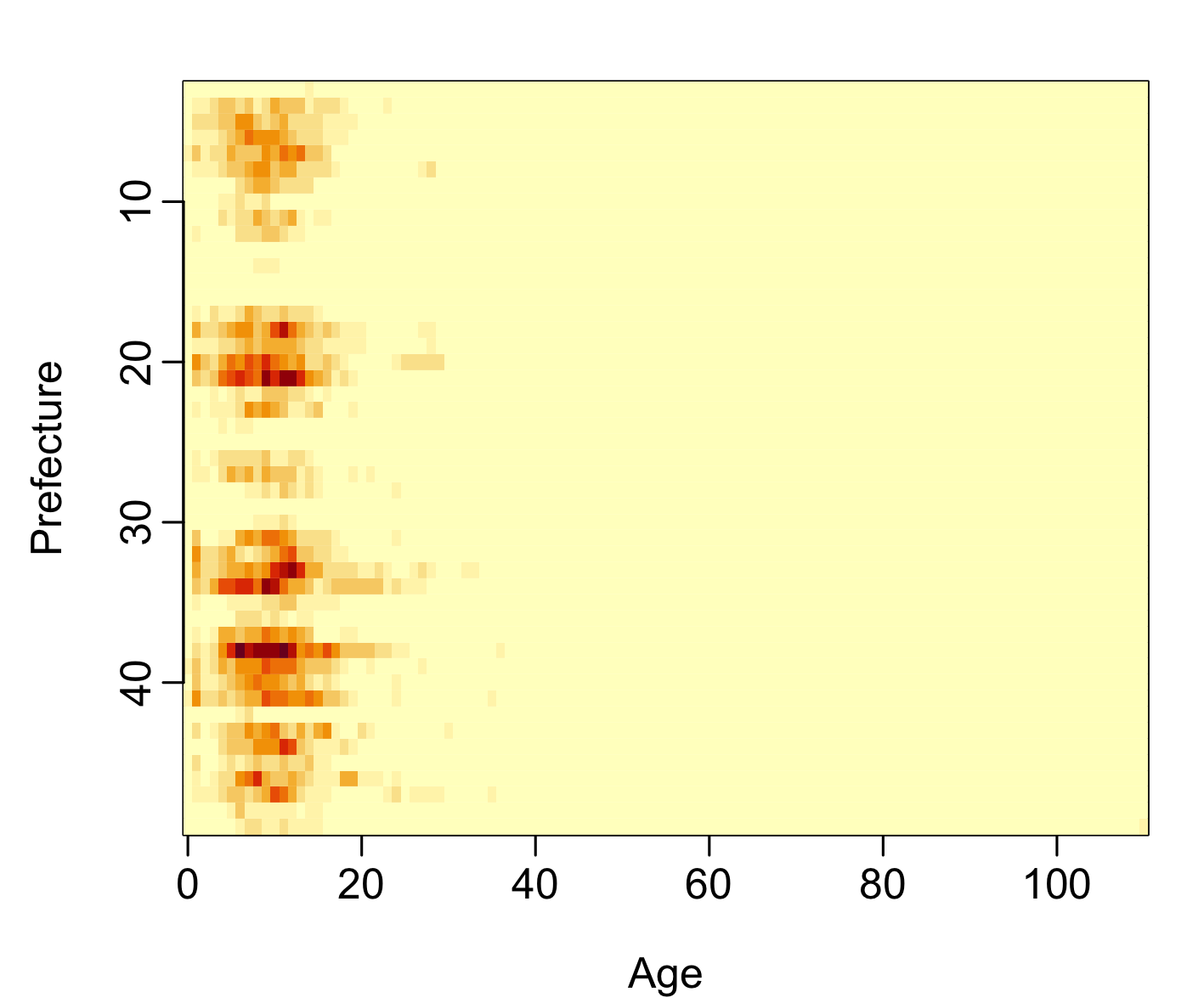}
\includegraphics[width=5.9cm]{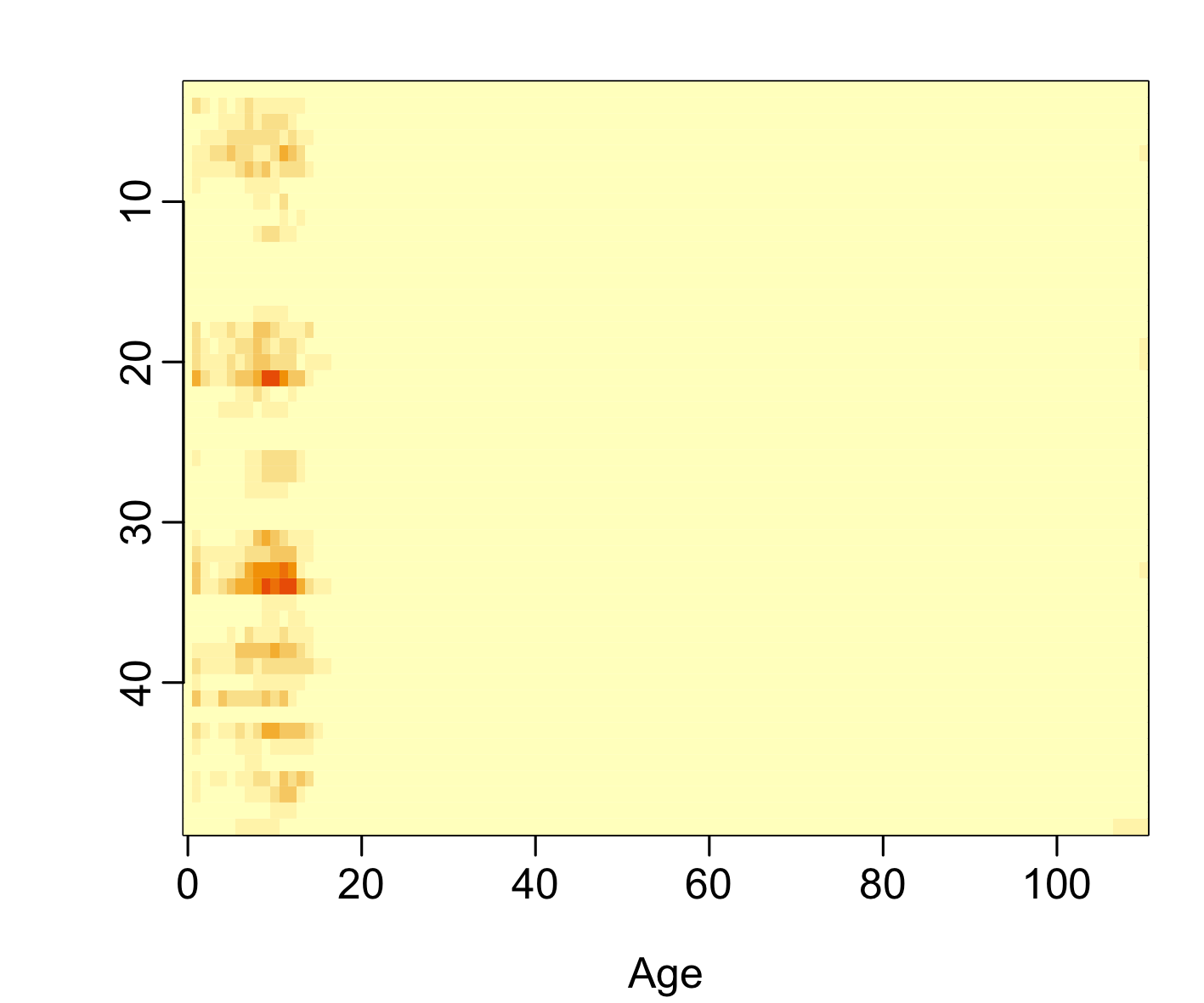}
\includegraphics[width=5.9cm]{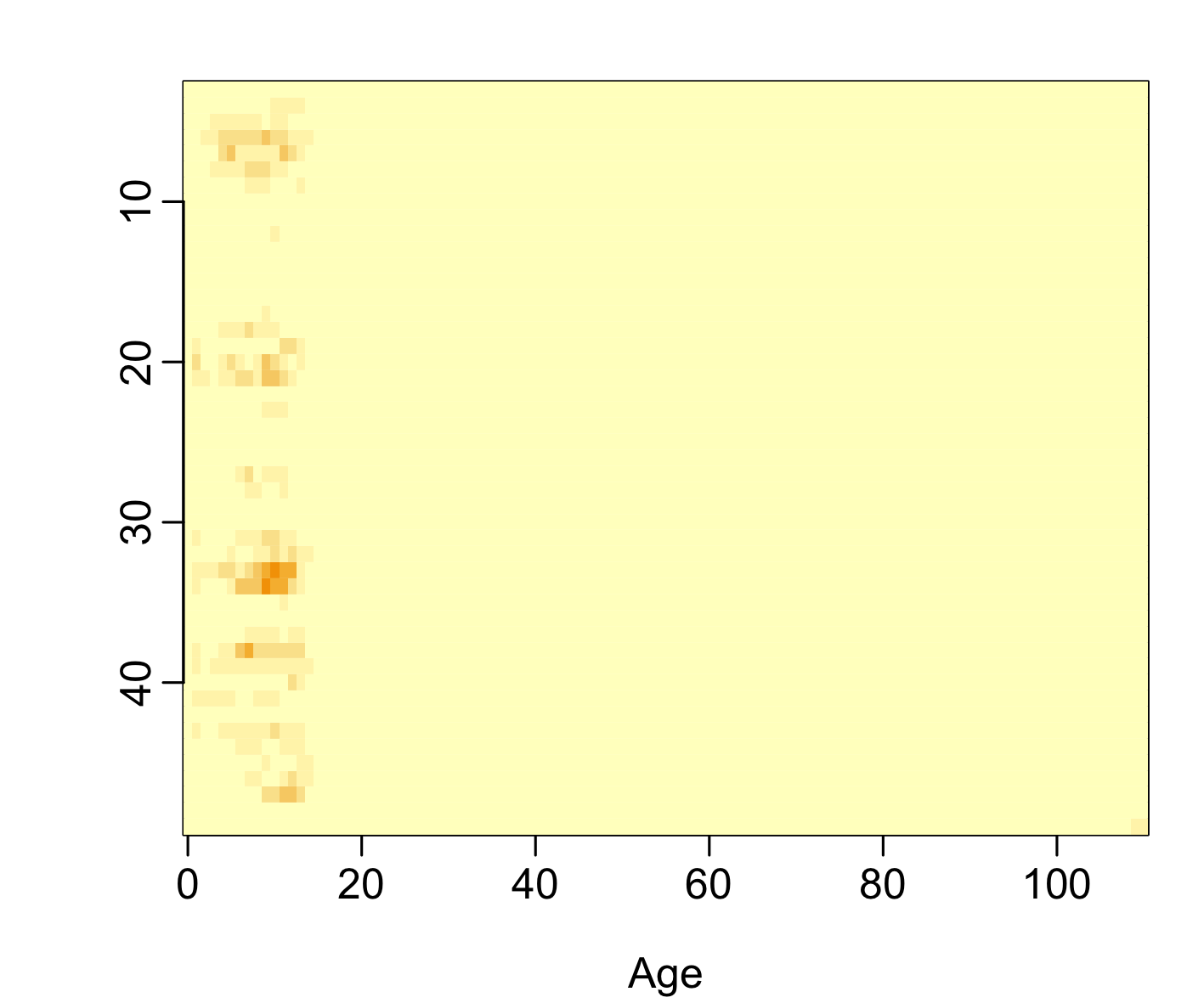}
\caption{\small{To illustrate regional heterogeneity, image plots show the symmetric KLD between life-table death counts of a prefecture and life-table death counts of the whole country. If the life-table death counts for a prefecture greatly differ from the national data, the image plot will show this in a darker red. The top panel shows mortality averaged over ages, while the bottom panel shows mortality averaged over years. Prefectures are numbered geographically from north to south.}}\label{fig:2}
\end{figure}

In Figure~\ref{fig:2}, we graph the symmetric KLD between life-table death counts for each prefecture to life-table death counts for the whole country, thus allowing relative mortality comparisons to be made. A divergent color palette is used, with red representing larger KLD values (i.e., greater difference between subnational and national data) and yellow denoting smaller KLD values. The prefectures are ordered geographically from north to south, so the most northerly prefecture (Hokkaido) is at the top of the panels, and the most southerly prefecture (Okinawa) is at the bottom.

The top row of the panels shows mortality for each prefecture and year, averaged over all ages. In 2011, in Iwate, Miyagi, and Fukushima, there was a large increase in mortality compared to other prefectures. These are northern coastal regions, and the elevated relative mortality is due to the tsunami of 11 March 2011 \citep[see, e.g.,][]{NI13} caused by a magnitude--9.0 earthquake. 

In 1995, there was an increase in mortality for Hyogo, which corresponds to the Kobe (Great Hanshin) earthquake (magnitude~6.9) of 17 January 1995. The mortality difference in Okinawa up to 2000 is also evident, suggesting a decline in the relative mortality advantages enjoyed by residents in this region.

The bottom of the panels shows the mortality for each prefecture and age, averaged over all years. Several striking features are apparent, including regional differences in mortality among children and adolescents; this may be due to socio-economic disparities and the accessibility of health services \citep{TMK23, OYK+25}.

\section{Cumulative distribution function transformation}\label{sec:3}

To model life-table death counts, there exists a range of transformations, including centered log-ratio transformation \citep{Aitchison86}, $\alpha$-transformation \citep{SH25b}, and CDF transformation \citep{SH24}. Among the three, the CDF transformation can handle not only zero counts but also the monotonicity constraint. In the CDF transformation, we first convert an age distribution of death for each year to a probability by dividing its radix $10^5$. Via the cumulative sum, we transform a probability density function into a CDF,
\begin{equation*}
D_{t,s}^{g}(u_x) = \sum^{x}_{i=1}d_{t,s}^{g}(u_i), \quad x=1,\dots,111, \quad t=1,\dots,T,
\end{equation*}
where $D_{t,s}^{g}(u_{111}) = 1$. In doing so, it enjoys the additional benefit of monotonicity \citep[see also][]{MS13}.

Then, we perform a logistic transformation,
\begin{equation*}
\X_{t,s}^g(u_x) = \text{logit}[D_{t,s}^g(u_x)] = \ln \Big[\frac{D_{t,s}^g(u_x)}{1-D_{t,s}^g(u_x)}\Big],
\end{equation*}
where $\ln(\cdot)$ denotes the natural logarithm. Since $D_{t,s}^g(u_{111})=1$, the last column is removed to avoid the undefinedness of the logistic transformation. By considering age as a continuum, the object $\X_{t,s}^g(u)$ is an example of a high-dimensional functional time series.

To transform $\X_{t,s}^g(u_x)$ back to the original scale, we first perform an inverse logit transformation, obtaining
\begin{equation*}
D_{t,s}^g(u_x) = \frac{\exp^{\X_{t,s}^g(u_x)}}{1+\exp^{\X_{t,s}^g(u_x)}},
\end{equation*}
where we add back the last column of ones. Then, we take the first-order differencing to obtain
\begin{equation*}
d_{t,s}^g(u_x) = \Delta_{i=1}^x D_{t,s}^g(u_i)\times 10^5,
\end{equation*}
where $\Delta$ represents the first-order differencing, where $\Delta_{i=1}^1D_{t,s}^g(u_1) = D_{t,s}^g(u_1)$. The constant~$10^5$ is the radix of the life-table death count. Since the data are observed discretely, we use cumulative sums and first-order differences to approximate integration and differentiation, respectively.

\section{Functional time-series forecasting models}\label{sec:4}

We obtain a set of unconstrained functional time series $(\X_{1,s}^g(u),\dots,\X_{T,s}^g(u))$ via the CDF transformation. Generally, we represent it as $\{\X_{t,s}^g(u)\}$, which is an arbitrary functional time series defined in a common probability space. It is assumed that the observations $\X_{t,s}^g(u)$ are elements of the Hilbert space $\mathcal{H}=\mathcal{L}^2(\mathcal{I})$ equipped with the inner product $\langle w, v\rangle = \int_{\mathcal{I}}w(u)v(u)du$, where $u$ represents a continuum and $\mathcal{I}\subset \mathds{R}$ denotes a function support range and $\mathds{R}$ is the real line. Each function is a square-integrable function that satisfies $\|\X_{t,s}^g\|^2 = \int_{\mathcal{I}}(\X_{t,s}^{g}(u))^2du <\infty$ and the associated norms.

\subsection{Univariate functional time-series (UFTS) forecasting method}\label{sec:4.1}

We assume $\X_{t,s}^g(u)$ has a finite mean and variance. Then, its non-negative definite covariance function is given by
\begin{align*}
c_s^g(u,v) &:= \text{Cov}[\X_{s}^g(u), \X_{s}^g(v)] \\
	&= \text{E}\left\{[\X_{s}^g(u) - \mu_s^g(u)][\X_{s}^g(v)-\mu_s^g(v)]\right\}, \qquad \forall u,v\in \mathcal{I},
\end{align*}
where $\mu_{s}^g(u) = \text{E}\{\X_{t,s}^g(u)\}$. By Mercer's lemma, there exists a set of orthonormal sequences $(\phi_k)$ of continuous functions in $\mathcal{L}^2(\mathcal{I})$ and a non-increasing sequence $\lambda_{k,s}^g$ of positive numbers such that
\begin{equation*}
c_s^g(u,v) = \sum^{\infty}_{k=1}\lambda_{k,s}^g \phi_{k,s}^g(u)\phi_{k,s}^g(v),
\end{equation*}
where $\lambda_{k,s}^g$ and $\phi_{k,s}^g(u)$ denote the $k$\textsuperscript{th} eigenvalue and eigenfunction, respectively. The Karhunen-Lo\`{e}ve expansion of a stochastic process $\X_{t,s}^g(u)$ can be expressed as
\begin{align*}
\X_{t,s}^g(u) &= \mu_s^g(u) + \sum_{k=1}^{\infty}\beta_{t,k,s}^g\phi_{k,s}^g(u) \\
&=\mu_s^g(u) + \sum_{k=1}^{K_s^g} \beta_{t,k,s}^{g}\phi_{k,s}^{g}(u) + e_{t,s}^{g}(u),
\end{align*}
where the principal component score $\beta_{t,k,s}^g$ is given by the projection of $[\X_{t,s}^{g}(u) - \mu_{s}^{g}(u)]$ in the direction of the $k$\textsuperscript{th} eigenfunction $\phi_{k,s}^g(u)$, that is, $\beta_{t,k,s}^{g} = \langle \X_{t,s}^g(u) - \mu_{s}^{g}(u), \phi_{k,s}^g(u)\rangle$; and $e_{t,s}^{g}(u)$ denotes the model error function with a mean of zero and a finite variance, and $K_s^g<T$ is the number of retained components. The eigenfunctions represent the fixed shapes of variation (e.g., longevity shift), while the scores capture the time-varying magnitude of these shifts.

There are a number of ways to select the retained number of principal components, such as the bootstrap approach proposed by \cite{HV06} and \cite{BYZ10}, description length approach proposed by \cite{PS13}, pseudo-Akaike information criterion \citep{Shibata81}, scree plot \citep{Cattell66}, and eigenvector variability plot \citep{TCC09}. We consider the eigenvalue ratio criterion (EVR) of \cite{LRS20}, defined as
\begin{equation}
K_s^g = \argmin_{k=1,\dots,k_{\max}}\left\{\frac{\widehat{\lambda}_{k+1,s}^g}{\widehat{\lambda}_{k,s}^g}\times \mathds{1}\Big\{\frac{\widehat{\lambda}_{k,s}^g}{\widehat{\lambda}_{1,s}^g}\geq \eta\Big\}+\mathds{1}\Big\{\frac{\widehat{\lambda}_{k,s}^g}{\widehat{\lambda}_{1,s}^g}<\eta\Big\}\right\}, \label{eq:LRS}
\end{equation}
where $\widehat{\lambda}_{k,s}^g$ denotes the $k$\textsuperscript{th} estimated eigenvalue at prefecture $s$ for gender $g$, $\mathds{1}\{\cdot\}$ is the binary indicator function, and $\eta = \frac{1}{\ln[\max(\widehat{\lambda}_{1,s}^g, T)]}$ is a small positive value. Instead of searching through $T$, we restrict our searching range by setting $k_{\max}=\#\{k|\widehat{\lambda}_{k,s}^g\geq \frac{1}{T}\sum^T_{k=1}\widehat{\lambda}_{k,s}^g, k\geq 1\}$. For comparison, we also consider $K_s^g=6$ as suggested in \cite{HBY13}. Note that when $K_s^g=1$, it reduces to \citeauthor{LC92}'s \citeyearpar{LC92} model.

Conditioning on the observed functions $\bm{\X}_{s}^g(u)=[\X_{1,s}^g(u),\dots,\X_{T,s}^g(u)]$ and the estimated functional principal components $\bm{B} = \big\{\phi_{1,s}^g(u),\dots,\phi_{K,s}^g(u)\big\}$, the $h$-step-ahead point forecast of $\X_{T+h,s}^g(u)$ can be expressed as
\begin{equation*}
\widehat{\X}_{T+h|T,s}^g(u) = \text{E}\big[\X_{T+h,s}^g(u)|\bm{\X}_{s}^g(u), \bm{B}\big]  = \widehat{\mu}_{s}^g(u) + \sum^{K_s^g}_{k=1}\widehat{\beta}_{T+h|T,k,s}^g\widehat{\phi}_{k,s}^g(u),
\end{equation*}
where $\widehat{\mu}_s^g(u) = \frac{1}{T}\sum^T_{t=1}\X_{t,s}^g(u)$ and $\widehat{\beta}_{T+h|T,k,s}^g$ denotes time-series forecasts of the $k$\textsuperscript{th} principal component scores. The forecasts of these scores can be obtained via a univariate time-series forecasting method, such as exponential smoothing \citep{HKO+08}. Computationally, we implement the \verb|ets| function in the forecast package \citep{HK08}, which selects the optimal parameters based on the corrected Akaike information criterion.

\subsection{Multivariate functional time-series (MFTS) forecasting method}\label{sec:4.2}

The univariate functional time-series method does not account for correlations between female and male data within the same region. However, explicitly modeling the correlation between multiple series may improve forecast accuracy. We consider a multivariate functional time series (MFTS) method to jointly model and forecast multiple series that may be correlated.

Let $\X_{t,s}^{F}(u)$ and $\X_{t,s}^{M}(u)$ represent female and male subnational age distribution of death counts. As our multiple functional time series have the same function support, we consider data where each observation consists of multiple functions, that is, $[\X_{t,s}^F(u), \X_{t,s}^{M}(u)]\in \mathds{R}^{2}$, where $u\in \mathcal{I}$.

Multivariate functional time series are stacked into a vector. Let $\mu^{F}_{s}(u)$ and $\mu^{M}_{s}(u)$ represent the mean function, respectively. For $u, v\in \mathcal{I}$, the theoretical cross-covariance function can be defined with elements
\begin{equation*}
c_{s}(u,v) := \text{Cov}[\X^F_{s}(u), \X^M_{s}(v)] 
		= \text{E}\left\{[\X^F_{s}(u) - \mu^{F}_{s}(u)][\X^{M}_{s}(u) - \mu^{M}_{s}(u)]\right\}.
\end{equation*}
The Karhunen-Lo\`{e}ve expansion of a stochastic process can be expressed as
\begin{equation*}
\bm{\X}_s(u) = \bm{\mu}_s(u) + \bm{\Phi}_s(u)\bm{\beta}_{t,s}^{\top},
\end{equation*}
where $\bm{\X}_s(u) = [\bm{\X}_s^F(u), \bm{\X}_s^M(u)]$, $\bm{\X}_s^F(u) = [\X_{1,s}^F(u),\dots, \X_{T,s}^F(u)]$ denote stacked historical functions, and $\bm{\mu}_s(u) = [\bm{\mu}_s^{F}(u), \bm{\mu}_s^{M}(u)]$. Moreover, $\bm{\beta}_{s} = (\bm{\beta}_{1,s},\dots,\bm{\beta}_{K,s})$ and $\bm{\beta}_{1,s} = (\beta_{1,1,s},\dots,\beta_{T,1,s})$ are the vectors of the principal component scores, and
\begin{equation*} 
\bm{\Phi}_s(u) = \left[\begin{array}{cccccc}
\phi_{1,s}^F(u) & \cdots & \phi_{K,s}^F(u) &  & \cdots &  \\
 & \cdots &  &  \phi_{1,s}^M(u) & \cdots & \phi_{K,s}^M(u) \end{array} \right]_{2\times (K\times 2)},
\end{equation*}
where $K$ denotes the retained number of principal components shared by females and males.

Conditioning on the previous functions $\bm{\X}_s(u)$ and the estimated functional principal components $\bm{\Phi}_s(u)$, the $h$-step-ahead point forecast of $\bm{\X}_{T+h,s}(u)$ can be expressed as
\begin{equation*}
\widehat{\bm{\X}}_{T+h|T,s}(u) = \text{E}\left[\bm{\X}_{T+h,s}(u)|\bm{\X}_s(u), \bm{\Phi}_s(u)\right] =\widehat{\mu}_s(u)+\widehat{\bm{\Phi}}_s(u)\widehat{\bm{\beta}}_{T+h|T,s}^{\top},
\end{equation*}
where $\widehat{\bm{\beta}}_{T+h|T,s}$ denotes the time-series forecasts of the principal component scores corresponding to the female and male series, $\widehat{\mu}_s(u)$ and $\widehat{\bm{\Phi}}_s(u)$ denote the estimated mean function and estimated functional principal components, respectively.

\subsection{Multilevel functional time-series (MLFTS) forecasting method}\label{sec:4.3}

The multilevel functional data model strongly resembles the two-way functional analysis of variance studied by \cite{MC06} and \cite{CF10}. The basic idea is to extract common patterns shared by multiple series $R_{t,s}(u)$ and series-specific patterns $U_{t,s}^g(u)$. The common trend and series-specific trend are modeled by projecting them onto the eigenvectors of the covariance functions of the aggregated and series-specific stochastic processes, respectively. For $t=1,\dots,T$, a curve can be expressed as
\begin{equation}
\X_{t,s}^{g}(u) = \mu^{g}_{s}(u)+R^{c}_{t,s}(u) + U_{t,s}^{g}(u), \label{eq:MLFTS_0}
\end{equation}
where $R^c_{t,s}(u)$ can be the simple average of the decentered female and male series. To ensure model and parameter identifiability, we implement a two-stage estimation procedure based on functional principal component decomposition.

Since the stochastic processes $R^{c}_{t,s}(u)$ and $U^g_{t,s}(u)$ are unknown in practice, the population eigenvalues and eigenfunctions can only be approximated at best through a set of realizations $\bm{R}^c_{s}(u)$ and $\bm{U}_{s}^g(u)$. From the covariance function of $\bm{R}^c_{s}(u)$, we can extract a set of functional principal components and their associated scores, along with a set of residual functions. From the covariance function of the residuals, we can then extract a second set of functional principal components and their associated scores. While the first functional principal component decomposition captures the common pattern across multiple series, the second functional principal component decomposition captures the series-specific residual trend.

The sample versions of the series-specific mean function, a common trend, and the series-specific residual trend for a set of functional time series can be estimated by
\begin{align}
\widehat{\mu}_s^g(u) &= \frac{1}{T}\sum^T_{t=1}\X_{t,s}^g(u) \label{eq:MLFTS_1}\\
R_{t,s}^{c}(u) &\approx \sum^{K}_{k=1}\beta_{t,k,s}^{c}\phi_{k,s}^c(u) \label{eq:MLFTS_2}\\
U_{t,s}^g(u)  & \approx \sum^{L}_{l=1}\gamma_{t,l,s}^g\psi_{l,s}^g(u),\label{eq:MLFTS_3}
\end{align}
where $\widehat{\mu}^g_s(u)$ represents the simple average of the female or male series, $\bm{\beta}_{k,s}^c = (\beta_{1,k,s}^{c},\dots, \beta_{T,k,s}^{c})$ represents the $k$\textsuperscript{th} sample principal component scores of $\bm{R}^c_{s}(u)$, and $\bm{\gamma}_{l,s}^{g} = (\gamma_{1,l,s}^g,\dots,\gamma_{T,l,s}^g)$ represents the $l$\textsuperscript{th} sample principal component scores of $\bm{U}_{s}^g(u)$.

Plugging~\eqref{eq:MLFTS_1} into~\eqref{eq:MLFTS_3} into~\eqref{eq:MLFTS_0}, we obtain
\begin{equation*}
\X_{t,s}^g(u) \approx \widehat{\mu}^g_{s}(u)+\sum^{K}_{k=1}\beta_{t,k,s}^{c}\phi_{k,s}^c(u)+\sum^{L}_{l=1}\gamma_{t,l,s}^g\psi_{l,s}^g(u)+e_{t,s}^g(u),
\end{equation*} 
where $e_{t,s}^g(u)$ denotes measurement error with a finite variance.

To select the retained number of components, we use the eigenvalue ratio criterion in~\eqref{eq:LRS}. A feature of the multilevel method is its ability to estimate the proportion of variability explained by aggregated data. For a given population $g$, a measure of the within-cluster variability is given by
\begin{equation*}
\frac{\sum_{k=1}^K\widehat{\lambda}_{k,s}^{c}}{\sum_{k=1}^K\widehat{\lambda}_{k,s}^{c}+\sum_{l=1}^L\widehat{\lambda}_{l,s}^g}.
\end{equation*}
When the common trend explains the primary mode of total variability, the within-cluster variability is close to 1.

Conditioning on observed data $\bm{\X}^g_{s}(u)$ and basis functions $\bm{\Phi}_s(u)$ and $\bm{\Psi}_s^g(u)=\{\widehat{\psi}_{1,s}^g(u)$,$\dots$,$\widehat{\psi}_{L,s}^g(u)\}$, the $h$-step-ahead forecasts can be obtained by
\begin{align*}
\widehat{\X}_{T+h|T,s}^g(u) &= \text{E}[\X_{T+h,s}^g(u)|\mu^{g}_{s}(u), \bm{\X}^g_{s}(u), \bm{\Phi}_s(u), \bm{\Psi}_s^g(u)] \\
&= \widehat{\mu}^g_s(u) + \sum^{K}_{k=1}\widehat{\beta}_{T+h|T,k,s}^{c}\widehat{\phi}_{k,s}^c(u)+\sum^{L}_{l=1}\widehat{\gamma}_{T+h|T,l,s}^g\widehat{\psi}_{l,s}^g(u),
\end{align*}
where $\widehat{\mu}_s^g(u) = \frac{1}{T}\sum^T_{t=1}\X_{t,s}^g(u)$, $\{\widehat{\beta}_{T+h|T,k,s}^{c}, k=1,\dots, K\}$ and $\{\widehat{\gamma}_{T+h|T,l,s}^g, l=1,\dots, L\}$ are the forecasted principal component scores, obtained from a univariate time-series forecasting method.

\subsection{Two-way functional analysis of variance (FANOVA)}\label{sec:4.4}

To model $\X_{t,s}^{g}(u)$, we consider a two-way FANOVA decomposition, expressed as:
\begin{equation*}
\X_{t,s}^g(u) = \mu(u) + \alpha_s(u)+\beta^g(u)+\epsilon_{t,s}^{g}(u),
\end{equation*}
where $\mu(u)$ denotes the functional grand effect, $\alpha_s(u)$ denotes the functional row effect, $\beta^g(u)$ denotes the functional column effect, and $\epsilon_{t,s}^g(u)$ denotes the error function. 

The FANOVA model can be estimated by means with
\begin{align*}
\widehat{\mu}(u) &= \frac{1}{S\times 2\times T}\sum^{S}_{s=1}\sum^2_{g=1}\sum^T_{t=1}\X_{t,s}^g(u) \\
\widehat{\alpha}_s(u) &= \frac{1}{2\times T }\sum^2_{g=1}\sum^T_{t=1}\X_{t,s}^g(u) - \widehat{\mu}(u) \\
\widehat{\beta}^g(u) &= \frac{1}{S\times T}\sum^{S}_{s=1}\sum^T_{t=1}\X_{t,s}^g(u) - \widehat{\mu}(u).
\end{align*}
Some identifiability constraints exist, $\forall u\in \mathcal{I}$, $\sum^{S}_{s=1}\alpha_s(u) = \sum^2_{g=1}\beta^g(u) = 0$, and $\sum^{S}_{s=1}\X_{t,s}^g(u) = \sum^2_{g=1}\X_{t,s}^g(u) = 0$, $\forall t$. 
The FANOVA model decomposes the FTS jointly for all prefectures and genders into three parameters: the overall mean $\widehat{\mu}(u)$, which represents the \textit{grand effect} measuring the overall baseline age distribution of deaths; $\widehat{\alpha}_s(u)$, which represents the \textit{functional prefecture effect} measuring the time-invariant functional difference attribute to each prefecture, relative to the grand mean; and $\widehat{\beta}^g(u)$, which represents the \textit{functional gender effect} measuring the time-invariant functional difference between genders, relative to the grand mean.

The residual functions $\widehat{\epsilon}_{t,s}^g(u) = \mathbf{X}_{t,s}^g(u) - \widehat{\mu}(u) - \widehat{\alpha}_s(u) - \widehat{\beta}^g(u)$ are time-varying, for $t=1,\dots,T$. Following the methodology of \cite{JCGS24}, we model and forecast these residual functions using the MFTS approach.

\subsection{Two-stage functional principal component analyses (HDFPCA)}

\cite{GSY19} propose a two-stage functional principal component decomposition to model and forecast HDFTS. One of the earliest methods for comparison \citep[see, e.g.,][]{HNT232, Shang25}, it consists of three parts.
\begin{enumerate}
\item[1)] Functional principal component analysis is performed separately on each set of functional time series, resulting in $S$ sets of principal component scores of low dimension $K$.
\item[2)] The first functional principal component scores from each of $S$ sets of functional time series are combined into an $S\times 1$ vector. We fit factor models to the scores to further reduce the dimension into an $r\times 1$ vector, where $r<<S$ is the number of factors in the second dimension-reduction stage. The same is carried out for the second, third, and so on until the $K$\textsuperscript{th} scores. The vector of the $P$ functional time series is reduced to an $r\times K$ matrix consisting of factors.
\item[3)] A univariate time series model can be fitted to each factor, and forecasts are generated. Forecast factors can be used to construct forecast functions.
\end{enumerate}

By no means are the methods in Section~\ref{sec:4} comprehensive, but they represent a pool of candidate FTS models for comparison purposes.

\section{Construction of prediction intervals}\label{sec:5} 

We split the data sample, consisting of 49 years from 1975 to 2023, into a training set consisting of 16 years, a validation set consisting of 16 years, and a testing set consisting of 17 years. While the choice of different data splits is arbitrary, it is customary in forecasting to allocate at least one-third of the total data for evaluation \citep{HA21}.

Using the data in the training sample, we implement an expanding-window forecast scheme to obtain $h$-step-ahead density forecasts in the validation set for $h=1,2,\dots,16$. For each horizon, we have a different number of curves in the validation set. For instance, when $h=1$, we have 16 years to evaluate the forecast errors, denoted by $\widehat{\epsilon}_{m,s}^g(u) = d_{m,s}^g(u) - \widehat{d}_{m,s}^g(u)$ for $m=1, 2,\dots, M=16$ (here $M$ denotes the total number of years available in the validation set for a specific horizon $h$, and $m$ serves as the year index); when $h=16$, we have two years to evaluate the residual functions between the samples in the validation set and their corresponding forecasts. From these residual functions, we compute the functional standard deviation, denoted by
\begin{equation}
\gamma(u) = \text{sd}[\widehat{\epsilon}_{1,s}^g(u),\dots,\widehat{\epsilon}_{M,s}^g(u)]. \label{eq:sd}
\end{equation}
We aim to determine $(\overline{\xi}_{\alpha},\underline{\xi}_{\alpha})$ such that $\alpha\times 100\%$ of the residuals satisfy
\begin{equation*} 
-\underline{\xi}_{\alpha}\gamma(u)\leq \widehat{\epsilon}_{m,s}^g(u)\leq \overline{\xi}_{\alpha}\gamma(u),
\end{equation*}
where $(\overline{\xi}_{\alpha},\underline{\xi}_{\alpha})$ are the tuning parameters. 

Typically, the constants $\overline{\xi}_{\alpha}$ and $\underline{\xi}_{\alpha}$ are chosen equal. By the law of large numbers, one may achieve
\begin{equation*}
\text{Pr}\left[-\xi_{\alpha}\gamma(u)\leq d_{T+h,s}^g(u) - \widehat{d}_{T+h,s}^g(u) \leq \xi_{\alpha}\gamma(u)\right]\approx\frac{1}{M}\sum^M_{m=1}\mathds{1}\left[-\xi_{\alpha}\gamma(u)\leq \widehat{\epsilon}_{m,s}^g(u)\leq \xi_{\alpha}\gamma(u)\right],
\end{equation*}
where $M$ denotes the number of years in the validation set.

To determine the optimal $\xi_{\alpha}$, the samples in the validation set are used to calibrate the prediction intervals so that the empirical coverage probabilities defined in~\eqref{eq:ECP} are close to their nominal coverage probabilities.

In Appendix~C, we also present the conformal prediction interval approach of \cite{SV08}, which does not require a tuning parameter. By considering a quantile of the absolute residual functions, the pointwise prediction interval for the out-of-sample testing set is centered on the point forecast, with its spread determined by the chosen quantile.

\section{Results}\label{sec:6}

\subsection{Expanding window scheme}\label{sec:6.1}

An expanding window analysis of a time-series model is commonly used to assess model and parameter stability and prediction accuracy over time. The expanding window analysis assesses the stability of a model's parameters by computing parameter estimates and their forecasts over an expanding window across the sample \citep[for details, see][pp. 313--314]{ZW06}.

Using the first 32 years from 1975 to 2006 in the Japanese subnational age- and sex-specific life-table death counts, we produce one- to 16-step-ahead forecasts. Through an expanding-window approach, we re-estimate the parameters in the time-series forecasting models using the first 33 years from 1975 to 2007. Forecasts from the estimated models are then produced for one- to 16-step-ahead forecasts. We iterate this process by increasing the sample size by one year until reaching the end of the data period in 2023. This process produces 16 one-step-ahead forecasts, 15 two-step-ahead forecasts, $\dots$, and one 16-step-ahead forecast. We compare these forecasts with holdout samples to assess out-of-sample forecast accuracy. In Figure~\ref{fig:expanding}, we display a diagram of the expanding window forecast scheme for the forecast horizon $h=1$, although we also consider other forecast horizons from $h = 2$ to $16$.
\begin{figure}[!htb]
\begin{center}
\begin{tikzpicture}
\draw[->] (0,0) -- (10,0) node[right] {Time};
\draw[fill=blue!20] (0,-0.5) rectangle (3,0.5) node[midway] {Train};
\draw[fill=red!20] (3,-0.5) rectangle (3.5,0.5) node[midway] {F};
\draw[fill=blue!20] (0,-1.5) rectangle (5,-0.5) node[midway] {Train};
\draw[fill=red!20] (5,-1.5) rectangle (5.5,-0.5) node[midway] {F};
\draw[fill=blue!20] (0,-2.5) rectangle (7,-1.5) node[midway] {Train};
\draw[fill=red!20] (7,-2.5) rectangle (7.5,-1.5) node[midway] {F};
\draw[fill=blue!20] (0,-3.5) rectangle (9,-2.5) node[midway] {Train};
\draw[fill=red!20] (9,-3.5) rectangle (9.5,-2.5) node[midway] {F};
\node[left] at (0,0) {1975:2006};
\node[left] at (0,-1) {1975:2007};
\node[left] at (0,-2) {\hspace{-0.8in}{$\vdots$}};
\node[left] at (0,-3) {1975:2022};
\draw[fill=blue!20] (6.5,1) rectangle (7,1.5);
\node[right] at (7,1.25) {Training Window};
\draw[fill=red!20] (6.5,0.5) rectangle (7,1);
\node[right] at (7,0.75) {Forecast (F)};
\end{tikzpicture}
\end{center}
\caption{\small A diagram of the expanding-window forecast scheme. The data begin in 1975 and end in 2023.}\label{fig:expanding}
\end{figure}

\subsection{Point forecast error metrics}\label{sec:6.2}

Since life-table death counts can be considered probability density functions, we use some density-evaluation measures. These measures include the discrete version of the KLD \citep{KL51}, the square root of the JSD \citep{Shannon48}. For two probability density functions, denoted by $d_{m+\xi}$ and $\widehat{d}_{m+\xi}$, the symmetric version of the KLD, also known as Jeffrey divergence, is defined as
\begin{align}
\text{KLD}(h) = \ & \text{D}_{\text{KL}}(d_{m+\xi}||\widehat{d}_{m+\xi}) + \text{D}_{\text{KL}}(\widehat{d}_{m+\xi}||d_{m+\xi}) \notag\\
	=\ & \frac{1}{111\times (17-h)}\sum^{16}_{\xi=h}\sum_{x=1}^{111}d_{m+\xi}(u_x)\cdot \left[\ln d_{m+\xi}(u_x)-\ln \widehat{d}_{m+\xi}(u_x)\right] + \notag\\
	 & \frac{1}{111\times (17-h)}\sum^{16}_{\xi=h}\sum_{x=1}^{111}\widehat{d}_{m+\xi}(u_x)\cdot \left[\ln \widehat{d}_{m+\xi}(u_x) - \ln d_{m+\xi}(u_x)\right], \label{eq:sym_KL}
\end{align}
which is non-negative, for $h=1,2,\dots,16$, where $m$ represents the year at the end of training period.

An alternative metric is the JSD, defined by
\begin{equation*}
\text{JSD}(h) = \frac{1}{2}D_{\text{KL}}(d_{m+\xi}||\delta_{m+\xi})+\frac{1}{2}D_{\text{KL}}(\widehat{d}_{m+\xi}||\delta_{m+\xi}),
\end{equation*}
where $\delta_{m+\xi}$ measures a common quantity between $d_{m+\xi}$ and $\widehat{d}_{m+\xi}$. Following the early work of \cite{SH20}, we consider the geometric mean $\delta_{m+\xi} = \sqrt{d_{m+\xi}\widehat{d}_{m+\xi}}$. To make the JSD a metric between two probability densities, we take the square root of the JSD \citep[see, e.g.,][]{FT04}.

\subsection{Comparison of point forecast accuracy}\label{sec:6.3}

In Table~\ref{tab:1}, we compare the point forecast accuracy across the functional time-series models, averaged across 47 prefectures. Based on the summary statistics of the KLD and JSD, the HDFPCA model provides the most accurate point forecasts for the female data. This result indicates that the two-stage principal component analyses can adequately summarize the main features in the female data without much loss of information. The female data are relatively stable in comparison to the male counterpart. For the male data, the MFTS and MLFTS models provide the most accurate point forecasts. For the UFTS, MFTS, MLFTS, and FANOVA models, the number of retained functional principal components was determined by the EVR criterion. For the HDFPCA model, we follow the default setting of \cite{GSY19}, where the number of retained components is six and two in the first and second stages, respectively.

\begin{center}
\renewcommand*{\arraystretch}{0.99}
\tabcolsep 0.065in
\begin{footnotesize}
\begin{longtable}{@{}llrrrrrrrrrrr@{}}
\caption{\small Averaged across 47 prefectures, we evaluate and compare the point forecast accuracy of the functional time-series models, measured by the KLD and JSD. The number of components is determined by the EVR criterion.}\label{tab:1} \\
\toprule
&  & \multicolumn{5}{c}{Female}   & \multicolumn{5}{c}{Male}   \\
\cmidrule{3-13}
Metric & $h$ & UFTS & MFTS  & MLFTS & FANOVA & HDFPCA & UFTS  & MFTS  & MLFTS & FANOVA & HDFPCA \\ 
\midrule
\endfirsthead
\toprule
&  & \multicolumn{5}{c}{Female}   & \multicolumn{5}{c}{Male}   \\
\cmidrule{3-13}
Metric & $h$ & UFTS & MFTS  & MLFTS & FANOVA & HDFPCA & UFTS  & MFTS  & MLFTS & FANOVA & HDFPCA \\ 
\midrule
\endhead
\midrule
\multicolumn{13}{r}{{Continued on next page}} \\
\endfoot
\endlastfoot
KLD  &  1 & 0.004 & 0.010 & 0.004 & 0.006 & 0.007 & 0.004 & 0.004 & 0.003 & 0.005 & 0.003 \\
     &  2 & 0.007 & 0.013 & 0.006 & 0.007 & 0.007 & 0.005 & 0.006 & 0.005 & 0.005 & 0.003 \\
     &  3 & 0.008 & 0.015 & 0.007 & 0.009 & 0.009 & 0.006 & 0.007 & 0.005 & 0.006 & 0.004 \\
     &  4 & 0.010 & 0.017 & 0.009 & 0.010 & 0.006 & 0.006 & 0.007 & 0.006 & 0.007 & 0.005 \\
     &  5 & 0.013 & 0.019 & 0.010 & 0.012 & 0.013 & 0.007 & 0.008 & 0.007 & 0.007 & 0.006 \\
     &  6 & 0.015 & 0.021 & 0.011 & 0.013 & 0.007 & 0.008 & 0.009 & 0.007 & 0.008 & 0.007 \\
     &  7 & 0.018 & 0.023 & 0.012 & 0.015 & 0.019 & 0.009 & 0.010 & 0.008 & 0.008 & 0.009 \\
     &  8 & 0.022 & 0.025 & 0.014 & 0.018 & 0.008 & 0.010 & 0.011 & 0.009 & 0.009 & 0.012 \\
     &  9 & 0.027 & 0.027 & 0.017 & 0.022 & 0.025 & 0.011 & 0.012 & 0.010 & 0.010 & 0.015 \\
     & 10 & 0.033 & 0.030 & 0.021 & 0.026 & 0.010 & 0.012 & 0.014 & 0.011 & 0.011 & 0.018 \\
     & 11 & 0.040 & 0.034 & 0.024 & 0.033 & 0.032 & 0.012 & 0.015 & 0.012 & 0.012 & 0.022 \\
     & 12 & 0.048 & 0.039 & 0.030 & 0.040 & 0.014 & 0.013 & 0.017 & 0.014 & 0.013 & 0.026 \\
     & 13 & 0.059 & 0.047 & 0.036 & 0.047 & 0.039 & 0.014 & 0.020 & 0.015 & 0.014 & 0.030 \\
     & 14 & 0.068 & 0.040 & 0.042 & 0.061 & 0.019 & 0.014 & 0.012 & 0.015 & 0.017 & 0.035 \\
     & 15 & 0.089 & 0.054 & 0.060 & 0.080 & 0.044 & 0.015 & 0.015 & 0.021 & 0.022 & 0.039 \\
     & 16 & 0.118 & 0.070 & 0.079 & 0.106 & 0.019 & 0.015 & 0.019 & 0.029 & 0.030 & 0.040 \\
     & 17 & 0.154 & 0.084 & 0.093 & 0.120 & 0.038 & 0.015 & 0.020 & 0.034 & 0.034 & 0.039 \\
\cmidrule{2-12} 
     & Mean & 0.043 & 0.033 & 0.028 & 0.037 & 0.019 & 0.010 & 0.012 & 0.012 & 0.013 & 0.019 \\
     & Median & 0.027 & 0.027 & 0.017 & 0.022 & 0.014 & 0.011 & 0.012 & 0.010 & 0.010 & 0.015 \\
\midrule
JSD & 1 & 0.034 & 0.049 & 0.034 & 0.039 & 0.042 & 0.030 & 0.032 & 0.028 & 0.034 & 0.029 \\
    & 2 & 0.041 & 0.055 & 0.040 & 0.043 & 0.038 & 0.035 & 0.036 & 0.033 & 0.036 & 0.030 \\
    & 3 & 0.045 & 0.059 & 0.043 & 0.046 & 0.050 & 0.037 & 0.039 & 0.036 & 0.038 & 0.032 \\
    & 4 & 0.049 & 0.063 & 0.046 & 0.049 & 0.037 & 0.040 & 0.041 & 0.038 & 0.040 & 0.035 \\
    & 5 & 0.054 & 0.066 & 0.049 & 0.053 & 0.060 & 0.042 & 0.043 & 0.040 & 0.041 & 0.039 \\
    & 6 & 0.058 & 0.069 & 0.052 & 0.056 & 0.040 & 0.044 & 0.045 & 0.041 & 0.043 & 0.043 \\
    & 7 & 0.063 & 0.071 & 0.055 & 0.060 & 0.070 & 0.047 & 0.047 & 0.043 & 0.044 & 0.049 \\
    & 8 & 0.069 & 0.073 & 0.059 & 0.065 & 0.044 & 0.050 & 0.049 & 0.045 & 0.046 & 0.055 \\
    & 9 & 0.076 & 0.075 & 0.063 & 0.070 & 0.082 & 0.052 & 0.051 & 0.048 & 0.047 & 0.062 \\
    & 10 & 0.083 & 0.079 & 0.069 & 0.076 & 0.051 & 0.054 & 0.053 & 0.050 & 0.049 & 0.069 \\
    & 11 & 0.092 & 0.083 & 0.075 & 0.085 & 0.092 & 0.056 & 0.054 & 0.052 & 0.052 & 0.075 \\
    & 12 & 0.102 & 0.088 & 0.082 & 0.096 & 0.058 & 0.058 & 0.056 & 0.055 & 0.054 & 0.081 \\
    & 13 & 0.116 & 0.094 & 0.092 & 0.106 & 0.100 & 0.060 & 0.059 & 0.058 & 0.057 & 0.087 \\
    & 14 & 0.128 & 0.097 & 0.102 & 0.121 & 0.066 & 0.061 & 0.057 & 0.061 & 0.064 & 0.094 \\
    & 15 & 0.150 & 0.112 & 0.122 & 0.141 & 0.104 & 0.062 & 0.062 & 0.070 & 0.072 & 0.097 \\
    & 16 & 0.176 & 0.126 & 0.141 & 0.164 & 0.068 & 0.064 & 0.067 & 0.081 & 0.084 & 0.095 \\
    & 17 & 0.202 & 0.134 & 0.152 & 0.175 & 0.094 & 0.064 & 0.069 & 0.087 & 0.088 & 0.093 \\
\cmidrule{2-12}  
    & Mean & 0.090 & 0.082 & 0.075 & 0.085 & 0.064 & 0.050 & 0.051 & 0.051 & 0.052 & 0.063 \\ 
    & Median & 0.076 & 0.075 & 0.063 & 0.070 & 0.060 & 0.052 & 0.051 & 0.048 & 0.047 & 0.062 \\ 
\bottomrule
\end{longtable}
\end{footnotesize}
\end{center}

\vspace{-.39in}

To assess the stability of the EVR criterion across prefectures, we examine the number of retained components ($K$) selected by the EVR rule for each of the 47 prefectures and across all 17 forecast horizons. The distribution of the selected dimensions is summarized in Table~\ref{tab:evr_stability_horizon}.
\begin{table}[!htbp]
\centering
\tabcolsep .2in
\caption{Frequency of selected dimensions ($K$) by the EVR criterion across 47 prefectures for each forecast horizon ($h$).}
\label{tab:evr_stability_horizon}
\scalebox{0.85}{
\begin{tabular}{@{}l cc cc cc cc cc@{}}
\toprule
 & \multicolumn{2}{c}{UFTS\_F} & \multicolumn{2}{c}{UFTS\_M} & \multicolumn{2}{c}{MFTS} & \multicolumn{2}{c}{MLFTS} & \multicolumn{2}{c}{FANOVA} \\
\cmidrule(lr){2-3} \cmidrule(lr){4-5} \cmidrule(lr){6-7} \cmidrule(lr){8-9} \cmidrule(lr){10-11}
$h$ & $K=1$ & $K=2$ & $K=1$ & $K=2$ & $K=1$ & $K=2$ & $K=1$ & $K=2$ & $K=1$ & $K=2$ \\
\midrule
$1$  & 1 & 46 & 13 & 34 & 47 & 0 & 47 & 0 & 47 & 0 \\
$2$  & 2 & 45 & 12 & 35 & 47 & 0 & 47 & 0 & 47 & 0 \\
$3$  & 1 & 46 & 9  & 38 & 47 & 0 & 47 & 0 & 47 & 0 \\
$4$  & 1 & 46 & 8  & 39 & 47 & 0 & 47 & 0 & 47 & 0 \\
$5$  & 1 & 46 & 8  & 39 & 47 & 0 & 47 & 0 & 47 & 0 \\
$6$  & 1 & 46 & 7  & 40 & 47 & 0 & 47 & 0 & 47 & 0 \\
$7$  & 1 & 46 & 6  & 41 & 47 & 0 & 47 & 0 & 47 & 0 \\
$8$  & 1 & 46 & 6  & 41 & 47 & 0 & 47 & 0 & 47 & 0 \\
$9$  & 2 & 45 & 7  & 40 & 47 & 0 & 47 & 0 & 47 & 0 \\
$10$ & 3 & 44 & 5  & 42 & 47 & 0 & 47 & 0 & 47 & 0 \\
$11$ & 3 & 44 & 6  & 41 & 47 & 0 & 47 & 0 & 47 & 0 \\
$12$ & 4 & 43 & 6  & 41 & 47 & 0 & 47 & 0 & 47 & 0 \\
$13$ & 4 & 43 & 6  & 41 & 47 & 0 & 47 & 0 & 47 & 0 \\
$14$ & 4 & 43 & 6  & 41 & 47 & 0 & 47 & 0 & 47 & 0 \\
$15$ & 4 & 43 & 6  & 41 & 47 & 0 & 47 & 0 & 47 & 0 \\
$16$ & 4 & 43 & 6  & 41 & 46 & 1 & 47 & 0 & 47 & 0 \\
$17$ & 4 & 43 & 8  & 39 & 46 & 1 & 47 & 0 & 47 & 0 \\
\bottomrule
\end{tabular}
}
\end{table}

As shown, the EVR criterion is highly stable across prefectures and horizons. For the multivariate and multilevel models (MFTS, MLFTS, and FANOVA), it almost exclusively selects $K=1$, indicating that the first component captures the dominant variation. 

\begin{figure}[!htb]
\centering
\subfloat[Female data]
{\includegraphics[width=8.6cm]{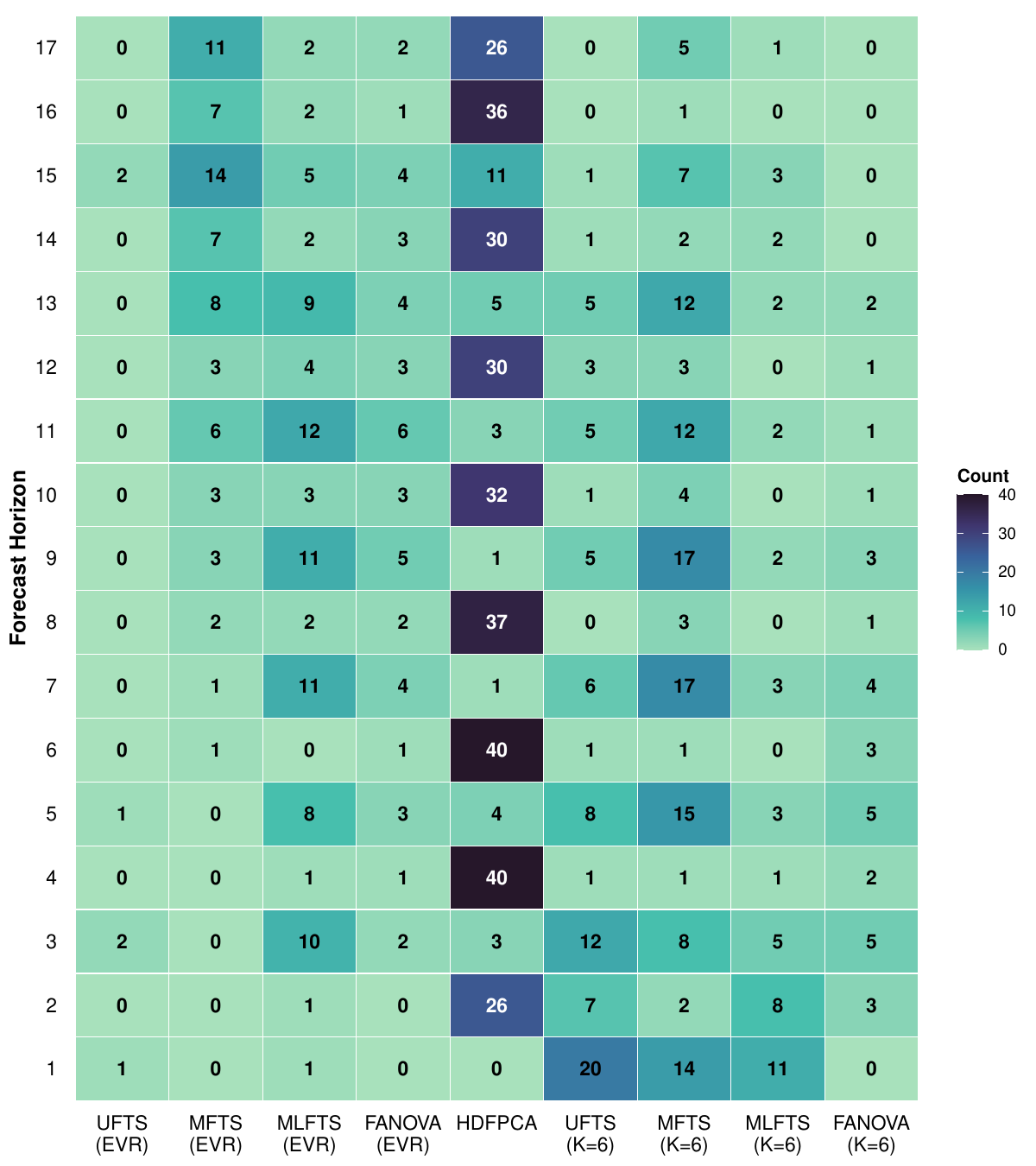}}
\quad
\subfloat[Male data]
{\includegraphics[width=8.6cm]{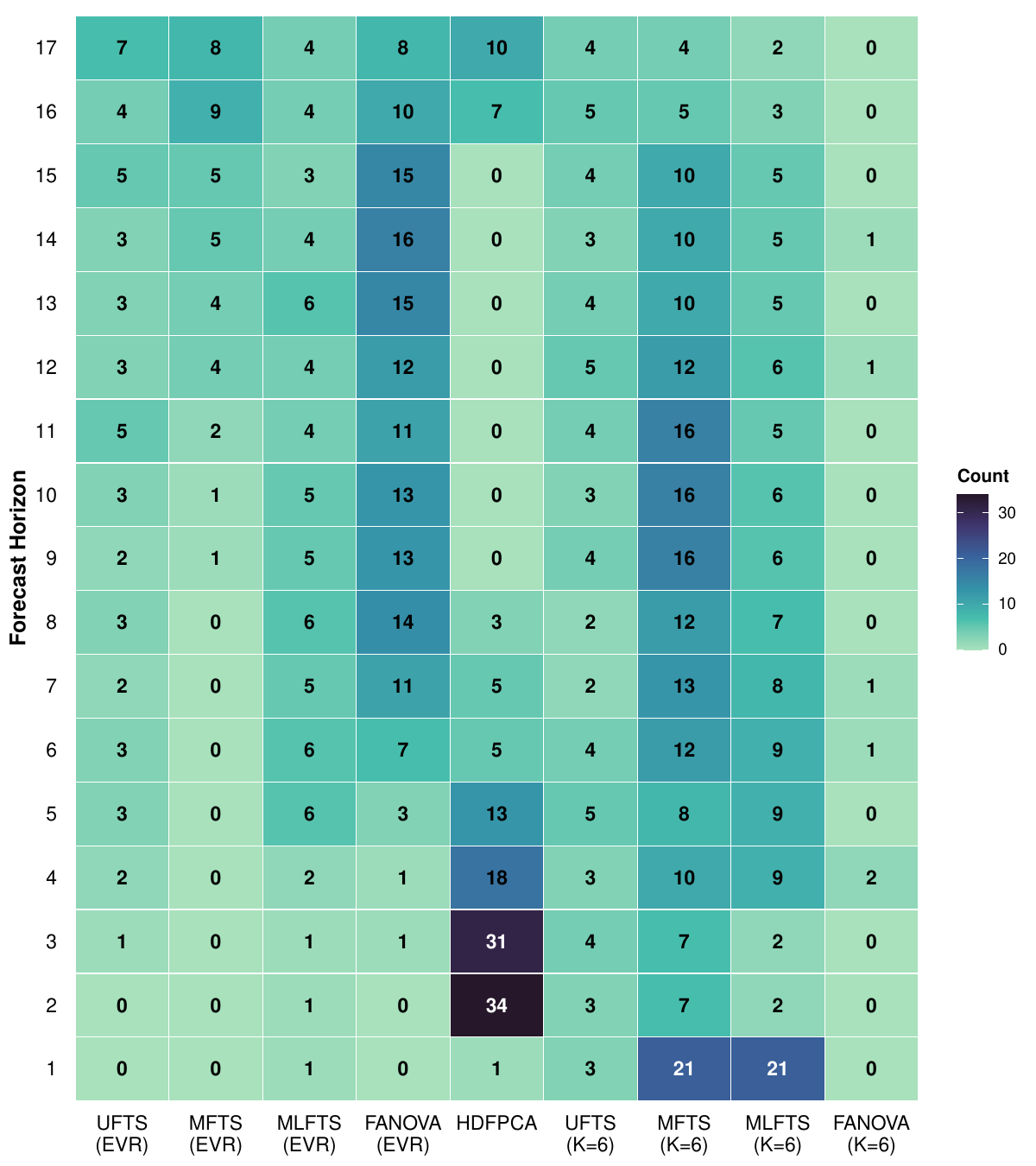}}
\caption{\small Point forecast accuracy (KLD) heatmaps for nine models across 17 horizons. Cell values represent the number of prefectures (out of 47) where each model performed best. A method that appears more than five times is highlighted in green, while those that appear fewer than five times are shown in red. If a method appears exactly five times, it is displayed in white.}\label{fig:4}
\end{figure}

Figure~\ref{fig:4} presents two heatmaps comparing nine functional time-series models across 17 forecast horizons for both female and male data. For each horizon, we count how often each method produces the smallest point forecast errors, as measured by the KLD. Appendix~A reports additional results for $K=6$ as a sensitivity analysis. Choosing $K=6$ has a slight advantage compared to the one based on the EVR criterion. For shorter forecast horizons, the UFTS, MFTS, and MLFTS models perform equally well for the female data. However, for relatively longer forecast horizons, the HDFPCA model is recommended. For male data, the MLFTS model is generally preferred; only at the shorter forecast horizon is $K=6$ advantageous. For relatively longer forecast horizons, the FANOVA with the EVR criterion is recommended. 

It is not surprising that the multi-population models (MFTS and MLFTS) outperform the univariate UFTS, as they exploit cross-sectional dependencies and share latent trends across regions and genders \citep{GSY19}. By pooling information across the 47 prefectures, they effectively regularize the estimates, mitigating volatility and data limitations in smaller populations. Moreover, the multi-population framework captures spatial and structural dependencies, enabling more robust predictive signals than independent modeling.

To formally assess whether the observed differences in point forecast accuracy obtained via the different presented methods are statistically significant, we further apply the Model Confidence Set (MCS) procedure \citep{MCS} based on the KLD across all forecast horizons and prefectures. The detailed MCS results and corresponding heatmap visualizations are reported in Appendix~D.

\subsection{Interval forecast error metrics}\label{sec:6.4}

To evaluate interval forecast accuracy, we consider the CPD between the empirical coverage probability (ECP) and nominal coverage probability, as well as the mean interval score of \cite{GR07}. For each year in the forecasting period, the $h$-step-ahead prediction intervals are computed at the $100(1-\alpha)\%$ nominal coverage probability. We consider the common case of the symmetric $100(1-\alpha)\%$ prediction intervals, with lower and upper bounds that are quantiles at $\alpha/2$ and $1-\alpha/2$, denoted by $\widehat{d}_{\nu+\xi,s}^{g,\text{lb}}(u)$ and $\widehat{d}_{\nu+\xi,s}^{g,\text{ub}}(u)$. For each horizon $h$, the ECP and CPD are defined as
\begin{align}
\text{ECP}_h &=  \frac{1}{111\times (17-h)}\times \sum_{\xi=h}^{16}\sum_{u=1}^{111}\mathds{1}\left\{\widehat{d}_{\nu+\xi,s}^{g, \text{lb}}(u)\leq d_{\nu+\xi,s}^g(u) \leq  \widehat{d}_{\nu+\xi,s}^{g,\text{ub}}(u)\right\}, \label{eq:ECP}\\
\text{CPD}_h &= \frac{1}{111\times (17-h)}\times \sum_{\xi=h}^{16}\sum_{u=1}^{111}\left[\mathds{1}\{d_{\nu+\xi,s}^g(u)>\widehat{d}_{\nu+\xi,s}^{g,\text{ub}}(u)\}+\mathds{1}\{d_{\nu+\xi,s}^g(u)<\widehat{d}_{\nu+\xi,s}^{g,\text{lb}}(u)\}\right],\notag
\end{align}
where $h=1,\dots,16$, $\nu$ denotes the end of the training set when we calibrate the CPD or interval score, and $\nu$ denotes the end of the validation set when we evaluate the interval forecast accuracy.

%, the mean and median ECP are defined as
%\begin{align*}
%\overline{\text{ECP}} &= \frac{1}{15}\text{ECP}_h \\
%\text{M[ECP]} &= \text{median}(\text{ECP}_h).
%\end{align*}
%Similarly

For different ages and years in the testing set, the mean and median CPD are defined as 
\begin{align*}
\overline{\text{CPD}} &= \frac{1}{16}\sum^{16}_{h=1}\text{CPD}_h, \\
\text{M[CPD]} &= \text{median}(\text{CPD}_h).
\end{align*}

As defined by \cite{GR07}, a scoring rule for the prediction intervals at age $u$ is
\begin{align*}
S_{\alpha,\xi}\left[\widehat{d}^{g,\text{lb}}_{\nu+\xi,s}(u), \widehat{d}^{g,\text{ub}}_{\nu+\xi,s}(u), d^g_{\nu+\xi,s}(u)\right]& =  \left[\widehat{d}^{g,\text{ub}}_{\nu+\xi,s}(u) - \widehat{d}^{g,\text{lb}}_{\nu+\xi,s}(u)\right] \\
&+\frac{2}{\alpha}\left[\widehat{d}^{g,\text{lb}}_{\nu+\xi,s}(u)-d_{n+\xi,s}^g(u)\right]\mathds{1}\left\{d_{\nu+\xi,s}^g(u) <\widehat{d}_{\nu+\xi,s}^{g,\text{lb}}(u)\right\} \\
&+\frac{2}{\alpha}\left[d_{\nu+\xi,s}^g(u) - \widehat{d}_{\nu+\xi,s}^{g,\text{ub}}(u)\right]\mathds{1}\left\{d_{\nu+\xi,s}^g(u) > \widehat{d}_{\nu+\xi,s}^{g,\text{ub}}(u)\right\},
\end{align*}
where the level of significance is customarily set to be $\alpha = 0.2$ or $0.05$. The interval score rewards a narrow prediction interval width if and only if $100(1-\alpha)\%$ of the holdout densities lie within the prediction interval.

For different ages and years in the testing set, the mean interval score is defined by
\begin{align*}
\overline{S}_{\alpha}(h) = \frac{1}{111\times (17-h)}\times \sum^{16}_{\xi=h}\sum^{111}_{u=1}S_{\alpha,\xi}\left[\widehat{d}_{\nu+\xi,s}^{g,\text{lb}}(u), \widehat{d}_{\nu+\xi,s}^{g,\text{ub}}(u), d_{\nu+\xi,s}^g(u)\right].
\end{align*}
Averaging over all forecast horizons, we obtain the overall mean interval score 
\begin{equation*}
\overline{S}_{\alpha} = \frac{1}{16}\sum^{16}_{h=1}\overline{S}_{\alpha}(h), \qquad
\text{M}[S_{\alpha}] = \text{median}[\overline{S}_{\alpha}(h)].
\end{equation*}

\subsection{Comparison of interval forecast accuracy}\label{sec:6.5}

In Table~\ref{tab:2}, averaged across 47 prefectures, we compare the interval forecast accuracy among the functional time-series models using the EVR criterion. From the summary statistics of the CPD and interval score at the level of significance $\alpha=0.2$, the MFTS offers the smallest CPD value for the female data, while the UFTS and MLFTS offer the smallest CPD values for the male data. In terms of the interval scores, the UFTS and MFTS provide the smallest interval scores for the female data, while the FANOVA gives the smallest interval scores for the male data. In Appendix~B, we also provide the results for $K=6$.
\begin{center}
\renewcommand*{\arraystretch}{1}
\tabcolsep 0.065in
\begin{footnotesize}
\begin{longtable}{@{}llrrrrrrrrrrr@{}}
\caption{\small Averaged across 47 prefectures, we evaluate and compare the interval forecast accuracy, as measured by the CPD and interval score, for the nine functional time-series models at the significance level $\alpha=0.2$. The number of components is determined by the EVR criterion.}\label{tab:2} \\
\toprule
&  & \multicolumn{5}{c}{Female}   & \multicolumn{5}{c}{Male}   \\
  \cmidrule{3-13}
Metric & $h$ & UFTS & MFTS  & MLFTS  & FANOVA & HDFPCA & UFTS  & MFTS  & MLFTS  & FANOVA & HDFPCA \\ 
  \midrule
\endfirsthead
\toprule
&  & \multicolumn{5}{c}{Female}   & \multicolumn{5}{c}{Male}   \\
  \cmidrule{3-12}
Metric & $h$ & UFTS & MFTS  & MLFTS  & FANOVA & HDFPCA & UFTS  & MFTS  & MLFTS  & FANOVA & HDFPCA \\ 
  \midrule
\endhead
\midrule
\multicolumn{12}{r}{{Continued on next page}} \\
\endfoot
\endlastfoot
$\overline{\text{CPD}}_{0.2}$ & 1 & 0.061 & 0.080 & 0.047 & 0.075 & 0.077 & 0.060 & 0.100 & 0.043 & 0.079 & 0.067 \\ 
	& 2 & 0.060 & 0.074 & 0.046 & 0.085 & 0.054 & 0.060 & 0.102 & 0.047 & 0.090 & 0.084 \\ 
	& 3 & 0.058 & 0.068 & 0.050 & 0.096 & 0.091 & 0.058 & 0.105 & 0.055 & 0.094 & 0.086 \\ 
	& 4 & 0.058 & 0.064 & 0.055 & 0.110 & 0.067 & 0.058 & 0.108 & 0.056 & 0.097 & 0.087 \\ 
	& 5 & 0.053 & 0.056 & 0.057 & 0.113 & 0.094 & 0.053 & 0.101 & 0.056 & 0.096 & 0.085 \\ 
	& 6 & 0.060 & 0.050 & 0.065 & 0.124 & 0.085 & 0.060 & 0.104 & 0.057 & 0.097 & 0.092 \\ 
	& 7 & 0.061 & 0.048 & 0.069 & 0.125 & 0.084 & 0.061 & 0.097 & 0.057 & 0.091 & 0.094 \\ 
	& 8 & 0.061 & 0.046 & 0.074 & 0.126 & 0.095 & 0.061 & 0.089 & 0.055 & 0.085 & 0.087 \\ 
	& 9 & 0.068 & 0.047 & 0.081 & 0.128 & 0.072 & 0.068 & 0.083 & 0.062 & 0.076 & 0.084 \\ 
	& 10 & 0.071 & 0.047 & 0.086 & 0.139 & 0.123 & 0.071 & 0.085 & 0.069 & 0.084 & 0.090 \\ 
	& 11 & 0.078 & 0.049 & 0.093 & 0.145 & 0.059 & 0.078 & 0.087 & 0.075 & 0.085 & 0.084 \\ 
	& 12 & 0.075 & 0.053 & 0.092 & 0.146 & 0.126 & 0.075 & 0.083 & 0.078 & 0.085 & 0.089 \\ 
	& 13 & 0.076 & 0.054 & 0.084 & 0.147 & 0.065 & 0.076 & 0.073 & 0.077 & 0.089 & 0.081 \\ 
	& 14 & 0.073 & 0.068 & 0.087 & 0.147 & 0.100 & 0.073 & 0.078 & 0.084 & 0.095 & 0.099 \\ 
	& 15 & 0.069 & 0.090 & 0.091 & 0.163 & 0.098 & 0.069 & 0.080 & 0.088 & 0.087 & 0.105 \\ 
	& 16 & 0.062 & 0.098 & 0.093 & 0.178 & 0.062 & 0.062 & 0.077 & 0.081 & 0.085 & 0.088 \\ 
\cmidrule{2-12}
	& Mean & 0.065 & 0.062 & 0.073 & 0.128 & 0.085 & 0.065 & 0.091 & 0.065 & 0.089 & 0.087 \\ 
	& Median & 0.062 & 0.055 & 0.078 & 0.127 & 0.084 & 0.062 & 0.088 & 0.060 & 0.088 & 0.086 \\
\midrule
$\overline{S}_{0.2}$ & 1 & 268 & 256 & 433 & 287 & 325 & 256 & 279 & 218 & 258 & 248 \\ 
	&  2 & 295 & 302 & 466 & 336 & 335 & 295 & 306 & 243 & 279 & 281 \\ 
	&  3 & 317 & 347 & 494 & 389 & 402 & 317 & 332 & 265 & 293 & 306 \\ 
	&  4 & 345 & 397 & 524 & 450 & 357 & 345 & 359 & 285 & 312 & 327 \\ 
	&  5 & 368 & 445 & 551 & 508 & 457 & 368 & 379 & 300 & 321 & 344 \\ 
	&  6 & 398 & 510 & 591 & 587 & 394 & 398 & 401 & 319 & 338 & 370 \\ 
	&  7 & 420 & 583 & 618 & 672 & 547 & 420 & 415 & 339 & 354 & 394 \\ 
	&  8 & 443 & 675 & 651 & 780 & 450 & 443 & 432 & 361 & 371 & 420 \\ 
	&  9 & 465 & 763 & 688 & 885 & 633 & 465 & 446 & 386 & 386 & 441 \\ 
	& 10 & 488 & 875 & 750 &1033 & 492 & 488 & 473 & 409 & 398 & 484 \\ 
	& 11 & 500 & 998 & 761 &1191 & 775 & 500 & 491 & 439 & 421 & 513 \\ 
	& 12 & 520 &1159 & 862 &1377 & 542 & 520 & 518 & 461 & 448 & 585 \\ 
	& 13 & 527 &1306 & 914 &1572 &1020 & 527 & 470 & 482 & 473 & 658 \\ 
	& 14 & 563 &1533 &1058 &1877 & 617 & 563 & 521 & 571 & 535 & 783 \\ 
	& 15 & 627 &1869 &1225 &2203 &1432 & 627 & 588 & 700 & 621 & 924 \\ 
	& 16 & 843 &2151 &1456 &2526 & 848 & 843 & 807 & 991 & 865 &1139 \\ 
\cmidrule{2-12}
	& Mean & 481 & 481 & 784 & 1281 & 705 & 481 & 491 & 453 & 440 & 534 \\ 
	& Median & 443 & 447 & 688 & 885 & 471 & 443 & 446 & 439 & 421 & 431 \\ 
\bottomrule
\end{longtable}
\end{footnotesize}
\end{center}

\vspace{-.3in}

\begin{figure}[!htb]
\centering
\subfloat[CPD for females]
{\includegraphics[width=9cm]{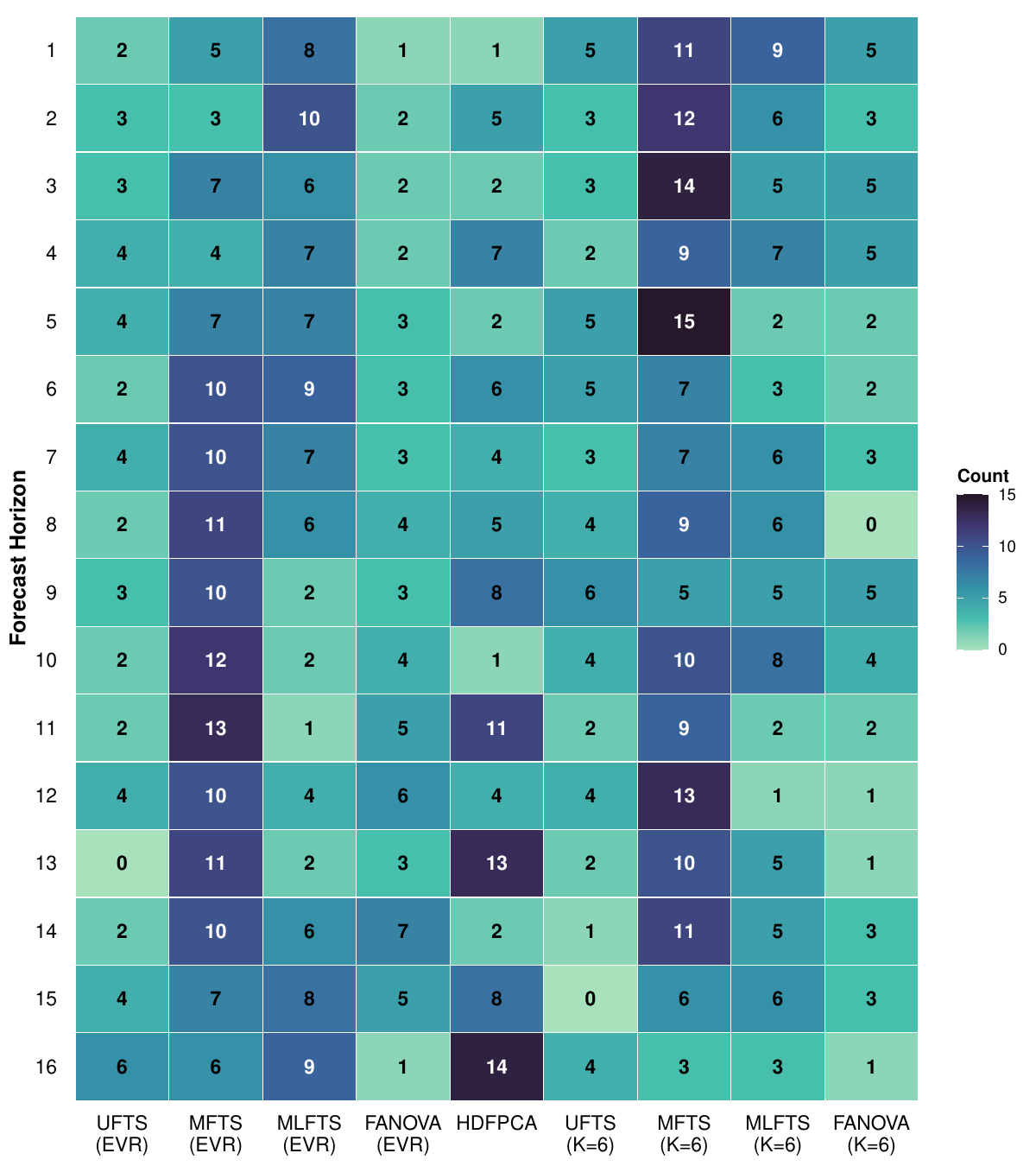}}
\subfloat[CPD for males]
{\includegraphics[width=9cm]{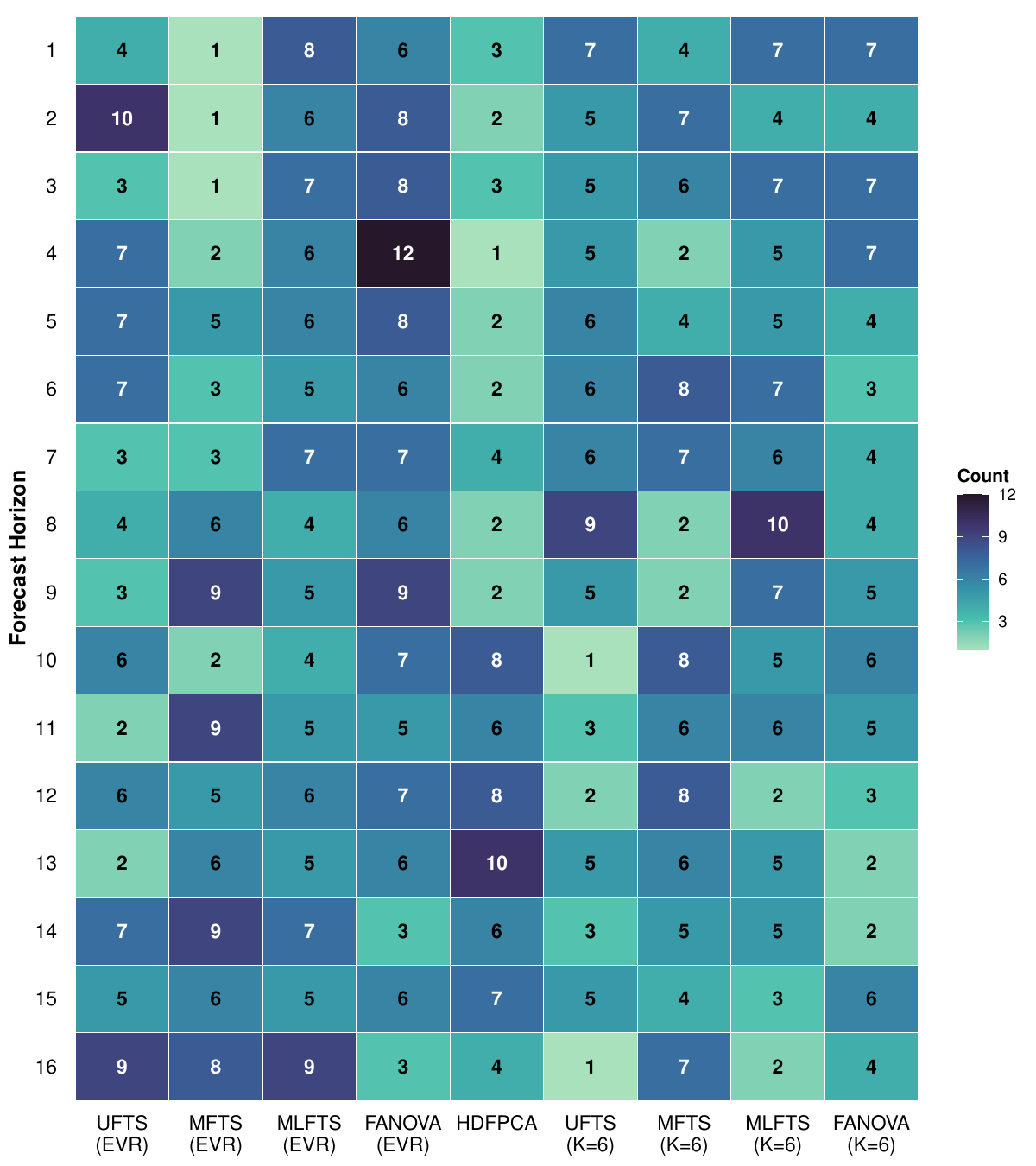}}
\\
\subfloat[Interval score for females]
{\includegraphics[width=9cm]{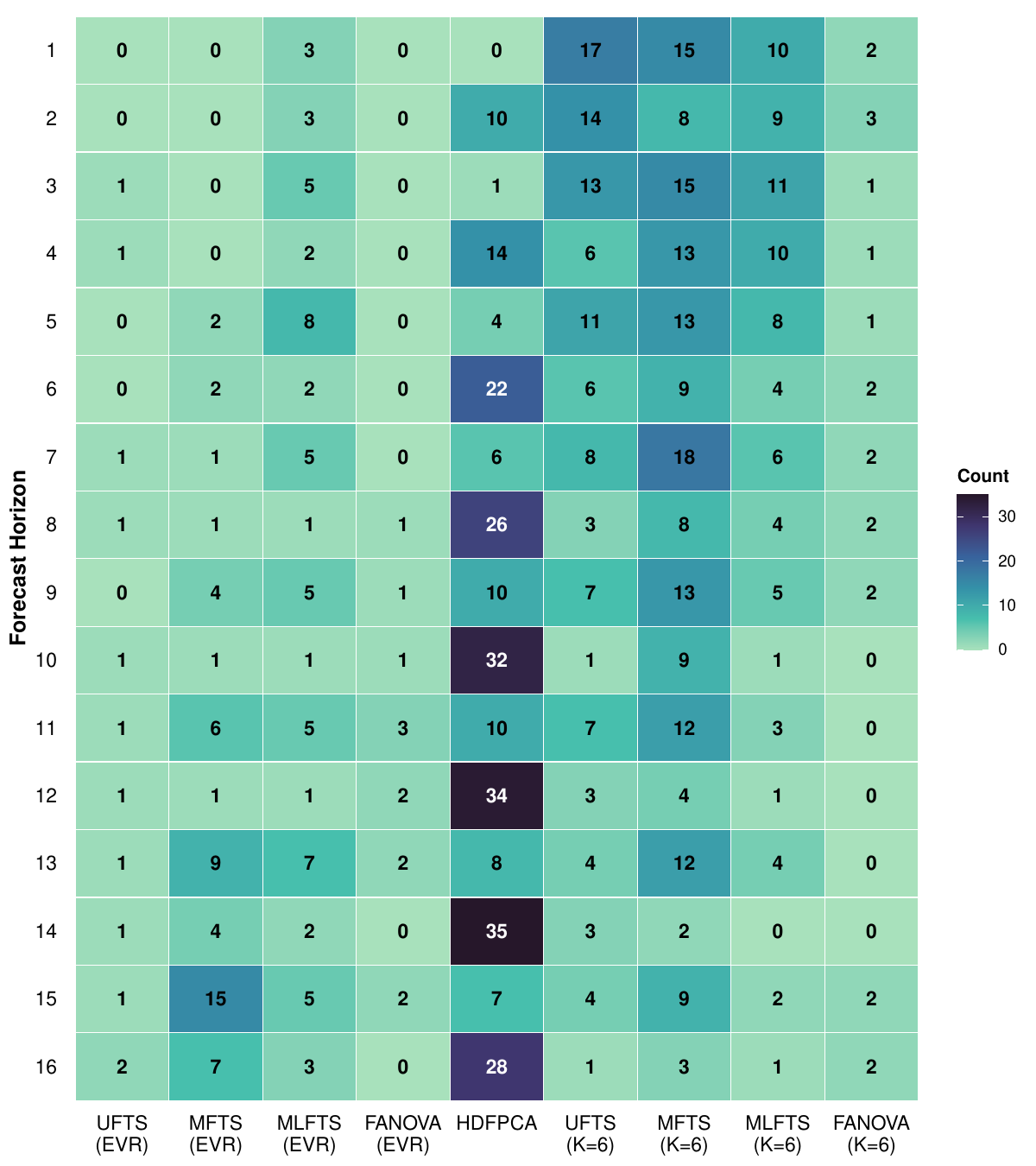}}
\subfloat[Interval score for males]
{\includegraphics[width=9cm]{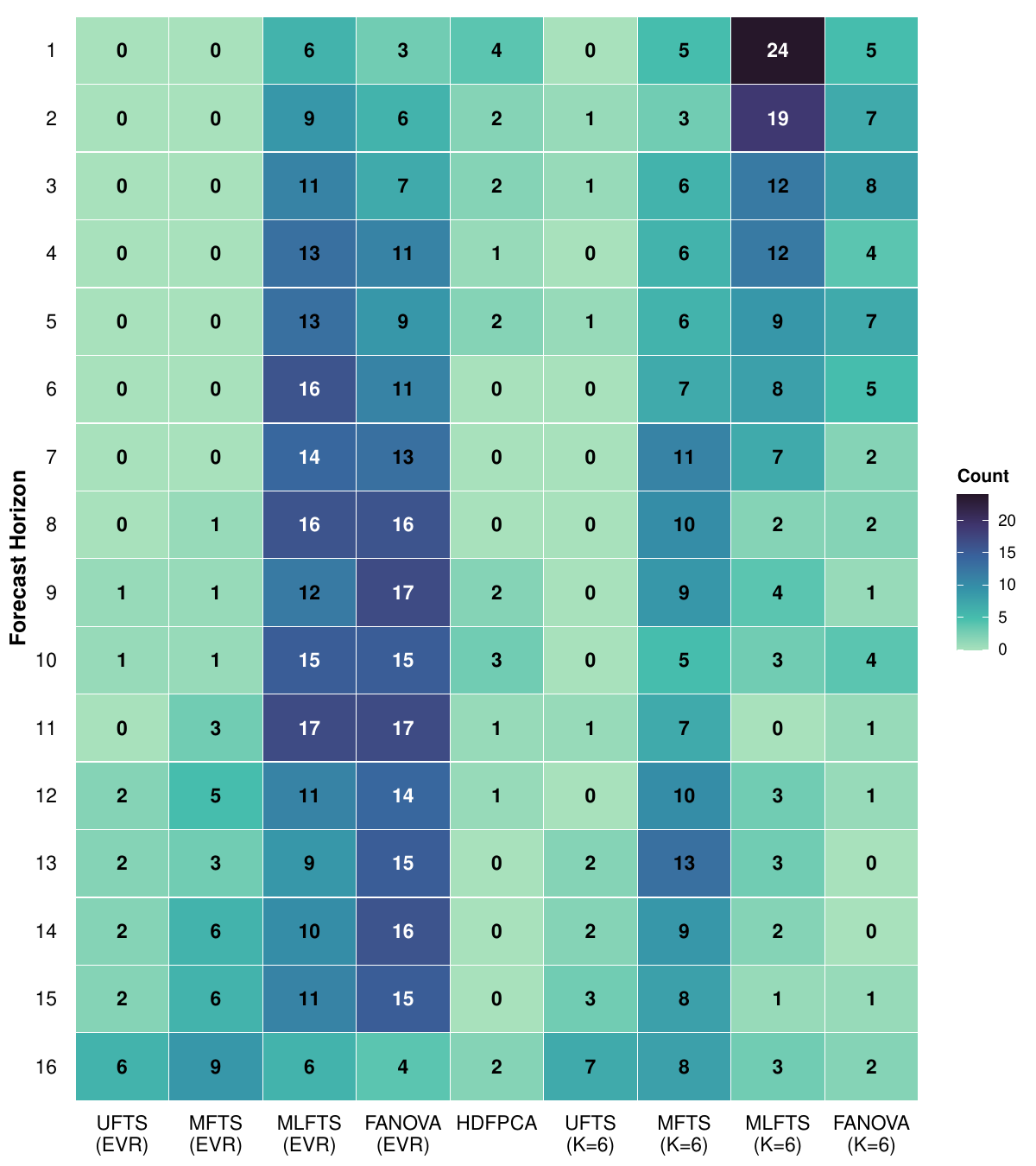}}
\caption{\small Interval forecast accuracy (CPD and interval score) at $\alpha=0.2$ for nine models across 16 horizons. Counts reflect the number of prefectures per horizon.}\label{fig:5}
\end{figure}

In Figure~\ref{fig:5}, we present two heatmaps comparing nine functional time-series models across 16 forecast horizons for both female and male testing data. For each horizon, we count how often each method produces the smallest interval forecast errors, as measured by the CPD and interval score. Based on the CPD results, we recommend the MFTS for the female data and the MLFTS for the male data, especially for $K=6$. By examining the interval score results, we recommend the MFTS for shorter horizons and the HDFPCA for longer horizons in the female data. For the male data, the MFTS generally provides the smallest interval scores.

\begin{figure}[!htb]
\centering
% --- Row 1: UFTS ---
\includegraphics[width=0.32\textwidth]{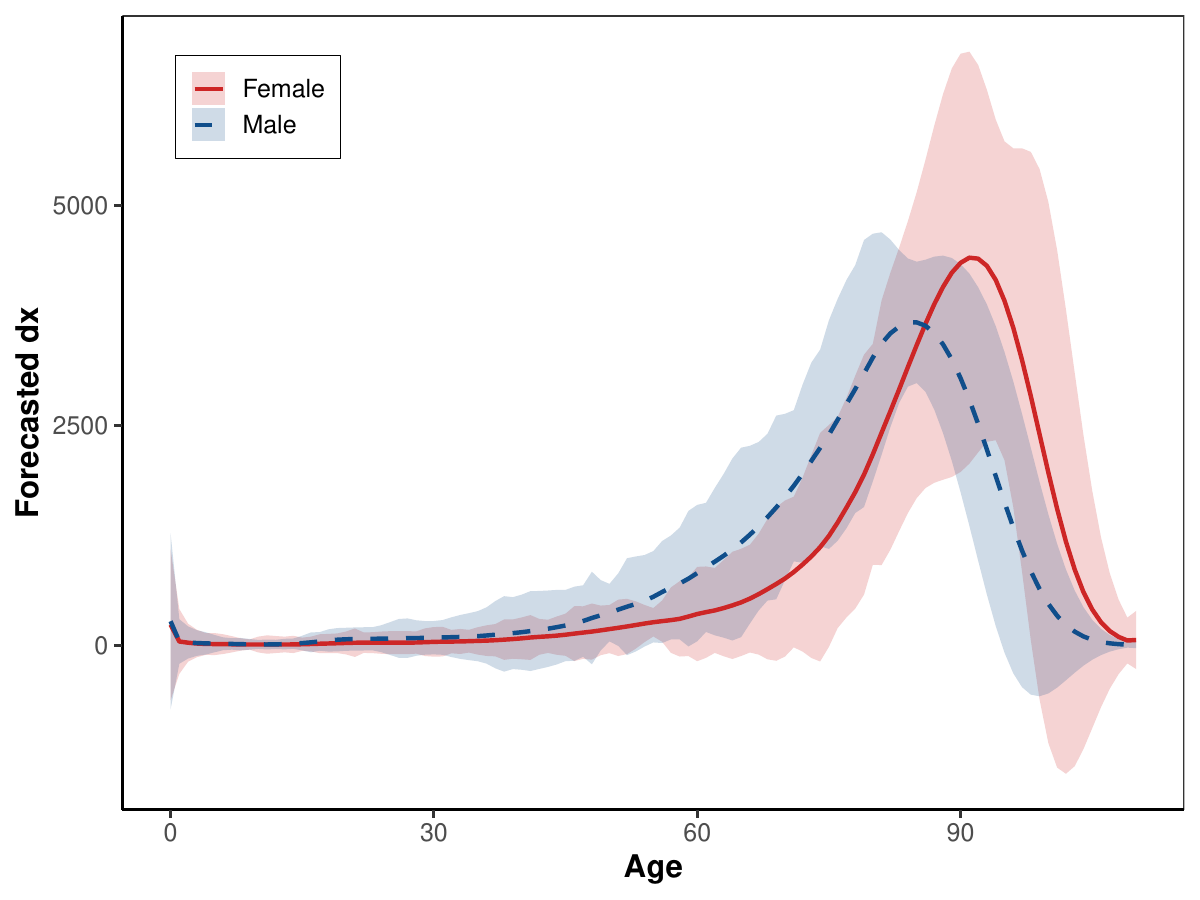} \hfill
\includegraphics[width=0.32\textwidth]{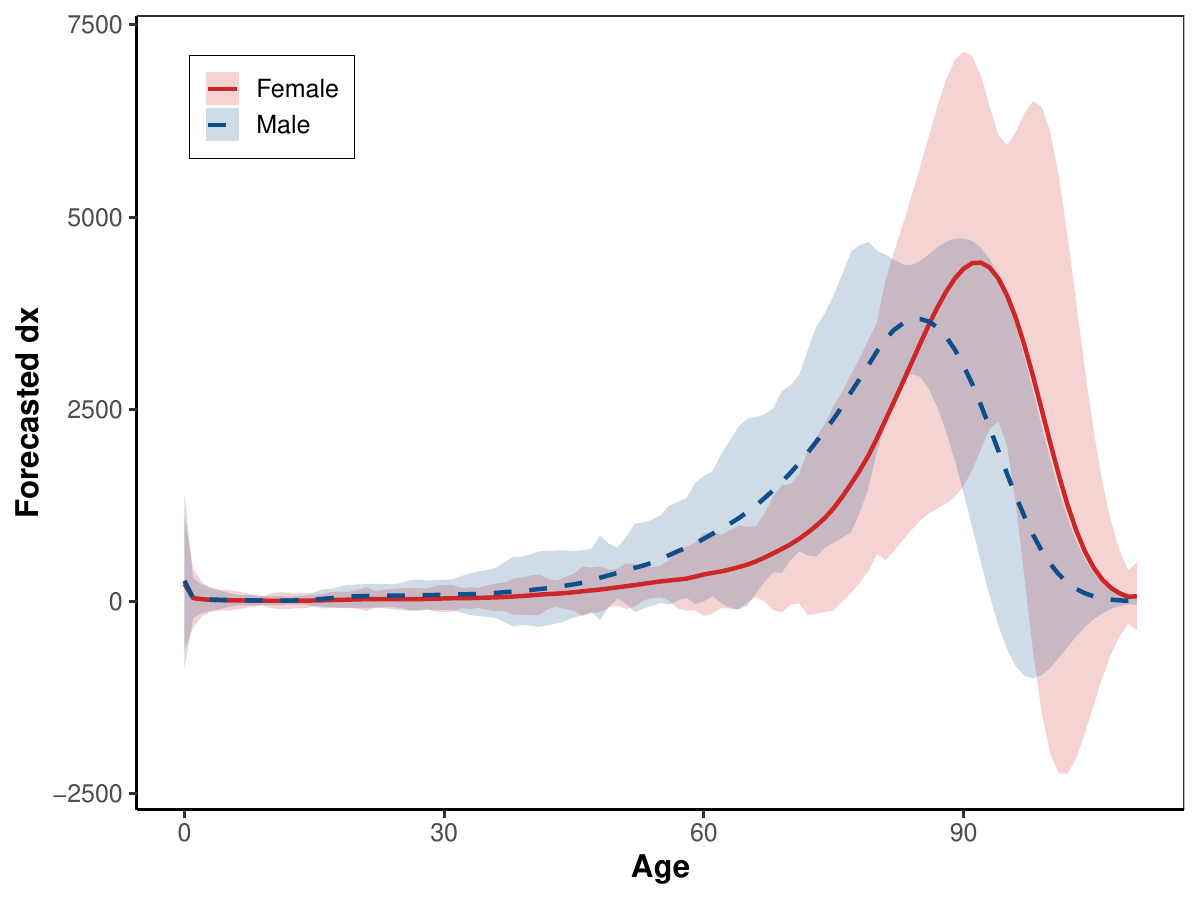} \hfill
\includegraphics[width=0.32\textwidth]{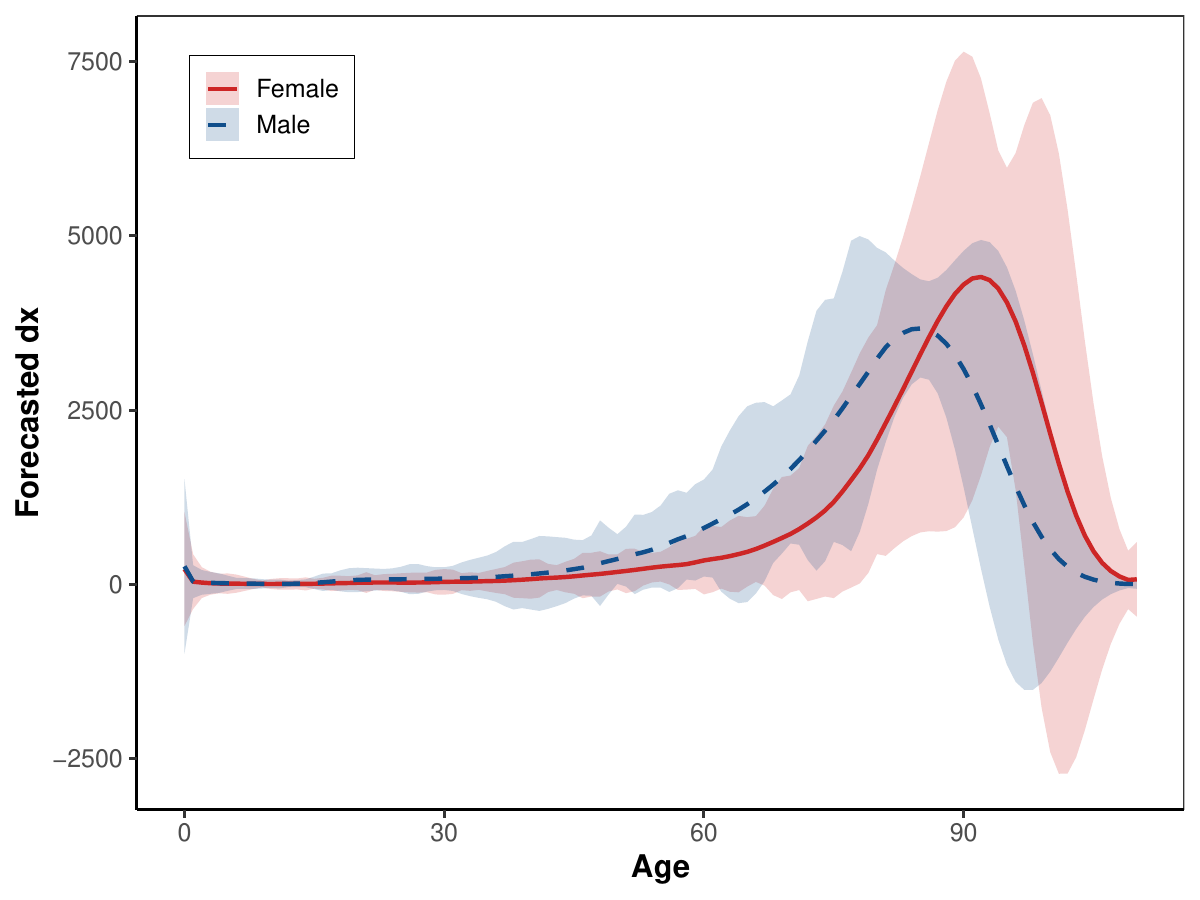}
\\[1ex] 

% --- Row 2: MFTS ---
\includegraphics[width=0.32\textwidth]{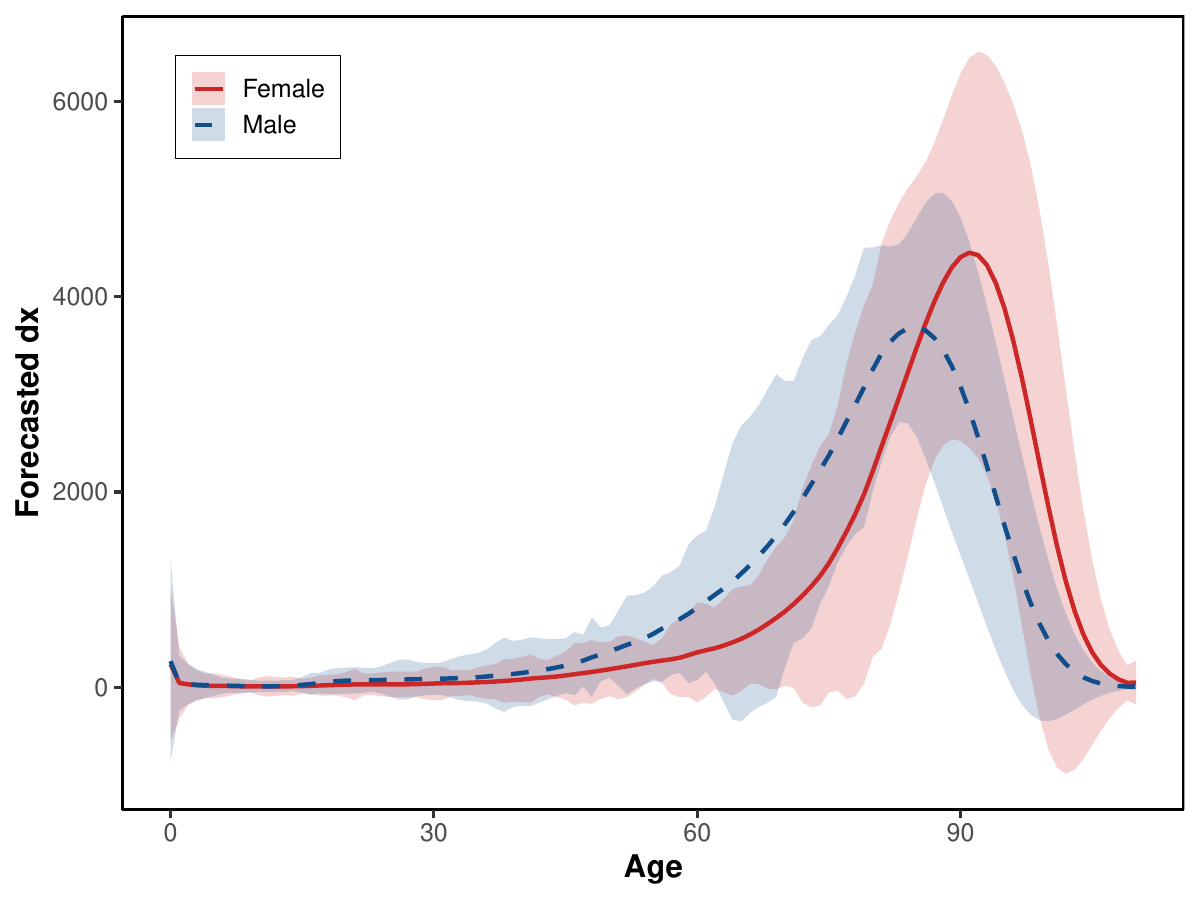} \hfill
\includegraphics[width=0.32\textwidth]{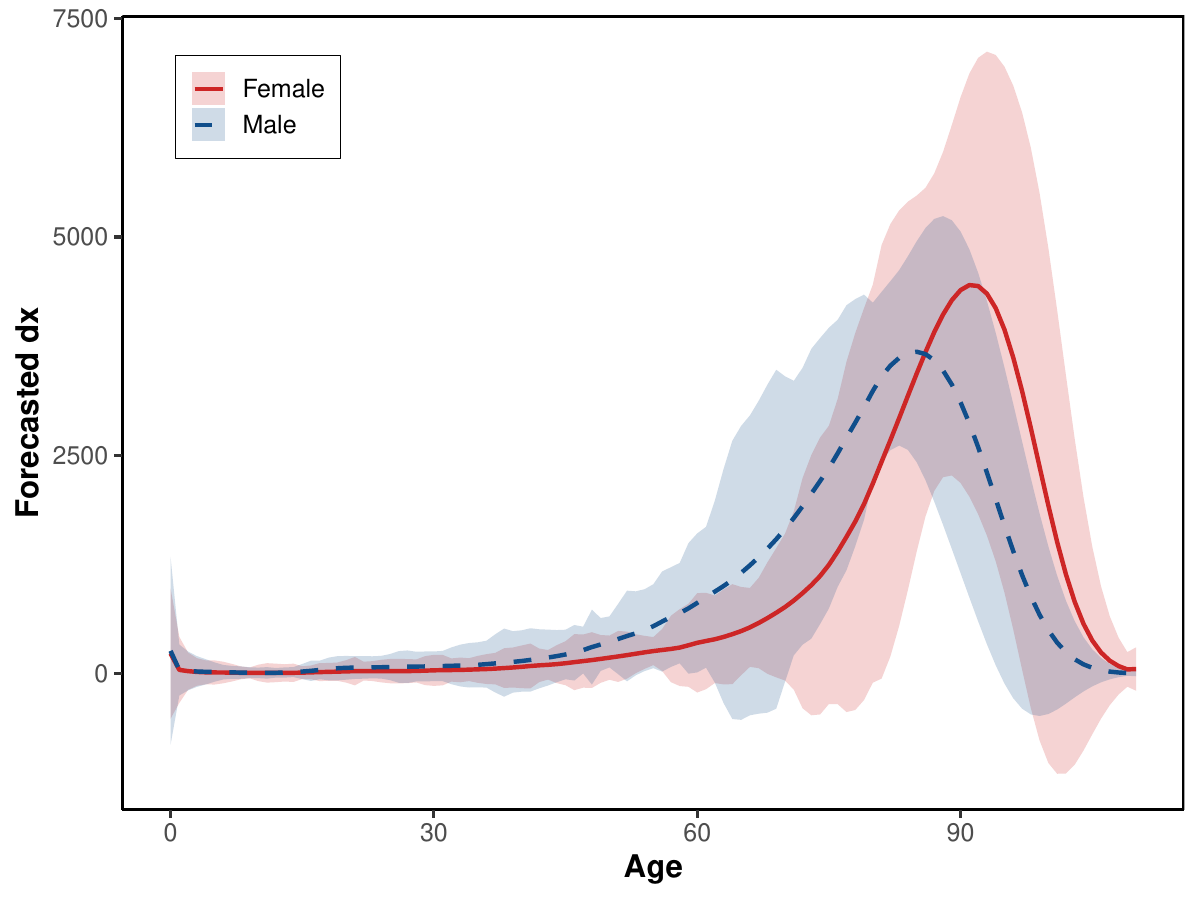} \hfill
\includegraphics[width=0.32\textwidth]{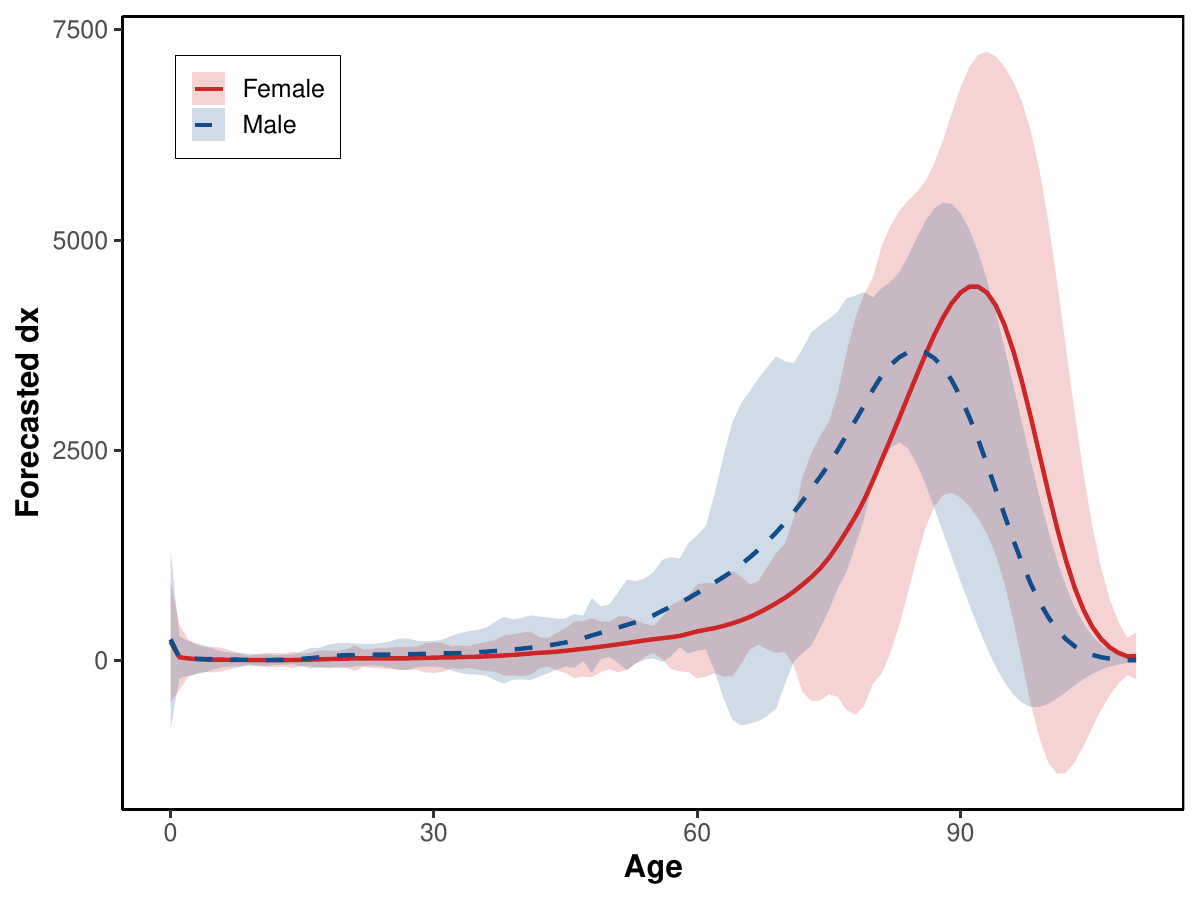}
\\[1ex]

% --- Row 3: MLFTS ---
\includegraphics[width=0.32\textwidth]{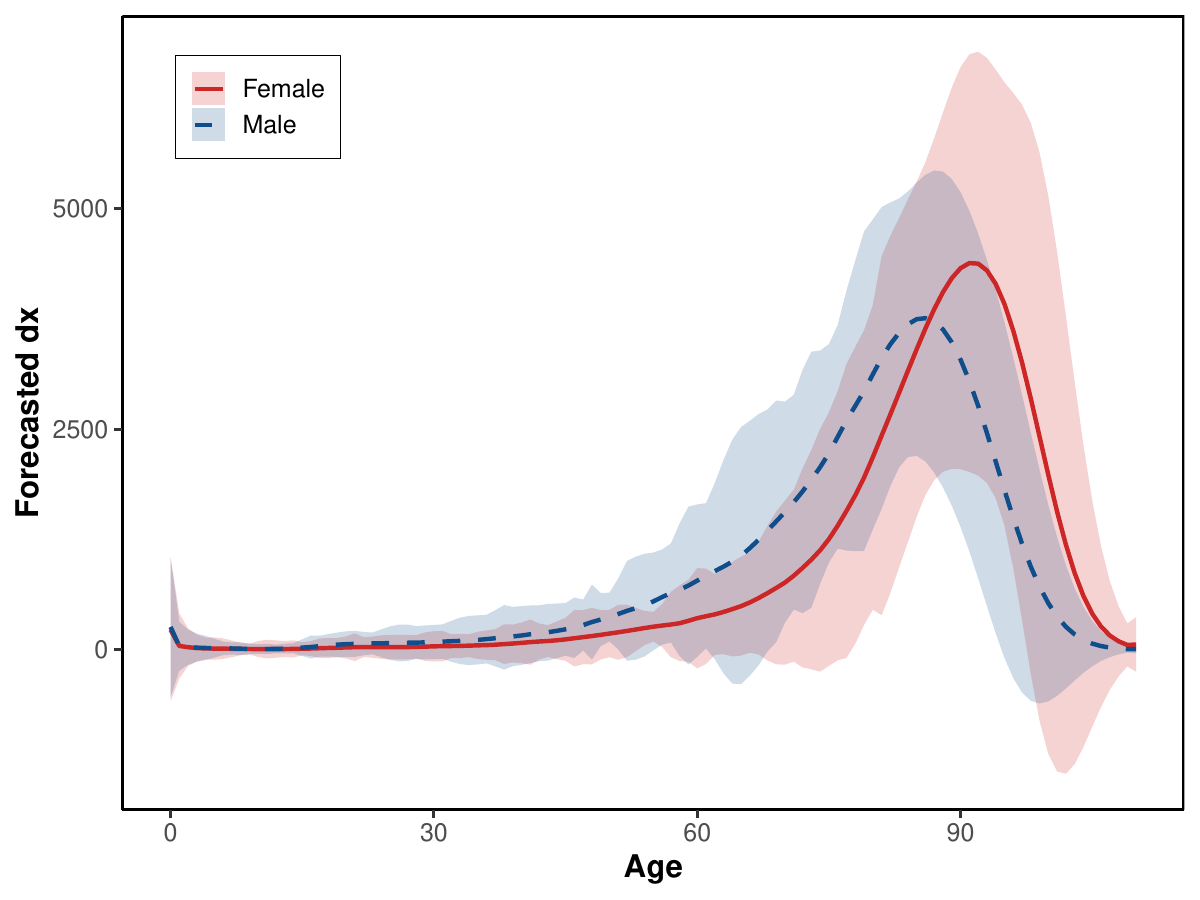} \hfill
\includegraphics[width=0.32\textwidth]{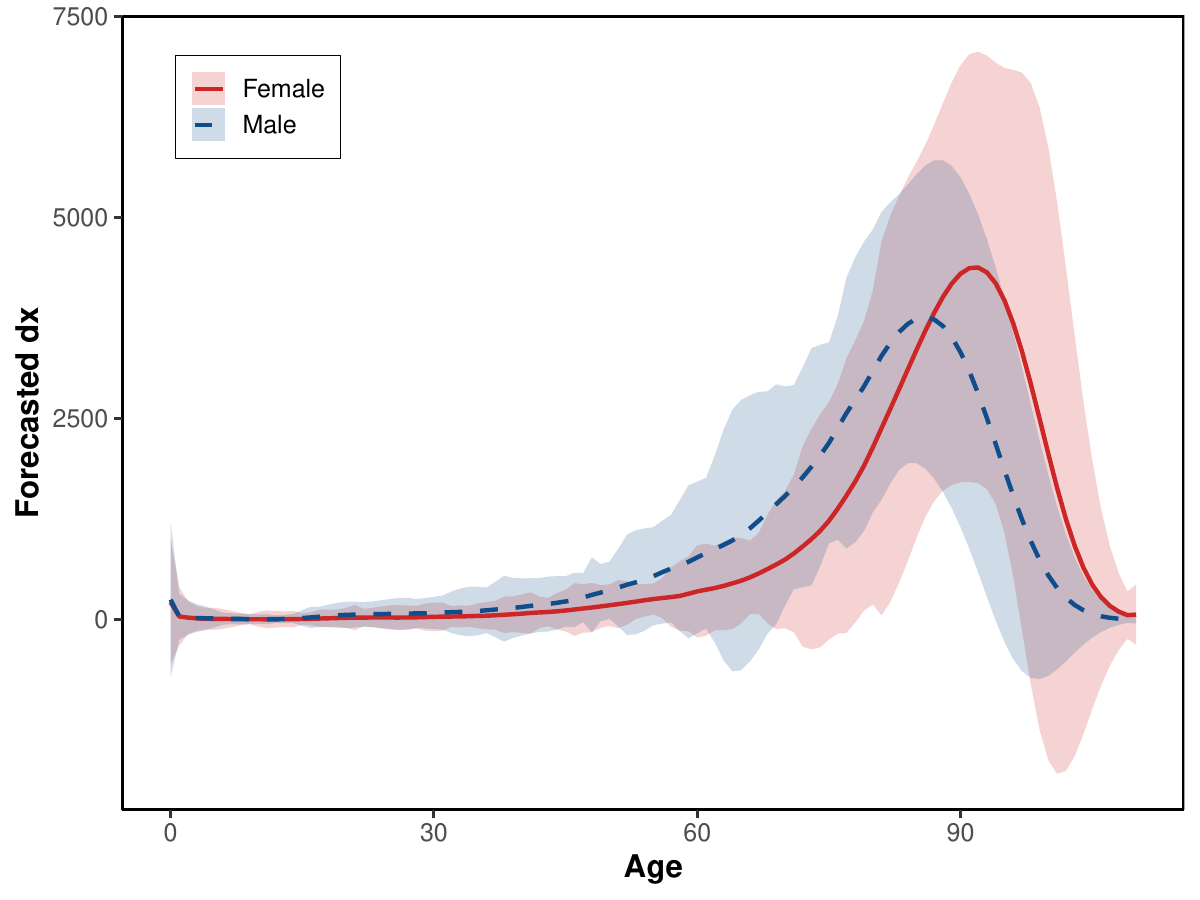} \hfill
\includegraphics[width=0.32\textwidth]{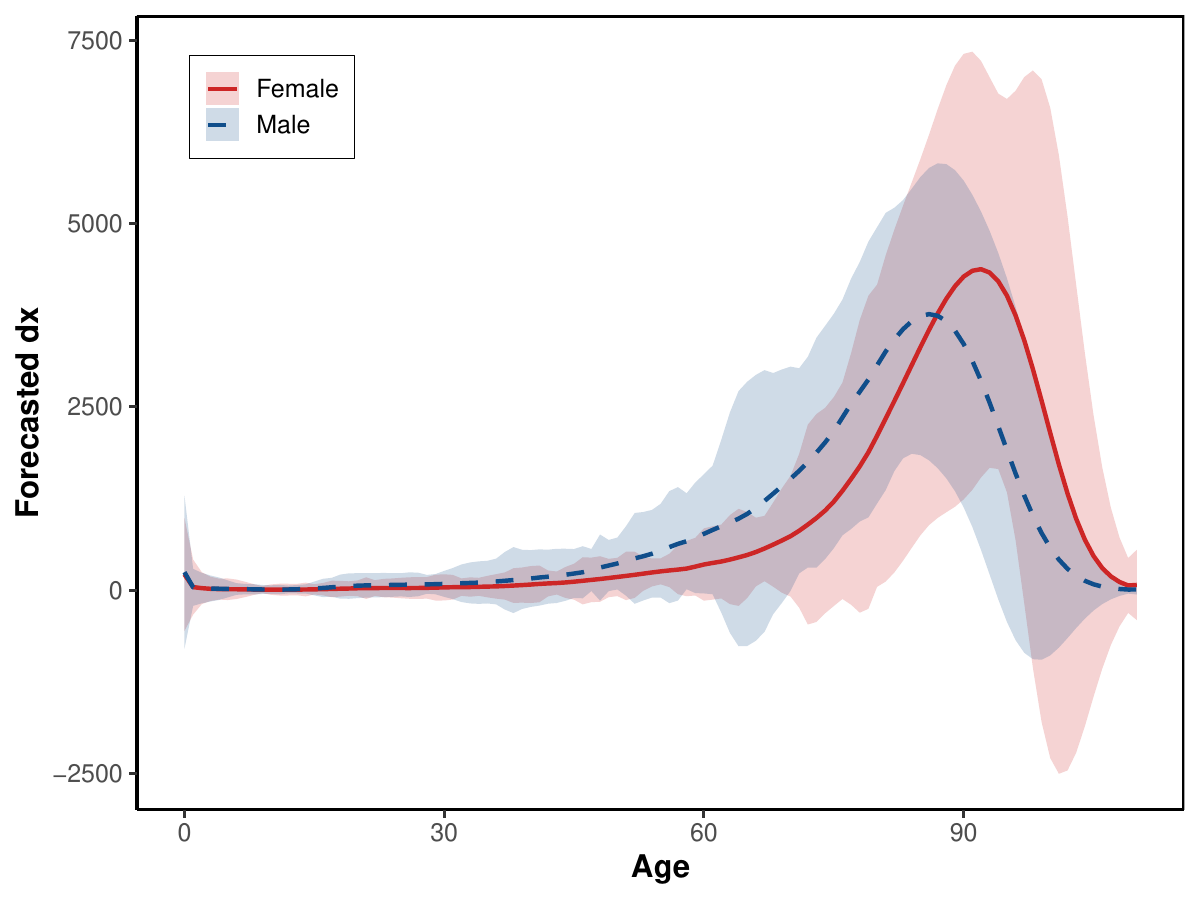}
\\[1ex]

% --- Row 4: FANOVA ---
\includegraphics[width=0.32\textwidth]{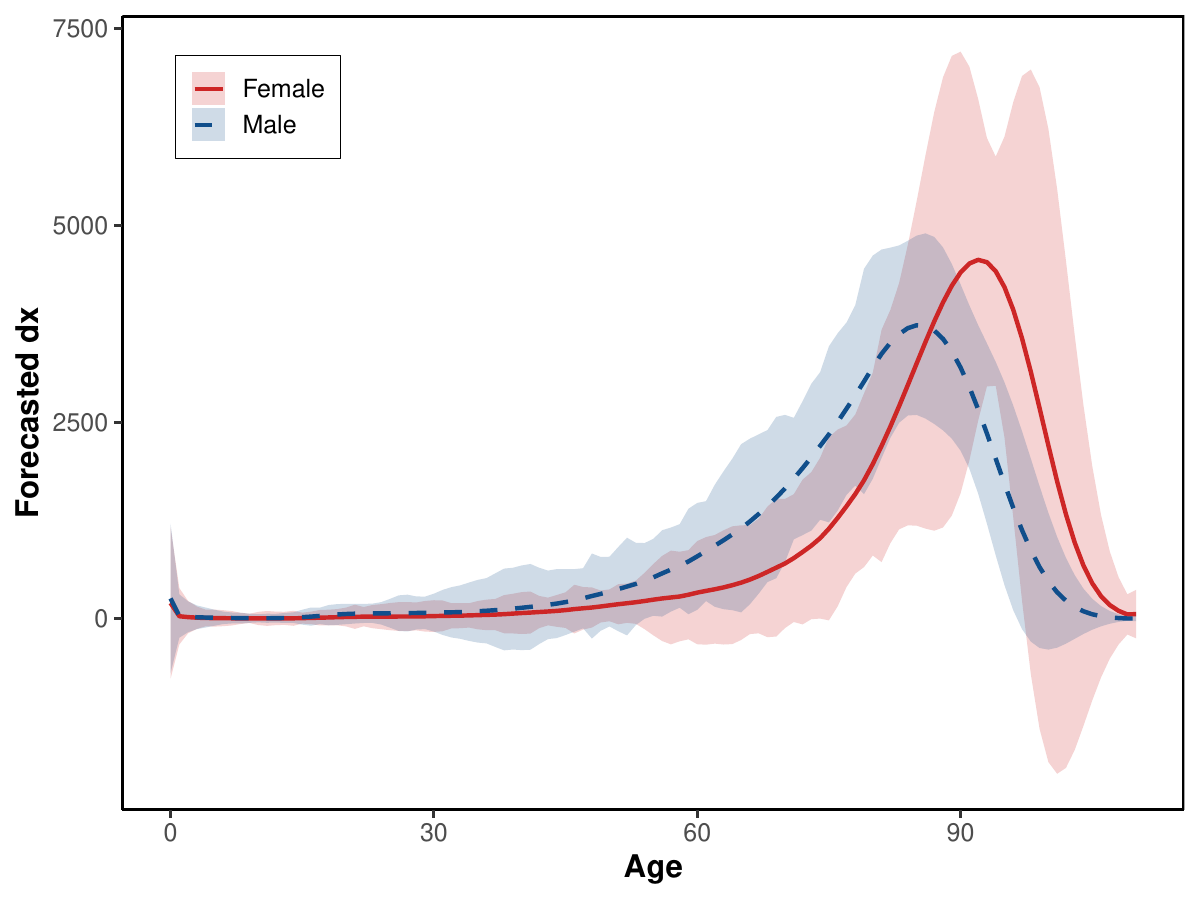} \hfill
\includegraphics[width=0.32\textwidth]{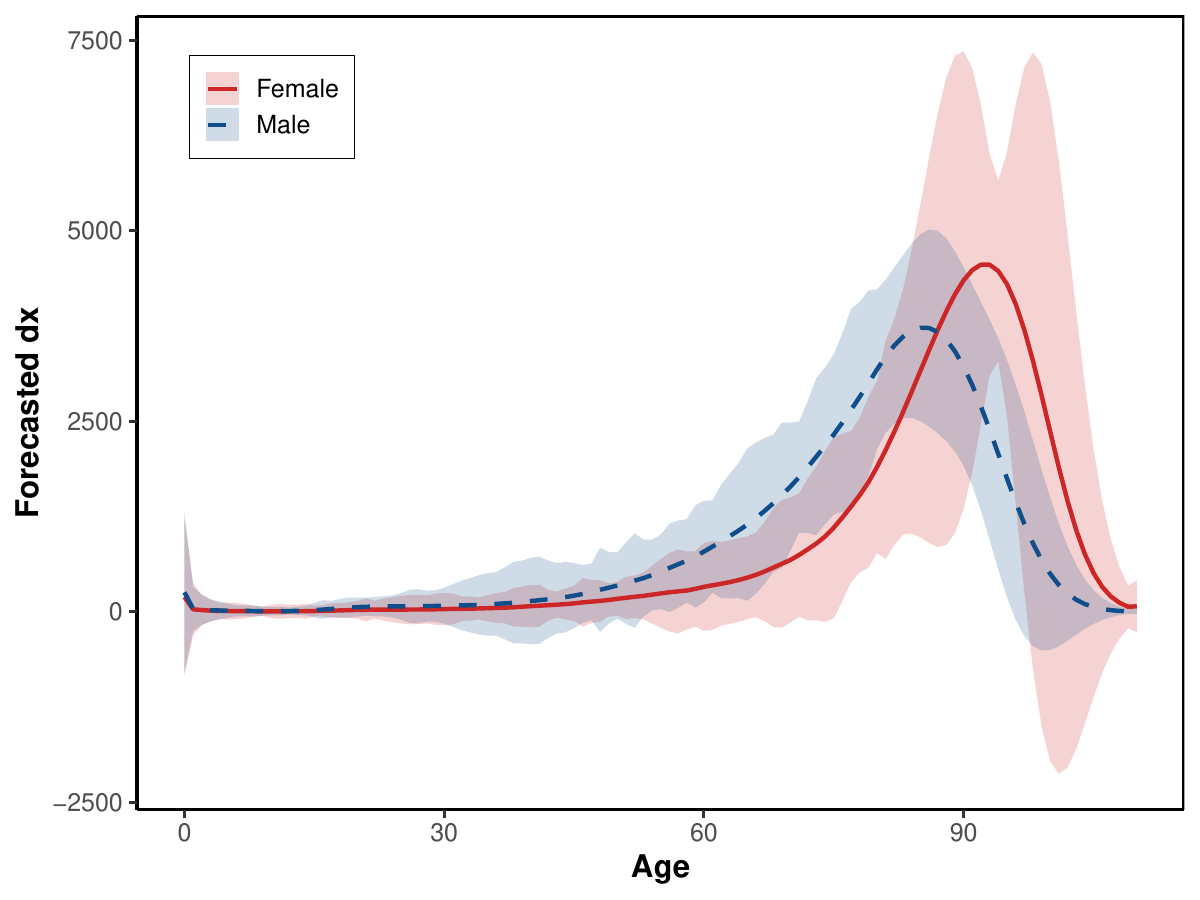} \hfill
\includegraphics[width=0.32\textwidth]{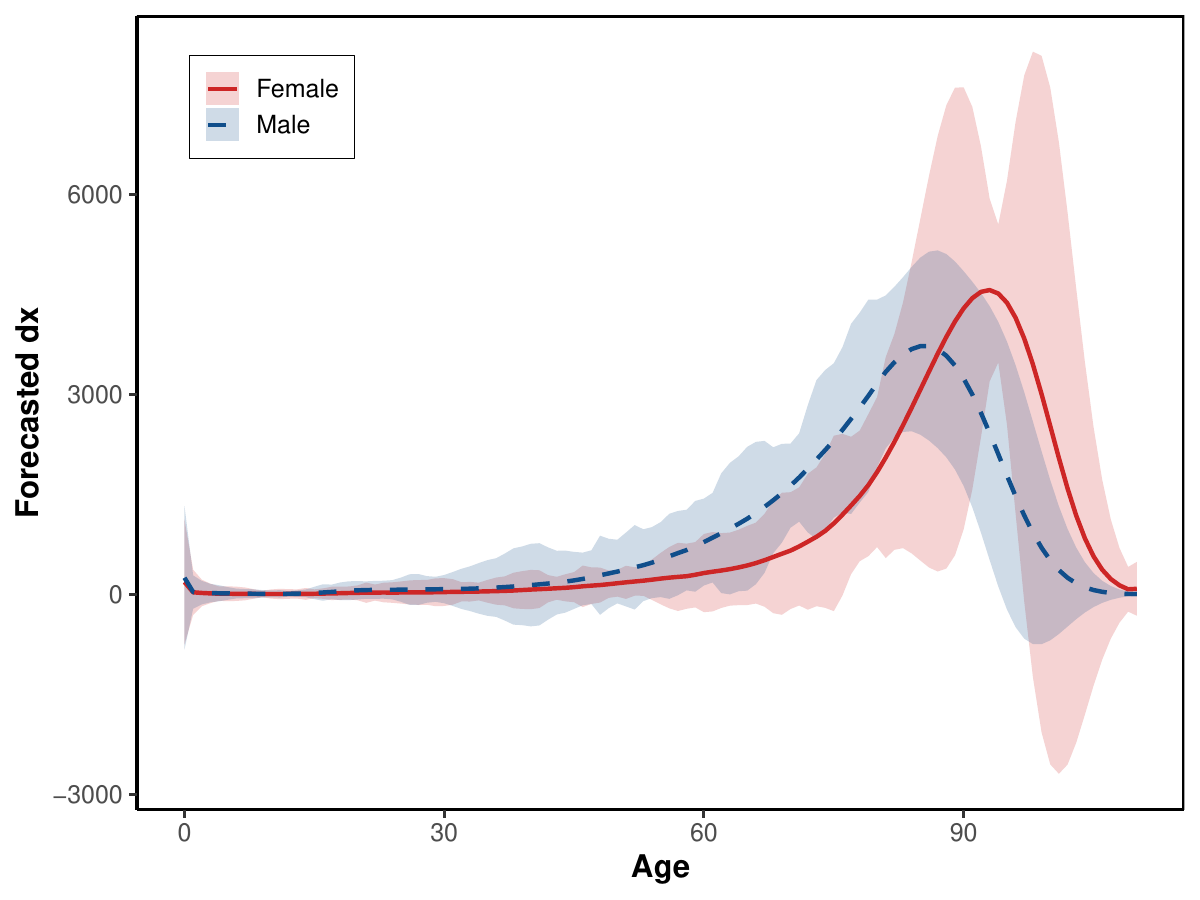}

\caption{\small Forecast age-specific death counts ($d_x$) for Hokkaido at horizons $h=1, 2, 3$ (left to right). Rows represent the UFTS, MFTS, MLFTS, and FANOVA methods respectively. Shaded areas denote the 80\% pointwise prediction intervals (Red: Female, Blue: Male).}
\label{fig:comparison_forecasts}
\end{figure}

As an illustration, Figure~\ref{fig:comparison_forecasts} displays the forecasted age-specific death counts ($d_x$) for Hokkaido at horizons $h=1, 2, 3$. We have included the 80\% pointwise prediction intervals (Red: Female, Blue: Male) for all considered methods. Across all rows, it can be observed that the prediction intervals for females are generally wider compared to those for males, reflecting the differing levels of uncertainty captured by the UFTS, MFTS, MLFTS, and FANOVA approaches.

The interval forecast results at the significance level of $\alpha = 0.05$ are available and can be viewed in a developed Shiny app at the following \href{https://cfjimenezv.shinyapps.io/Shiny_App_CDF/}{link}.

\subsection{Forecasting age-specific survival probabilities and life expectancy}\label{Convert_d_x}

\begin{figure}[!htb]
\centering
\includegraphics[width=0.485\textwidth]{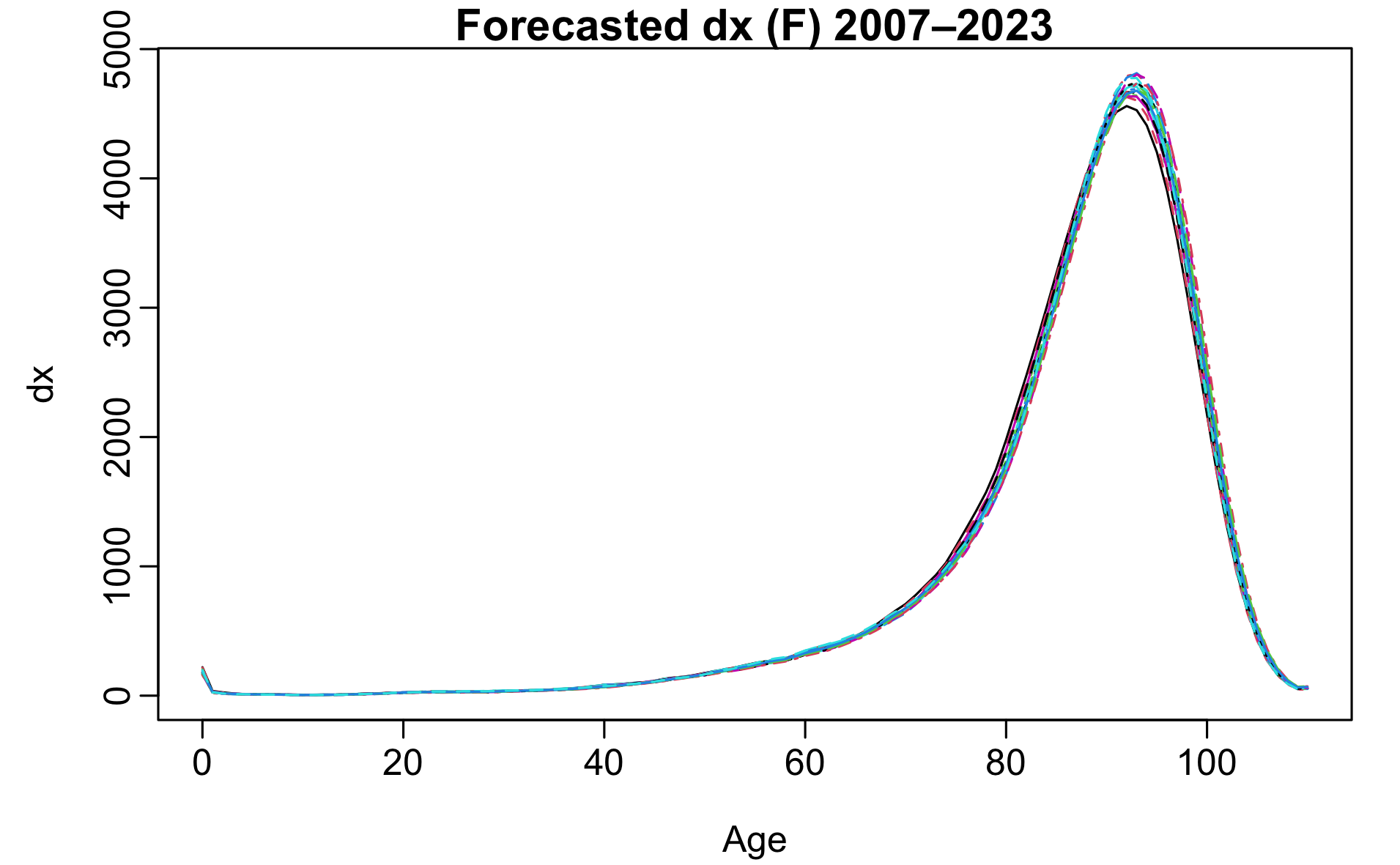}
\quad
\includegraphics[width=0.485\textwidth]{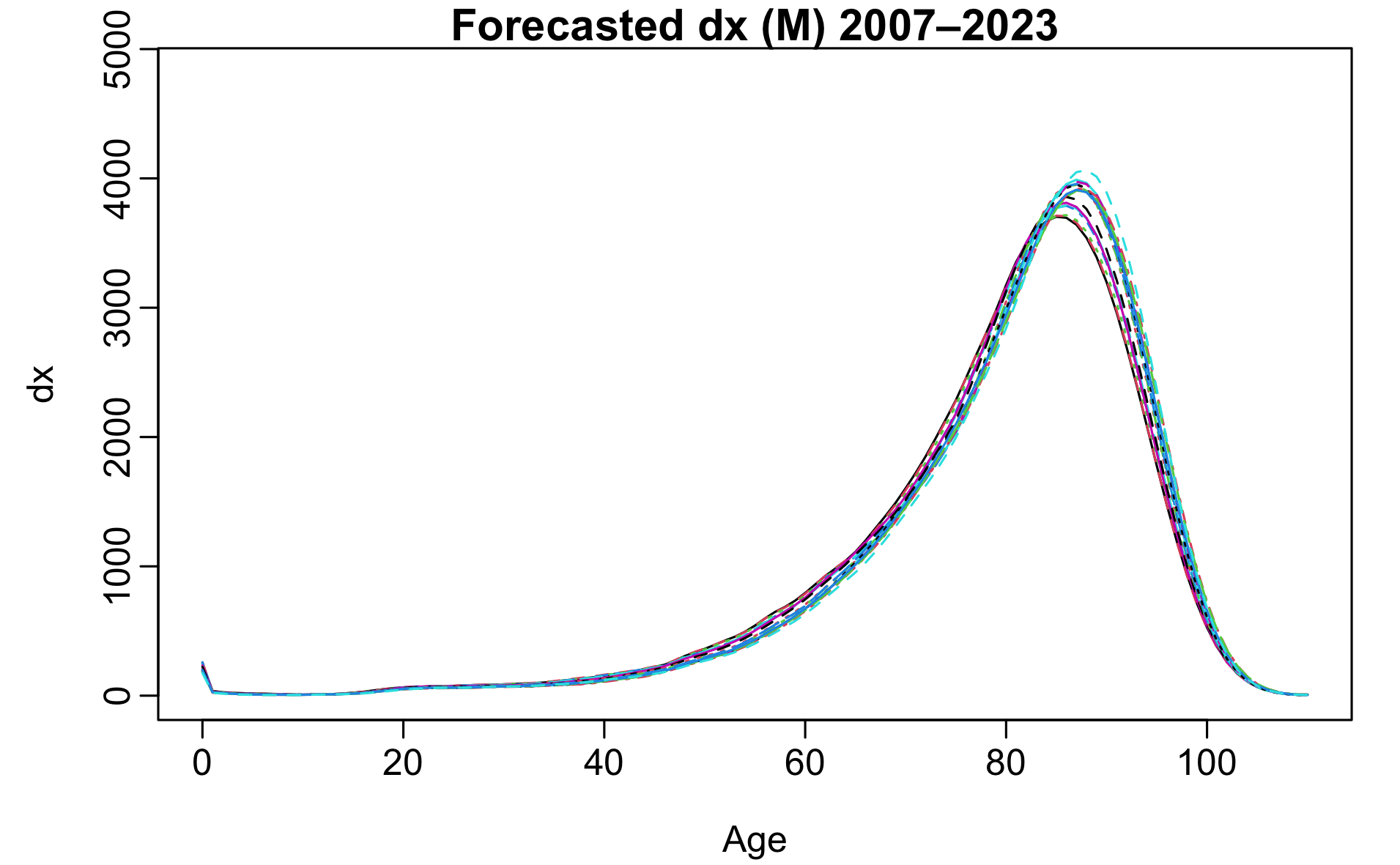}
\\
\includegraphics[width=0.485\textwidth]{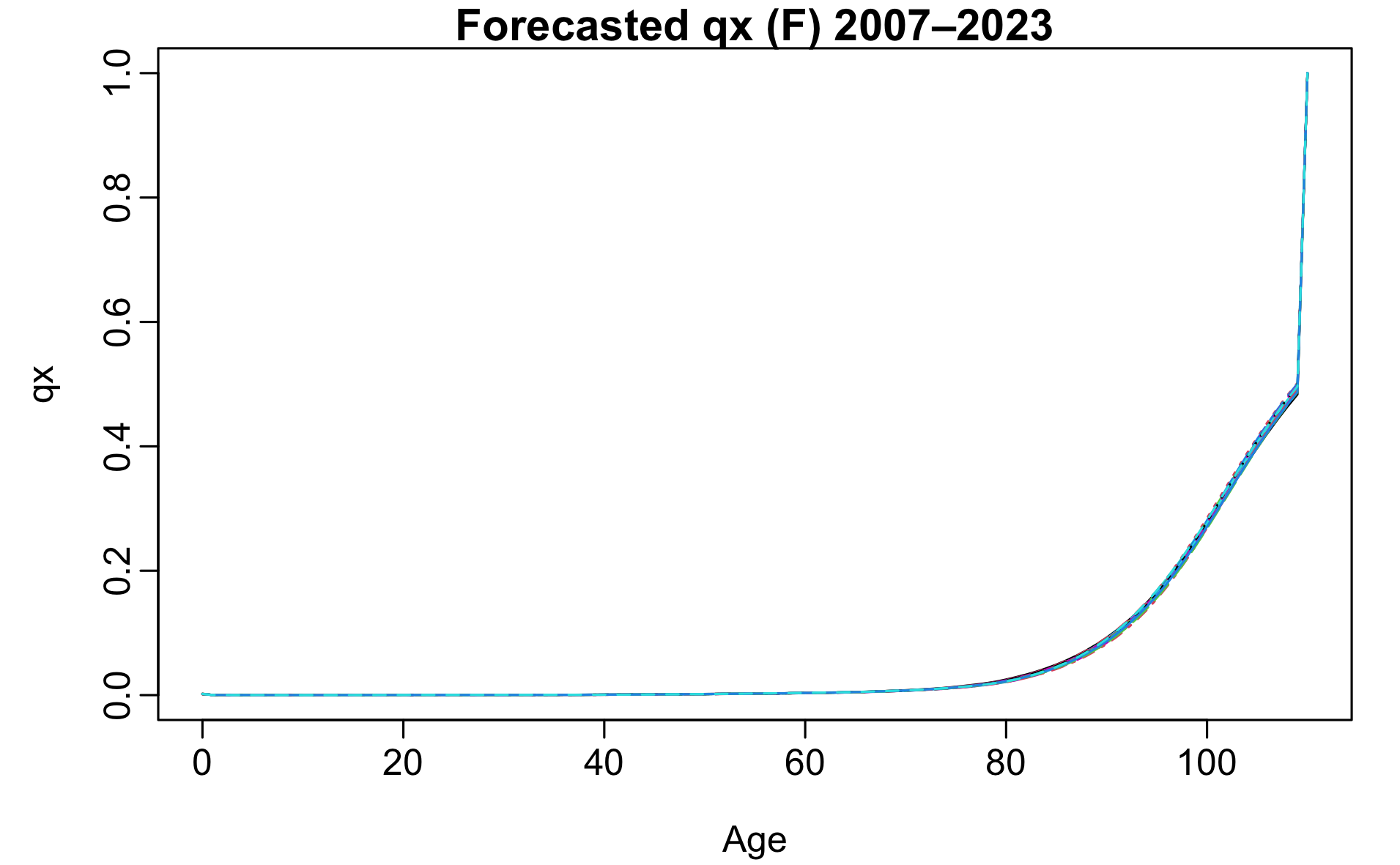}
\quad
\includegraphics[width=0.485\textwidth]{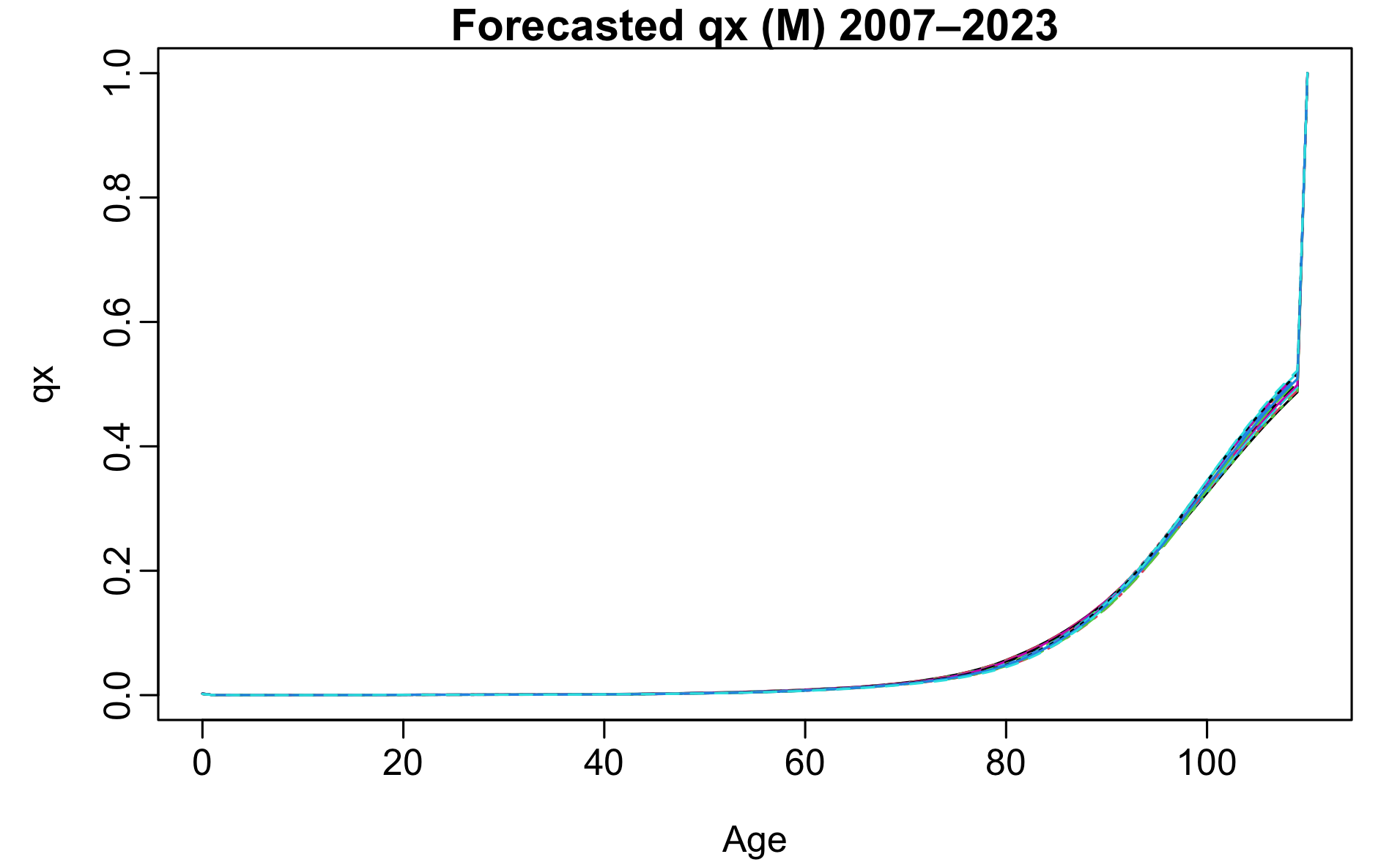}
\\
\includegraphics[width=0.485\textwidth]{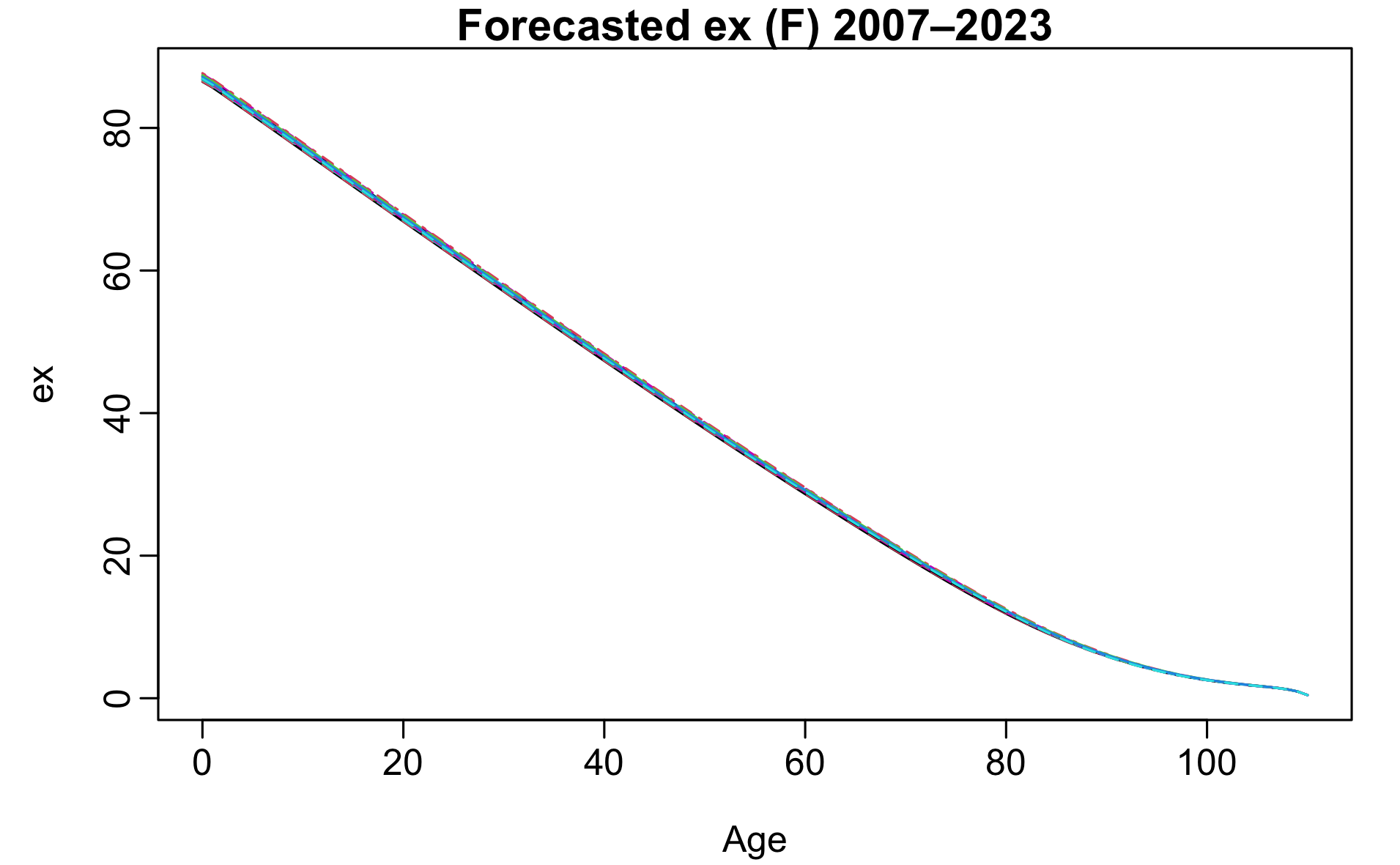}
\quad
\includegraphics[width=0.485\textwidth]{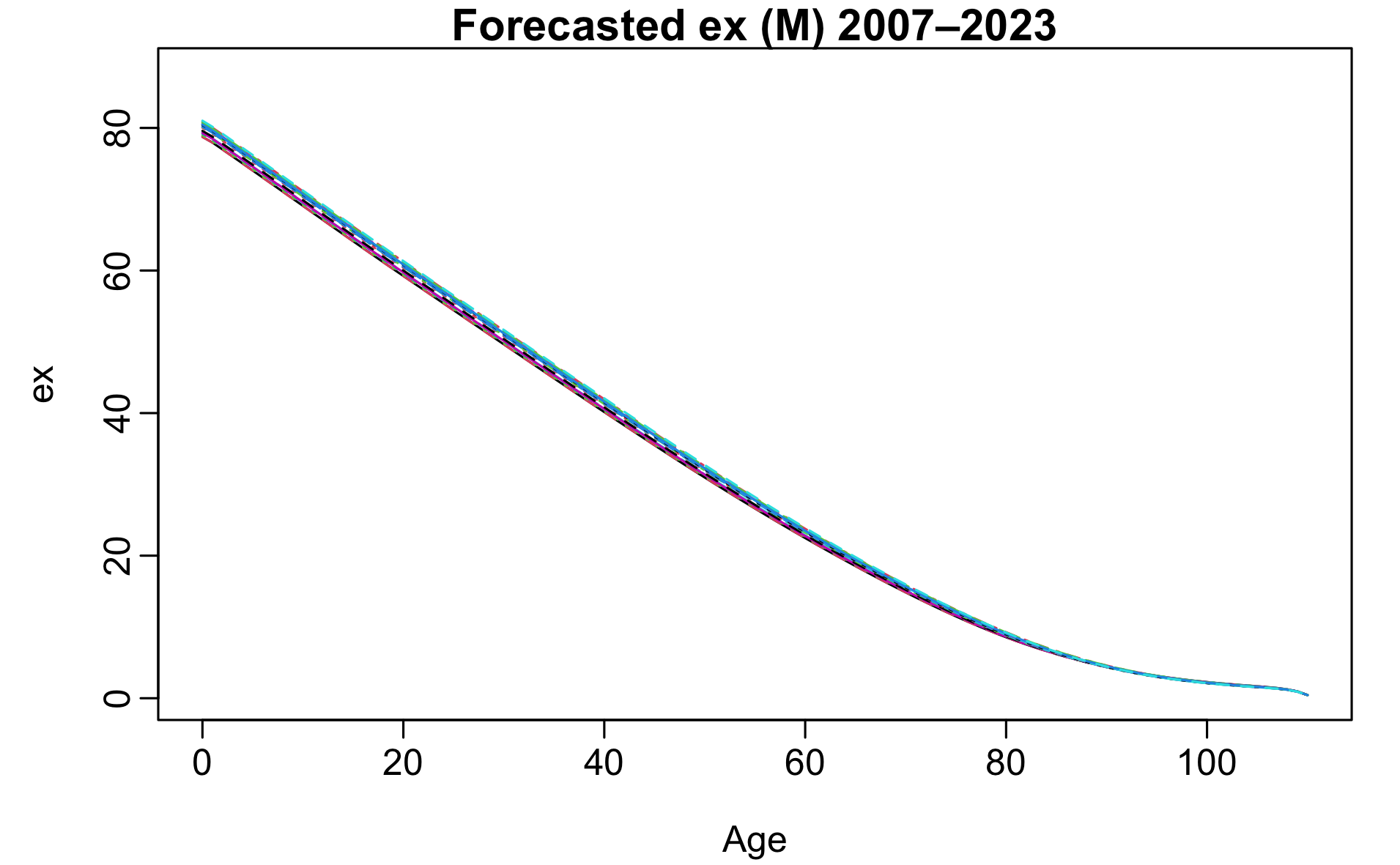}
\caption{\small Forecast life-table metrics for Hokkaido, 2007--2023. Panels show age-specific deaths ($d_x$), probability of dying ($q_x$), and life expectancy ($e_x$) for females (F) and males (M).}\label{fig:forecast_plots}
\end{figure}

The enhanced forecast accuracy of subnational life-table death counts ($d_{x}$) enables precise estimation of key demographic indicators, such as the age-specific probability of dying ($q_x$) and life expectancy ($e_x$). These metrics are derived by constructing a complete life table from the forecasted death counts ($\widehat{d}_x$). Specifically, we apply the standard life-table relationships using the \texttt{LifeTable} and \texttt{convertFx} functions from the \texttt{MortalityLaws} package \citep{Pascariu25} in \Rlogo.\ The life-table radix ($l_0$) is set to 100,000 for all forecasted tables. For illustration, Figure~\ref{fig:forecast_plots} shows the forecast distributions of deaths, probabilities of dying, and life expectancy for the prefecture Hokkaido for the years 2007 to 2023, based on data from 1975 to 2006.

\subsection{Model diagnostics}\label{sec:model_diagnostics}

We illustrate model diagnostics using the FANOVA residuals to assess the weak functional white noise assumption, focusing on stationarity and conditional heteroskedasticity. Stationarity was tested using the $\mathbf{T}$-statistic-based method for functional time series \citep[\texttt{T\_stationary} in \texttt{ftsa} package in \Rlogo,][]{HS25}, separately for female and male residuals (matrices $110 \times 49$ per prefecture). Conditional heteroskedasticity was assessed using the functional conditional heteroskedasticity test \citep[\texttt{fCH\_test} in \texttt{FTSgo} package in \Rlogo,][]{RWZ20}. Table~\ref{tab:stationarity} summarizes the $p$-values across the 47 prefectures. Since all $p$-values exceed 0.05, the residuals can be considered approximately stationary. Small $p$-values indicate some heteroskedasticity, but as our methods model the conditional mean, variance modeling is beyond the scope of this study; functional GARCH-type models could be explored in future work.
\begin{table}[!htb]
\centering
\tabcolsep 0.24in
\caption{\small Summary of $p$-values from stationarity and conditional heteroskedasticity tests. From the stationarity test, the series exhibit stationarity at the 5\% level of significance. From the conditional heteroskedasticity test, the series exhibits heteroskedasticity.}
\begin{tabular}{@{}lrrrrrr@{}}
\toprule
Gender & Min & 1st Qu. & Median & Mean & 3rd Qu. & Max \\
\midrule
\multicolumn{6}{l}{\hspace{-.23in}{\underline{Stationary test}}} & \\
Female & 0.060 & 0.078 & 0.085 & 0.0853 & 0.091 & 0.108 \\
Male   & 0.057 & 0.0705 & 0.079 & 0.0784 & 0.0855 & 0.099 \\
\\
\multicolumn{7}{l}{\hspace{-.23in}{\underline{Conditional heteroskedasticity test}}} \\
Female & 0.0000 & 8.22e-15 & 1.64e-12 & 7.67e-04 & 2.58e-10 & 3.60e-02 \\
Male   & 3.85e-13 & 3.35e-09 & 6.56e-06 & 1.61e-02 & 2.13e-04 & 4.88e-01 \\
\bottomrule
\end{tabular}
\label{tab:stationarity}
\end{table}

\section{Conclusion}\label{sec:7}

We present several graphical tools, including image plots, to visualize the differences and dependence in life-table death counts between each prefecture and Japan as a whole. 

Via the CDF transformation, we consider a suite of functional time-series models to model and forecast the subnational age distribution of death counts for 47 Japanese prefectures. The CDF transformation is ideal for handling zero values in subnational data. Within the CDF transformation, we also present two general strategies for constructing pointwise prediction intervals.

By dividing the datasets into training and testing samples, we evaluate and compare point and interval forecast accuracy among the functional time-series models. We confirm the importance of joint modeling techniques, in which the MFTS, MLFTS, FANOVA, and HDFPCA models generally perform well on the aggregate level across 47 prefectures. We present a novel heatmap showing the frequency (of 47 prefectures), in which one method performs the best for each forecast horizon. Individual forecast errors for various horizons, obtained from all methods for each prefecture, are available in a developed shiny app at the following \href{https://cfjimenezv.shinyapps.io/Shiny_App_CDF/}{link}. The forecast life-table death counts are not only important to compute age-specific survival probability and life expectancy in terms of demography, but they are also useful to determine annuity prices for various ages and maturities as studied in \cite{SH20}. While the primary focus of this paper is on statistical accuracy, the improvement in forecast accuracy represents a preliminary step that could eventually lead to more informed annuity pricing; in turn, these calculations help inform policymakers about the sustainability of pension systems.

There are several ways in which the methodology can be extended, and we briefly list seven. 
\begin{inparaenum}[1)]
\item The functional standard deviation in~\eqref{eq:sd} was computed coordinate-wise. There is a range of functional depth that can be implemented to compute other variants of standard deviations. 
\item Instead of symmetric prediction intervals, asymmetric ones can be considered, searching for two tuning parameters to adjust the lower and upper bounds. 
\item The data set was divided almost equally into training, validation, and testing samples. Other proportions may be possible, leading to a more accurate selection of the tuning parameter $\xi_{\alpha}$ and more accurate interval forecasts. 
\item We implemented a suite of functional time-series models for evaluation and comparison. Other time-series extrapolation models may also be considered, such as the robust functional principal component analysis by \cite{OFH+25}. Doing so extends the Mahalanobis distance to Bayes spaces and introduces a regularized approach for improved principal component estimation in the presence of outliers.
\item We focus on modeling and forecasting the conditional mean. For modeling and forecasting conditional heteroskedasticity, the functional generalized autoregressive conditional heteroskedasticity (GARCH) model may be more adequate \citep[see, e.g.,][]{AHP17, CFH+19, RWZ20, RWZ23, KSZ25}. Future work can incorporate the GARCH framework into some functional models considered.
\item Instead of recommending the best forecasting model, one could consider a forecast combination approach by averaging forecasts from different models.
\item While our results demonstrate significant improvements in statistical forecast accuracy at the subnational level, the precise economic sensitivity of annuity pricing to these specific forecast gains remains an area for future empirical research.
\end{inparaenum}

\section*{Supplementary Materials}
 This information can be found in the unblinded version.

\section*{Acknowledgment}

This information can be found in the unblinded version.

\clearpage
\section*{Appendix A: Point forecast accuracy when $K=6$}
\renewcommand{\thetable}{A\arabic{table}}
\setcounter{table}{0}

In Table~\ref{tab:appendix_1}, averaged across 47 prefectures, we compare the point forecast accuracy among the functional time-series models, when the number of retained components is set to six as in \cite{HBY13}. Based on the summary statistics of the KLD and JSD, the MFTS provides the most accurate point forecasts for the female and male data.

\begin{center}
\renewcommand*{\arraystretch}{0.99}
\tabcolsep 0.1in
\begin{longtable}{@{}llrrrrrrrr@{}}
\caption{\small Averaged across 47 prefectures, we evaluate and compare the point forecast accuracy of the functional time-series models, measured by the KLD and JSD. The number of components is determined by setting $K=6$.}\label{tab:appendix_1} \\
\toprule
&  & \multicolumn{4}{c}{Female}   & \multicolumn{4}{c}{Male}   \\
\cmidrule{3-10}
Metric & $h$ & UFTS & MFTS  & MLFTS & FANOVA & UFTS  & MFTS  & MLFTS & FANOVA \\
\midrule
\endfirsthead
\toprule
&  & \multicolumn{4}{c}{Female}   & \multicolumn{4}{c}{Male}   \\
\cmidrule{3-10}
Metric & $h$ & UFTS & MFTS  & MLFTS & FANOVA & UFTS  & MFTS  & MLFTS & FANOVA \\
\midrule
\endhead
\midrule
\multicolumn{10}{r}{{Continued on next page}} \\
\endfoot
\endlastfoot
KLD & 1 & 0.004 & 0.004 & 0.004 & 0.006 & 0.003 & 0.003 & 0.003 & 0.004 \\
    & 2 & 0.006 & 0.008 & 0.006 & 0.007 & 0.004 & 0.004 & 0.004 & 0.005 \\
    & 3 & 0.008 & 0.009 & 0.008 & 0.008 & 0.005 & 0.005 & 0.005 & 0.005 \\
    & 4 & 0.010 & 0.011 & 0.009 & 0.009 & 0.006 & 0.006 & 0.006 & 0.006 \\
    & 5 & 0.012 & 0.013 & 0.010 & 0.011 & 0.006 & 0.007 & 0.006 & 0.007 \\
    & 6 & 0.014 & 0.014 & 0.011 & 0.013 & 0.007 & 0.007 & 0.007 & 0.007 \\
    & 7 & 0.017 & 0.016 & 0.013 & 0.015 & 0.008 & 0.008 & 0.008 & 0.008 \\
    & 8 & 0.021 & 0.018 & 0.016 & 0.018 & 0.009 & 0.009 & 0.009 & 0.009 \\
    & 9 & 0.026 & 0.021 & 0.019 & 0.022 & 0.010 & 0.010 & 0.010 & 0.011 \\
    & 10 & 0.032 & 0.025 & 0.023 & 0.027 & 0.011 & 0.012 & 0.011 & 0.011 \\
    & 11 & 0.039 & 0.029 & 0.027 & 0.033 & 0.012 & 0.013 & 0.013 & 0.013 \\
    & 12 & 0.047 & 0.035 & 0.034 & 0.041 & 0.012 & 0.015 & 0.014 & 0.014 \\
    & 13 & 0.058 & 0.042 & 0.040 & 0.048 & 0.013 & 0.017 & 0.015 & 0.015 \\
    & 14 & 0.067 & 0.038 & 0.047 & 0.063 & 0.014 & 0.011 & 0.015 & 0.018 \\
    & 15 & 0.088 & 0.053 & 0.067 & 0.083 & 0.014 & 0.014 & 0.020 & 0.023 \\
    & 16 & 0.116 & 0.071 & 0.089 & 0.110 & 0.015 & 0.018 & 0.027 & 0.030 \\
    & 17 & 0.151 & 0.085 & 0.106 & 0.125 & 0.015 & 0.020 & 0.031 & 0.033 \\
\cmidrule{1-10}
    & Mean  & 0.042 & 0.029 & 0.031 & 0.038 & 0.010 & 0.011 & 0.012 & 0.013 \\
    & Median & 0.026 & 0.021 & 0.019 & 0.022 & 0.010 & 0.010 & 0.010 & 0.010 \\
\midrule
JSD & 1 & 0.032 & 0.034 & 0.033 & 0.037 & 0.027 & 0.026 & 0.026 & 0.031 \\
    & 2 & 0.040 & 0.042 & 0.039 & 0.040 & 0.033 & 0.032 & 0.032 & 0.034 \\
    & 3 & 0.044 & 0.046 & 0.043 & 0.043 & 0.035 & 0.034 & 0.034 & 0.036 \\
    & 4 & 0.048 & 0.049 & 0.047 & 0.047 & 0.038 & 0.037 & 0.037 & 0.038 \\
    & 5 & 0.052 & 0.053 & 0.050 & 0.051 & 0.040 & 0.039 & 0.039 & 0.040 \\
    & 6 & 0.057 & 0.056 & 0.053 & 0.055 & 0.043 & 0.041 & 0.041 & 0.042 \\
    & 7 & 0.062 & 0.059 & 0.057 & 0.059 & 0.045 & 0.043 & 0.043 & 0.044 \\
    & 8 & 0.067 & 0.062 & 0.061 & 0.065 & 0.048 & 0.045 & 0.045 & 0.047 \\
    & 9 & 0.074 & 0.066 & 0.066 & 0.070 & 0.051 & 0.047 & 0.048 & 0.049 \\
    & 10 & 0.081 & 0.070 & 0.072 & 0.077 & 0.053 & 0.049 & 0.050 & 0.051 \\
    & 11 & 0.090 & 0.076 & 0.079 & 0.087 & 0.055 & 0.051 & 0.052 & 0.054 \\
    & 12 & 0.101 & 0.083 & 0.088 & 0.098 & 0.057 & 0.053 & 0.055 & 0.057 \\
    & 13 & 0.115 & 0.090 & 0.098 & 0.109 & 0.059 & 0.056 & 0.058 & 0.060 \\
    & 14 & 0.128 & 0.094 & 0.109 & 0.124 & 0.060 & 0.055 & 0.061 & 0.066 \\
    & 15 & 0.149 & 0.111 & 0.130 & 0.144 & 0.061 & 0.060 & 0.069 & 0.074 \\
    & 16 & 0.175 & 0.128 & 0.150 & 0.169 & 0.063 & 0.066 & 0.079 & 0.085 \\
    & 17 & 0.201 & 0.137 & 0.164 & 0.181 & 0.063 & 0.069 & 0.085 & 0.089 \\
\cmidrule{1-10}
    & Mean & 0.089 & 0.074 & 0.079 & 0.086 & 0.049 & 0.047 & 0.050 & 0.053 \\
    & Median & 0.074 & 0.066 & 0.066 & 0.070 & 0.051 & 0.047 & 0.048 & 0.049 \\
\bottomrule
\end{longtable}
\end{center}

\newpage
\section*{Appendix B: Interval forecast accuracy when $K=6$}
\renewcommand{\thetable}{B\arabic{table}}
\setcounter{table}{0}

In Table~\ref{tab:appendix_2_K6}, averaged across 47 prefectures, we compare the interval forecast accuracy among the functional time-series models. From the summary statistics of the CPD and interval scores at the level of significance $\alpha=0.2$, the MFTS offers the smallest CPD value for the female data, while the MLFTS offers the smallest CPD values for the male data. In terms of interval scores, the MFTS yields the smallest for the female data, while FANOVA yields the smallest for the male data.
\begin{center}
\renewcommand*{\arraystretch}{0.96}
\tabcolsep 0.1in
\begin{longtable}{@{}llrrrrrrrr@{}}
\caption{\small Averaged across 47 prefectures, we evaluate and compare the interval forecast accuracy, as measured by the CPD and interval score, among the functional time-series models at the level of significance $\alpha=0.2$. The number of components is determined by setting $K=6$. For either females or males, we highlight the method with the smallest overall CPD and interval scores.}\label{tab:appendix_2_K6}\\
\toprule  
  & & \multicolumn{4}{c}{Female}   & \multicolumn{4}{c}{Male}   \\
  Metric & $h$ & UFTS & MFTS & MLFTS & FANOVA & UFTS & MFTS & MLFTS & FANOVA \\
\midrule
\endfirsthead
\toprule  
  & & \multicolumn{4}{c}{Female}   & \multicolumn{4}{c}{Male}   \\
  Metric & $h$ & UFTS & MFTS & MLFTS & FANOVA & UFTS & MFTS & MLFTS & FANOVA \\
\midrule
\endhead
\midrule
\multicolumn{10}{r}{{Continued on next page}} \\
\endfoot
\endlastfoot
CPD$_{0.2}$ & 1 & 0.052 & 0.042 & 0.045 & 0.058 & 0.054 & 0.049 & 0.046 & 0.051 \\
    & 2 & 0.059 & 0.038 & 0.046 & 0.061 & 0.056 & 0.058 & 0.050 & 0.055 \\
    & 3 & 0.066 & 0.041 & 0.057 & 0.068 & 0.055 & 0.060 & 0.052 & 0.057 \\
    & 4 & 0.073 & 0.045 & 0.061 & 0.079 & 0.058 & 0.068 & 0.056 & 0.061 \\
    & 5 & 0.072 & 0.045 & 0.059 & 0.084 & 0.058 & 0.070 & 0.054 & 0.062 \\
    & 6 & 0.080 & 0.052 & 0.065 & 0.095 & 0.061 & 0.073 & 0.058 & 0.071 \\
    & 7 & 0.084 & 0.049 & 0.066 & 0.097 & 0.059 & 0.073 & 0.054 & 0.069 \\
    & 8 & 0.089 & 0.052 & 0.068 & 0.105 & 0.061 & 0.078 & 0.055 & 0.071 \\
    & 9 & 0.090 & 0.054 & 0.073 & 0.106 & 0.070 & 0.077 & 0.060 & 0.067 \\
    & 10 & 0.101 & 0.055 & 0.082 & 0.115 & 0.071 & 0.077 & 0.068 & 0.078 \\
    & 11 & 0.108 & 0.056 & 0.087 & 0.128 & 0.075 & 0.084 & 0.076 & 0.118 \\
    & 12 & 0.109 & 0.054 & 0.087 & 0.127 & 0.074 & 0.081 & 0.071 & 0.129 \\
    & 13 & 0.113 & 0.054 & 0.076 & 0.127 & 0.071 & 0.075 & 0.075 & 0.128 \\
    & 14 & 0.119 & 0.064 & 0.085 & 0.129 & 0.067 & 0.076 & 0.081 & 0.155 \\
    & 15 & 0.120 & 0.091 & 0.093 & 0.135 & 0.065 & 0.080 & 0.084 & 0.156 \\
    & 16 & 0.109 & 0.103 & 0.104 & 0.140 & 0.062 & 0.062 & 0.088 & 0.153 \\
\midrule
    & Mean & 0.089 & 0.056 & 0.073 & 0.112 & 0.065 & 0.071 & 0.063 & 0.094 \\
    & Median & 0.090 & 0.053 & 0.067 & 0.108 & 0.062 & 0.073 & 0.058 & 0.071 \\
\midrule
$\overline{S}_{0.2}$ & 1 & 233 & 253 & 234 & 248 & 239 & 221 & 208 & 225 \\ 
                     & 2 & 281 & 292 & 275 & 301 & 268 & 252 & 237 & 251 \\  
                     & 3 & 326 & 325 & 310 & 352 & 294 & 278 & 261 & 275 \\  
                     & 4 & 374 & 357 & 346 & 406 & 324 & 305 & 285 & 299 \\ 
                     & 5 & 421 & 388 & 376 & 460 & 348 & 327 & 304 & 321 \\  
                     & 6 & 485 & 433 & 421 & 533 & 381 & 354 & 330 & 350 \\  
                     & 7 & 558 & 480 & 468 & 614 & 403 & 372 & 353 & 369 \\  
                     & 8 & 649 & 531 & 533 & 721 & 428 & 392 & 380 & 395 \\ 
                     & 9 & 734 & 587 & 602 & 825 & 450 & 408 & 404 & 418 \\ 
                     & 10 & 845 & 659 & 689 & 961 & 472 & 436 & 429 & 438 \\  
                     & 11 & 969 & 756 & 800 &1117 & 486 & 460 & 455 & 457 \\ 
                     & 12 &1134 & 848 & 896 &1305 & 513 & 485 & 471 & 487 \\  
                     & 13 &1283 & 872 & 999 &1507 & 516 & 451 & 489 & 513 \\  
                     & 14 &1508 &1035 &1220 &1804 & 552 & 504 & 572 & 573 \\  
                     & 15 &1834 &1226 &1466 &2122 & 615 & 578 & 714 & 656 \\ 
                     & 16 &2130 &1463 &1754 &2410 & 840 & 797 & 942 & 903 \\ 
\midrule 
    & Mean & 885 & 667 & 841 & 1376 & 484 & 465 & 502 & 460 \\  
    & Median & 734 & 587 & 689 & 1305 & 472 & 436 & 455 & 457 \\  
\bottomrule
\end{longtable}
\end{center}

\newpage
\section*{Appendix C: Conformal prediction intervals}
\renewcommand{\thetable}{C\arabic{table}}
\setcounter{table}{0}

In machine learning, a popular methodology known as conformal prediction \citep{SV08} is used to construct probabilistic forecasts calibrated on out-of-sample errors. From the absolute value of $[\widehat{\varepsilon}_{1,s}^g(u),\dots,\widehat{\varepsilon}_{M,s}^g(u)]$, we compute its $100(1-\alpha)\%$ quantile for a level of significance $\alpha$, denoted by $q_{\alpha}(u)$. The prediction interval can be obtained as 
\begin{equation*}
\left[\widehat{d}^g_{T+h|T,s}(u)-q_{\alpha}(u), \quad \widehat{d}^g_{T+h|T,s}(u)+q_{\alpha}(u)\right],
\end{equation*}
where $\widehat{d}^g_{T+h|T,s}(u)$ denotes the $h$-step-ahead point forecasts for the data in the testing set.

The conformal prediction approach performs well when the validation set has a reasonably large sample size \citep{DDR24}. For smaller sample sets, we recommend using the standard deviation method described in Section~\ref{sec:5}. As shown in Table~\ref{tab:appendix_3}, the MLFTS achieves the lowest CPD values under the EVR criterion, whereas the MFTS provides the smallest CPD values for $K=6$ in the female dataset. For the male dataset, MLFTS is the preferred method. Regarding interval scores, MFTS attains the lowest values under the EVR criterion, while HDFPCA performs best for $K=6$ in the female data. For males, UFTS and FANOVA are favored under the EVR criterion, with MFTS performing best for $K=6$.
\begin{center}
\renewcommand*{\arraystretch}{0.934}
\tabcolsep 0.062in
\begin{small}
\begin{longtable}{@{}lllrrrrrrrrrr@{}}
\caption{\small Averaged across 47 prefectures, we evaluate and compare the interval forecast accuracy, measured by the CPD and interval score, for the functional time-series models at the significance level $\alpha=0.2$. The number of components is determined by the EVR criterion or $K=6$; for each, we highlight the method with the smallest overall CPD and interval score values.}\label{tab:appendix_3} \\
\toprule
& & \multicolumn{5}{c}{EVR} & & \multicolumn{4}{c}{$K=6$} \\
Metric & Sex & $h$ & UFTS & MFTS & MLFTS & FANOVA & HDFPCA & UFTS & MFTS & MLFTS & FANOVA \\
\midrule
\endfirsthead
\toprule
& & \multicolumn{5}{c}{EVR} & & \multicolumn{4}{c}{$K=6$} \\
Metric & Sex & $h$ & UFTS & MFTS & MLFTS & FANOVA & HDFPCA & UFTS & MFTS & MLFTS & FANOVA \\
\midrule
\endhead
\midrule
\multicolumn{12}{r}{{Continued on next page}} \\
\endfoot
\endlastfoot
%----------------- CPD Female -----------------
CPD$_{0.2}$ & F & 1 & 0.090 & 0.109 & 0.063 & 0.090 & 0.102 & 0.067 & 0.056 & 0.054 & 0.067 \\
& & 2 & 0.105 & 0.106 & 0.065 & 0.105 & 0.052 & 0.080 & 0.057 & 0.056 & 0.080 \\
& & 3 & 0.120 & 0.099 & 0.066 & 0.120 & 0.118 & 0.091 & 0.057 & 0.062 & 0.091 \\
& & 4 & 0.136 & 0.094 & 0.067 & 0.136 & 0.044 & 0.103 & 0.059 & 0.066 & 0.103 \\
& & 5 & 0.141 & 0.084 & 0.062 & 0.141 & 0.130 & 0.109 & 0.060 & 0.064 & 0.109 \\
& & 6 & 0.159 & 0.083 & 0.063 & 0.159 & 0.052 & 0.124 & 0.062 & 0.068 & 0.124 \\
& & 7 & 0.161 & 0.080 & 0.071 & 0.161 & 0.133 & 0.133 & 0.064 & 0.073 & 0.133 \\
& & 8 & 0.163 & 0.074 & 0.071 & 0.163 & 0.060 & 0.138 & 0.063 & 0.071 & 0.138 \\
& & 9 & 0.172 & 0.077 & 0.075 & 0.172 & 0.130 & 0.147 & 0.068 & 0.077 & 0.147 \\
& & 10 & 0.183 & 0.080 & 0.081 & 0.183 & 0.054 & 0.161 & 0.075 & 0.084 & 0.161 \\
& & 11 & 0.192 & 0.081 & 0.096 & 0.192 & 0.135 & 0.171 & 0.084 & 0.095 & 0.171 \\
& & 12 & 0.199 & 0.095 & 0.106 & 0.199 & 0.055 & 0.178 & 0.095 & 0.108 & 0.178 \\
& & 13 & 0.198 & 0.095 & 0.107 & 0.198 & 0.168 & 0.176 & 0.099 & 0.112 & 0.176 \\
& & 14 & 0.221 & 0.115 & 0.129 & 0.221 & 0.082 & 0.188 & 0.115 & 0.132 & 0.188 \\
& & 15 & 0.256 & 0.158 & 0.162 & 0.256 & 0.241 & 0.219 & 0.158 & 0.164 & 0.219 \\
& & 16 & 0.299 & 0.190 & 0.186 & 0.299 & 0.138 & 0.256 & 0.191 & 0.201 & 0.256 \\
\cmidrule{1-12}
& & Mean & 0.184 & 0.103 & 0.096 & 0.184 & 0.117 & 0.152 & 0.092 & 0.101 & 0.152 \\
& & Median & 0.177 & 0.094 & 0.078 & 0.177 & 0.118 & 0.136 & 0.071 & 0.081 & 0.136 \\
\cmidrule{1-12}
%----------------- CPD Male -----------------
& M & 1 & 0.094 & 0.122 & 0.062 & 0.094 & 0.090 & 0.058 & 0.064 & 0.063 & 0.058 \\
& & 2 & 0.113 & 0.127 & 0.072 & 0.113 & 0.111 & 0.070 & 0.077 & 0.068 & 0.070 \\
& & 3 & 0.119 & 0.130 & 0.076 & 0.119 & 0.115 & 0.076 & 0.083 & 0.070 & 0.076 \\
& & 4 & 0.125 & 0.137 & 0.083 & 0.125 & 0.117 & 0.083 & 0.091 & 0.077 & 0.083 \\
& & 5 & 0.123 & 0.137 & 0.087 & 0.123 & 0.122 & 0.090 & 0.098 & 0.080 & 0.090 \\
& & 6 & 0.133 & 0.146 & 0.097 & 0.133 & 0.137 & 0.105 & 0.109 & 0.090 & 0.105 \\
& & 7 & 0.134 & 0.139 & 0.099 & 0.134 & 0.142 & 0.111 & 0.107 & 0.090 & 0.111 \\
& & 8 & 0.126 & 0.132 & 0.095 & 0.126 & 0.139 & 0.117 & 0.112 & 0.089 & 0.117 \\
& & 9 & 0.116 & 0.127 & 0.093 & 0.116 & 0.139 & 0.118 & 0.110 & 0.087 & 0.118 \\
& & 10 & 0.119 & 0.133 & 0.096 & 0.119 & 0.154 & 0.119 & 0.117 & 0.092 & 0.119 \\
& & 11 & 0.114 & 0.129 & 0.100 & 0.114 & 0.160 & 0.118 & 0.118 & 0.099 & 0.118 \\
& & 12 & 0.110 & 0.130 & 0.102 & 0.110 & 0.177 & 0.122 & 0.118 & 0.103 & 0.122 \\
& & 13 & 0.100 & 0.121 & 0.093 & 0.100 & 0.172 & 0.121 & 0.114 & 0.091 & 0.121 \\
& & 14 & 0.104 & 0.130 & 0.105 & 0.104 & 0.187 & 0.125 & 0.125 & 0.099 & 0.125 \\
& & 15 & 0.115 & 0.143 & 0.118 & 0.115 & 0.196 & 0.133 & 0.136 & 0.114 & 0.133 \\
& & 16 & 0.134 & 0.183 & 0.151 & 0.134 & 0.234 & 0.156 & 0.179 & 0.151 & 0.156 \\
\cmidrule{1-12}
& & Mean & 0.120 & 0.135 & 0.097 & 0.120 & 0.142 & 0.104 & 0.107 & 0.092 & 0.104 \\
& & Median & 0.119 & 0.133 & 0.096 & 0.119 & 0.138 & 0.118 & 0.113 & 0.091 & 0.118 \\
\cmidrule{1-12}
%----------------- Interval Score Female -----------------
\cmidrule{1-12}
$\overline{S}_{0.2}$ & F & 1 & 286 & 423 & 247 & 286 & 331 & 253 & 256 & 237 & 253 \\
& & 2 & 338 & 459 & 285 & 338 & 323 & 308 & 297 & 280 & 308 \\
& & 3 & 393 & 489 & 319 & 393 & 411 & 360 & 331 & 316 & 360 \\
& & 4 & 457 & 521 & 354 & 457 & 349 & 419 & 366 & 353 & 419 \\
& & 5 & 514 & 548 & 386 & 514 & 470 & 474 & 398 & 386 & 474 \\
& & 6 & 593 & 573 & 429 & 593 & 393 & 551 & 438 & 436 & 551 \\
& & 7 & 671 & 601 & 477 & 671 & 549 & 632 & 481 & 486 & 632 \\
& & 8 & 784 & 627 & 536 & 784 & 451 & 741 & 521 & 549 & 741 \\
& & 9 & 897 & 660 & 600 & 897 & 624 & 855 & 570 & 617 & 855 \\
& & 10 & 1050 & 715 & 676 & 1050 & 477 & 1001 & 633 & 707 & 1001 \\
& & 11 & 1217 & 778 & 766 & 1217 & 735 & 1170 & 710 & 804 & 1170 \\
& & 12 & 1396 & 877 & 862 & 1396 & 513 & 1356 & 802 & 908 & 1356 \\
& & 13 & 1588 & 865 & 954 & 1588 & 968 & 1563 & 820 & 1003 & 1563 \\
& & 14 & 1872 & 1015 & 1181 & 1872 & 570 & 1843 & 994 & 1237 & 1843 \\
& & 15 & 2185 & 1199 & 1387 & 2185 & 1368 & 2153 & 1201 & 1469 & 2153 \\
& & 16 & 2528 & 1343 & 1552 & 2528 & 619 & 2475 & 1358 & 1665 & 2475 \\
\cmidrule{1-12}
& & Mean & 1092 & 731 & 740 & 1092 & 603 & 1095 & 630 & 742 & 1095 \\
& & Median & 840 & 644 & 638 & 840 & 495 & 928 & 546 & 677 & 928 \\
\cmidrule{1-12}
%----------------- Interval Score Male -----------------
& M & 1 & 259 & 280 & 221 & 259 & 251 & 224 & 222 & 209 & 224 \\
& & 2 & 281 & 309 & 247 & 281 & 286 & 249 & 253 & 238 & 249 \\
& & 3 & 296 & 335 & 268 & 296 & 314 & 272 & 278 & 262 & 272 \\
& & 4 & 312 & 359 & 287 & 312 & 334 & 291 & 303 & 284 & 291 \\
& & 5 & 320 & 378 & 301 & 320 & 353 & 311 & 325 & 302 & 311 \\
& & 6 & 337 & 399 & 319 & 337 & 383 & 337 & 350 & 325 & 337 \\
& & 7 & 351 & 412 & 337 & 351 & 413 & 361 & 367 & 347 & 361 \\
& & 8 & 366 & 424 & 359 & 366 & 438 & 385 & 383 & 372 & 385 \\
& & 9 & 379 & 435 & 378 & 379 & 462 & 404 & 396 & 390 & 404 \\
& & 10 & 394 & 457 & 397 & 394 & 504 & 420 & 420 & 412 & 420 \\
& & 11 & 414 & 475 & 426 & 414 & 548 & 440 & 444 & 439 & 440 \\
& & 12 & 440 & 507 & 445 & 440 & 626 & 468 & 472 & 456 & 468 \\
& & 13 & 456 & 454 & 454 & 456 & 699 & 485 & 432 & 464 & 485 \\
& & 14 & 500 & 498 & 545 & 500 & 839 & 528 & 481 & 551 & 528 \\
& & 15 & 562 & 555 & 661 & 562 & 977 & 586 & 543 & 657 & 586 \\
& & 16 & 609 & 600 & 744 & 609 & 1094 & 640 & 595 & 730 & 640 \\
\cmidrule{1-12}
& & Mean & 393 & 419 & 396 & 393 & 516 & 395 & 387 & 414 & 395 \\
& & Median & 372 & 418 & 378 & 372 & 450 & 400 & 375 & 381 & 400 \\
\bottomrule
\end{longtable}
\end{small}
\end{center}

\vspace{-.8in}

\section*{Appendix D: Model Confidence Set (MCS) Analysis}
\label{sec:appendixD}
\renewcommand{\thetable}{D\arabic{table}}
\setcounter{table}{0}
\renewcommand{\thefigure}{D\arabic{figure}}
\setcounter{figure}{0}

We performed a rigorous statistical comparison of the point forecast accuracy across all competing models using the MCS procedure \citep{MCS}. The MCS procedure identifies a set of superior models, $\mathcal{M}^*$, where the null hypothesis of Equal Predictive Ability cannot be rejected at a given significance level $\alpha$.

The analysis was based on the point forecast errors measured by the \textbf{KLD} for nine FTS models over 17 forecast horizons ($h=1$ to $h=17$). The procedure was applied to the forecast errors for 47 Japanese prefectures separately for male and female mortality data. We used a significance level of $\alpha = 0.10$, a block-bootstrap approach with $B=2000$ resamples, and the $T_{\max}$ statistic.

\begin{figure}[!htb]
\centering
\subfloat[Female data]
{\includegraphics[width=8cm]{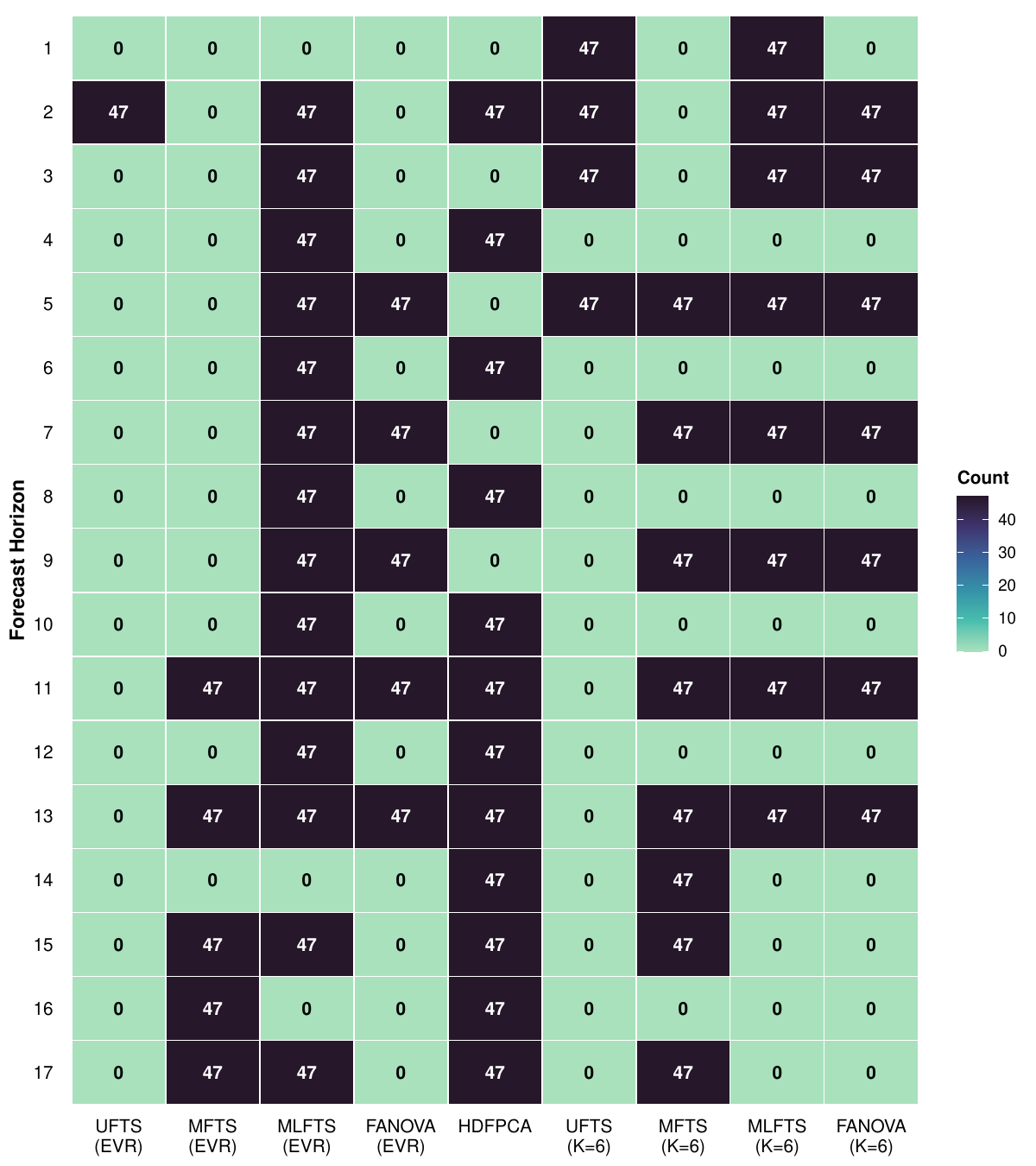}}
\quad
\subfloat[Male data]
{\includegraphics[width=8cm]{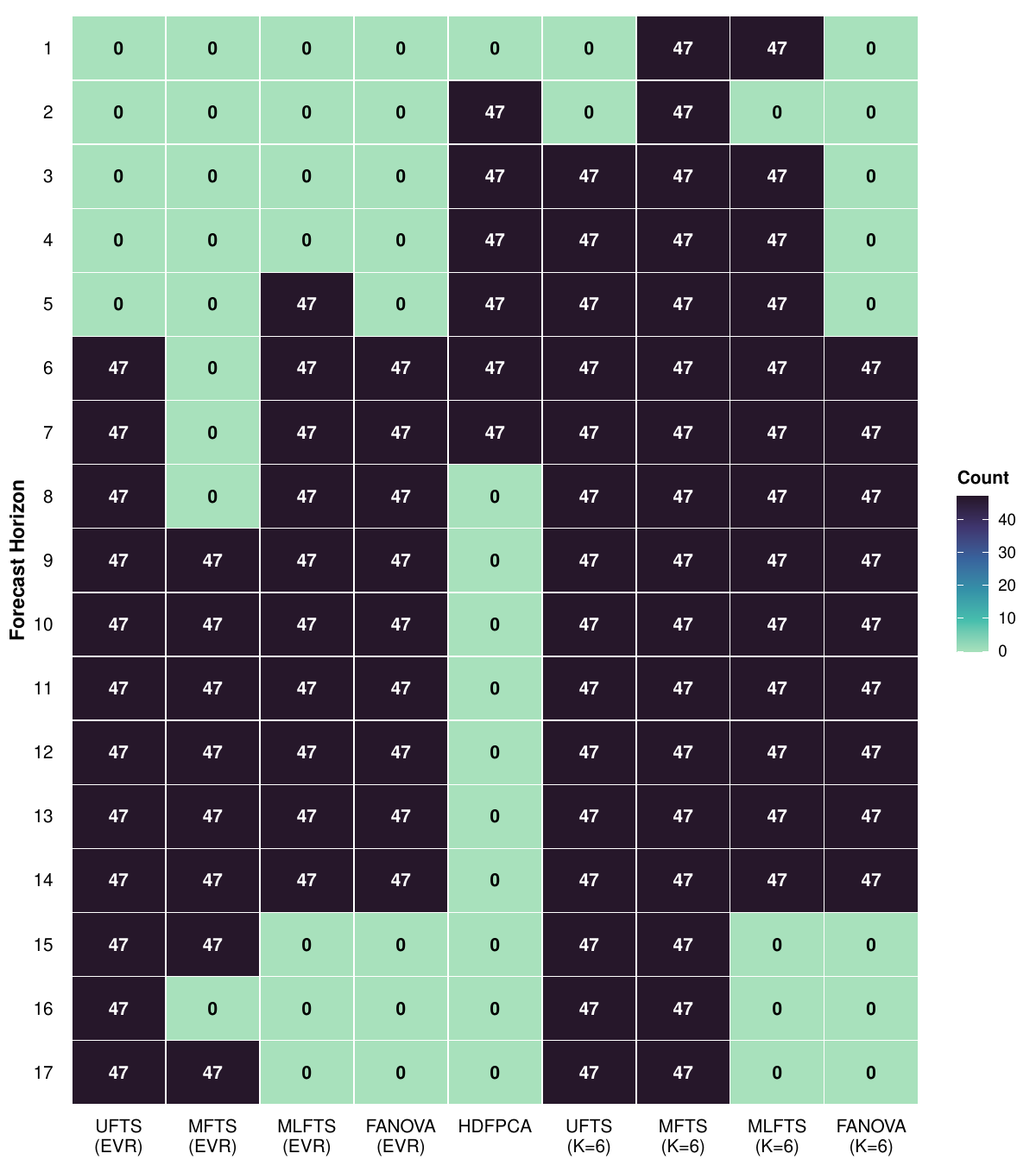}}
\caption{\small Summary of MCS procedure results for point forecast accuracy (KLD). Cell values indicate the number of prefectures where a model was included in the superior set $\mathcal{M}^*$ for horizons $h=1$ to $17$.}\label{fig:AD}
\end{figure}
The heatmaps in Figure~\ref{fig:AD} visualize the results of the MCS procedure. In each cell, the count represents the number of prefectures (out of 47) for which the corresponding model was included in the statistically superior set $\mathcal{M}^*$ at that horizon, demonstrating the robust predictive ability of the models across the regional series. For the female data, the MLFTS method consistently achieved statistical superiority across most prefectures at shorter horizons. Conversely, the HDFPCA method proved most robust at longer horizons. For the male dataset, the MFTS method was consistently included in the superior set $\mathcal{M}^*$ for both short and long forecast horizons.

%\newpage
\bibliographystyle{agsm}
\bibliography{CDF_subnational}

\end{document}